\documentclass[aps,rmp,preprint,groupedaddress,floatfix]{revtex4-1}
\usepackage{epsfig}
\usepackage{natbib}
\usepackage{genetics}

\begin{document}

\newcommand{\mon}{\begin{displaymath}}
\newcommand{\moff}{\end{displaymath}}
\renewcommand{\b}[1]{\mbox{\boldmath ${#1}$}}
\newcommand{\pd}[2]{\frac{\partial {#1}}{\partial {#2}}}
\newcommand{\od}[2]{\frac{d {#1}}{d {#2}}}
\newcommand{\inti}{\int_{-\infty}^{\infty}}
\newcommand{\eon}{\begin{equation}}
\newcommand{\eoff}{\end{equation}}
\newcommand{\eaon}{\begin{eqnarray}}
\newcommand{\eaoff}{\end{eqnarray}}
\newcommand{\e}[1]{\times 10^{#1}}
\newcommand{\chem}[2]{{}^{#2} \mathrm{#1}}
\renewcommand{\sb}{s}
\newcommand{\s}{s}
\newcommand{\eq}[1]{Eq. (\ref{#1})}
\newcommand{\ev}[1]{\langle #1 \rangle}
\newcommand{\mat}[1]{\bf{\mathcal{#1}}}
\newcommand{\fig}[1]{Fig. \ref{#1}}
\newcommand{\degrees}{\,^{\circ}\mathrm{C}}
\renewcommand{\log}{\ln}
\renewcommand{\sec}[1]{section \ref{#1}}
\newcommand{\pr}[1]{P_{#1}}
\newcommand{\like}{\mathcal{L}}
\newcommand{\binom}[2]{{#1 \choose #2}}

\newcommand{\kav}{k_{\text{av}}}
\newcommand{\defeq}{\equiv}
\newcommand{\ud}{U_d}
\newcommand{\un}{U_n}
\newcommand{\thetad}{\theta_d}
\newcommand{\thetan}{\theta_n}
\renewcommand{\L}{\zeta}
\newcommand{\pck}[2]{P_c^{#1, #1 \to #2}}
\newcommand{\pckk}[3]{P_c^{#1, #2 \to #3}}
\newcommand{\pc}{P_c}
\newcommand{\probtimes}[2]{Q^{#1}_{#2}}
\newcommand{\probtimescond}[2]{R^{#1}_{#2}}
\newcommand{\g}[1]{G_{#1}}
\newcommand{\cjk}[2]{C_{#1}^{#2}}
\newcommand{\cjl}[2]{C_{#1}^{#2}}
\newcommand{\problap}[2]{\tilde{Q}^{#1}_{#2}}
\newcommand{\aklm}[3]{A_{#3}^{#1, #2}}
\newcommand{\st}{\tau}
\newcommand{\pst}[3]{\phi_{#1}^{#2}(#3)}
\newcommand{\psttotal}[1]{\phi (\st = #1)}
\newcommand{\hdistn}{H}
\newcommand{\ppid}{\rho}
\newcommand{\ppin}{\rho}
\newcommand{\probtau}{\phi}
\newcommand{\probt}{\psi}
\newcommand{\ptauthree}{\chi}
\newcommand{\psthree}{\zeta}
\newcommand{\minpair}{\pi_{3m}}
\newcommand{\minpaird}{\pi_{3m}^d}
\newcommand{\infinity}{\infty}
\renewcommand{\Pi}{\pi}
\newcommand{\boxed}{}
\renewcommand{\otimes}{\star}

\renewcommand{\baselinestretch}{1.0}

\title{The Structure of Genealogies in the Presence of Purifying Selection:  A ``Fitness-Class Coalescent''}

\author{Aleksandra M. Walczak$^{1,*}$}
\author{Lauren E. Nicolaisen$^{2,*}$}
\author{Joshua B. Plotkin$^{3}$}
\author{Michael M. Desai$^{2}$}
\affiliation{\mbox{${}^{1}$CNRS-Laboratoire de Physique Th\'eorique de l'\'Ecole Normale Sup\'erieure,} \\ \mbox{${}^2$Department of Organismic and Evolutionary Biology, Department of Physics, and} \mbox{FAS Center for Systems Biology, Harvard University} \\ \mbox{${}^3$Department of Biology, University of Pennsylvania} \\ \mbox{${}^*$These authors contributed equally to this work}}

\begin{abstract}
Compared to a neutral model, purifying selection distorts the structure of genealogies and hence alters the patterns of sampled genetic variation.  Although these distortions may be common in nature, our understanding of how we expect purifying selection to affect patterns of molecular variation remains incomplete.  Genealogical approaches such as coalescent theory have proven difficult to generalize to situations involving selection at many linked sites, unless selection pressures are extremely strong.  Here, we introduce an effective coalescent theory (a ``fitness-class coalescent'') to describe the structure of genealogies in the presence of purifying selection at many linked sites.  We use this effective theory to calculate several simple statistics describing the expected patterns of variation in sequence data, both at the sites under selection and at linked neutral sites.  Our analysis combines our earlier description of the allele frequency spectrum in the presence of purifying selection \citep{allelebased} with the structured coalescent approach of \citet{nordborg97}, to trace the ancestry of individuals through the distribution of fitnesses within the population. Alternatively, we can derive our results using an extension of the coalescent approach of \citet{hudsonkaplan94}.  We find that purifying selection leads to patterns of genetic variation that are related but not identical to a neutrally evolving population in which population size has varied in a specific way in the past.
\end{abstract}

\date{\today}

\maketitle

Running Head:  Coalescent Theory with Purifying Selection


Corresponding Author:

Michael M. Desai

Departments of Organismic and Evolutionary Biology and of Physics

FAS Center for Systems Biology

Harvard University

435.20 Northwest Labs

52 Oxford Street

Cambridge, MA 02138

617-496-3613

mdesai@oeb.harvard.edu

\clearpage

\newpage

\section{Introduction}

Purifying selection acting simultaneously at many linked sites (``background selection'')
can substantially alter the patterns of molecular variation at these sites, and at linked
neutral sites \citep{hillrobertson66, kaplanhudson88, hudsonkaplan94, hudsonkaplan95,
mcveancharlesworth00, hahn08, gordocharlesworth02, seger10, ofallonadler10, EtheridgePMID19341750, EtheridgePMID20685218}.  In recent
years, evidence from sequence data points to the general importance of weak selective
forces among many linked variants in microbial and viral populations, and on short
distance scales in the genomes of sexual organisms \citep{hahn08, comeron08, seger10}.  In
these situations, existing theory does not fully explain patterns of molecular evolution
\citep{hahn08}.

It is difficult to incorporate negative selection at many linked sites into genealogical
frameworks such as coalescent theory, since these frameworks typically rely on
characterizing the space of possible genealogical trees \emph{before} considering the
possibility of mutations at various locations on these trees.  When selection operates,
the probabilities of particular trees cannot be defined independently of the mutations,
and the approach breaks down \citep{wakeleybook, tavarebook}.

Despite this difficulty, a number of productive approaches have been developed to predict
how negative selection influences patterns of molecular variation and to infer selection
pressures from data.  \citet{charlesworth93} showed that strong purifying selection
reduces the effective population size relevant for linked neutral sites
\citep{charlesworth94, charlesworth95}.  However, weaker selection also distorts patterns of
variation, in a way that cannot be completely described by a neutral model with any effective
population size \citep{mcveancharlesworth00, comeronkreitman02} -- a phenomenon often
referred to as Hill-Robertson interference \citep{hillrobertson66}.  Several theoretical
frameworks have been developed to analyze this situation.  The ancestral selection graph
of \citet{neuhauserkrone97} and \citet{kroneneuhauser97} provides an elegant formal
solution to the problem, but unfortunately it requires extensive numerical calculations
\citep{przeworski99}.  These limit the intuition we can draw from this method, and make it
impractical as the basis for inference from most modern sequence data.  An alternative
approach is based on the structured coalescent of \citet{nordborg97}, which views the
population as subdivided into different fitness classes and traces the genealogies of
individuals as they move between classes. This approach was first introduced by
\citet{kaplanhudson88} and \citet{hudson90} and further developed by
\citet{hudsonkaplan94} and \citet{hudsonkaplan95}.  It has been the basis for
computational methods developed by \citet{gordocharlesworth02} and \citet{seger10} and
analytical approaches such as those of \citet{bartonetheridge04}, \citet{hermisson02}, and
\citet{ofallonadler10}.

In this paper, we build on the structured coalescent framework by introducing the idea of
a ``fitness-class coalescent.''  Rather than considering the coalescence process in real
time, we treat each fitness class as a ``generation'' and trace how individuals have
descended by mutations through fitness classes, moving from one ``generation'' to the next
by subsequent mutations.  We show that the coalescent probabilities in this fitness-class
coalescent can be computed using an approach based on the Poisson Random Field method of
\citet{sawyerhartl92}, or equivalently can be exactly derived as an extension of the
structured coalescent approach of \citet{hudsonkaplan94}.

Our fitness-class coalescent theory can be precisely mapped to a coalescence theory in which certain quantities (e.g. coalescence times) have different meanings than in the traditional theory.  We can then invert this mapping to determine the structure of genealogies and calculate statistics describing expected patterns of genetic variation. This approach requires certain approximations, but it also has several advantages.  Most importantly, we are able to derive relatively simple analytic expressions for coalescent probabilities and distributions of simple statistics such as heterozygosity.  Consistent with earlier work, we find that the effects of purifying selection are broadly similar to an effective population size that changes as time recedes into the past.  Our analysis makes this analysis precise and quantitative:  we can compute the exact form of this time-varying effective population size.  We also show that this intuition has important limitations: for example, different pairs of individuals have different time-varying effective population size histories, meaning that in principle it may be possible to distinguish selection from changing population size.  Our approach also makes it possible to calculate the diversity created at the selected sites themselves, which may be important when selection is common.

We begin in the next section by describing the fitness-class coalescent idea which underlies our approach.  We then describe the details of our model and analyze two alternative ways to implement the fitness-class coalescent.  The first relies on the framework developed in \citet{allelebased} to calculate the frequency distribution of distinct lineages within each fitness class.  This provides a simple intuitive framework for computing the structure of genealogies, but is algebraically involved.  The second approach is based on tracing paths in the order that events occur as described by \citet{hudsonkaplan94}, and implemented numerically by \citet{gordocharlesworth02}.  This approach has the advantage of algebraic simplicity, and it provides a correspondence between our analytical results and earlier structured coalescent methods.  However, it is unwieldy to generalize to other types of selection and is less intuitive in certain respects.  We show how both approaches can be used to analyze the structures of genealogies, and we calculate various statistics describing genetic variation in these populations, which we compare to numerical simulations.  We finally discuss the relationship between our results, neutral theory, and earlier work on selection, and we explore how various approximations limit our approach.  The most important of these approximations is that we neglect Muller's ratchet.  We discuss this and related approximations briefly in the next section, and justify their regime of validity in more detail in the Discussion.

\section{The fitness-class coalescent}

We begin in this section by outlining the main ideas underlying our approach. We begin our analysis by considering the balance between mutations at
many linked sites and negative selection against the mutants, which leads to an equillibrium distribution of fitnesses within a population \citep{haigh78}.  We illustrate this in \fig{fig1}, for the case in which all deleterious mutations have the same fitness cost.  Each individual is characterized by the number $k$ of deleterious mutations it contains.  Each fitness class $k$ contains many genetically distinct lineages, each of which arose from mutations in more-fit individuals, as illustrated in \fig{fig2}.

\citet{hudsonkaplan94} observed that individual lineages move between fitnesses by mutations, and that when two individuals are in the same fitness class they could be from the same lineage and hence coalesce.  Our fitness-class coalescent exploits this observation to define an effective genealogical process that completely bypasses the ancestral process in real time.  Instead, we treat each fitness class as a ``generation,'' and we count time in deleterious mutations:  each deleterious mutation moves us from one ``generation'' to the next.  In this way, we can trace the ancestry of individuals through the fitness distribution.  For example, there is some probability that two individuals chosen from fitness class $k$ are genetically identical (i.e. come from the same lineage).  If not, they each arose from mutations within fitness class $k-1$.  If both those mutations occurred in individuals in the same lineage in fitness class $k-1$, we say the two individuals ``coalesced'' in class $k-1$.  If not, they came from different mutations from class $k-2$, and could have coalesced there, and so on.  In this way, we can construct a fitness-class coalescent tree describing the relatedness of two individuals, as illustrated in \fig{fig2}.

In this paper we show that the probability that two randomly chosen individuals who are currently in fitness classes $k$ and $k'$ coalesce in class $k-\ell$, $\pckk{k}{k'}{k-\ell}$, is approximately \eon \pckk{k}{k'}{k-\ell} = \frac{1}{2 n_{k-\ell} s_{k-\ell}} \aklm{k}{k'}{\ell}, \label{coalprob} \eoff where $n_k$ is the population size of fitness class $k$, $s_k$ is an effective selection pressure against these individuals, and \eon \aklm{k}{k'}{\ell} = \frac{{k' \choose k-\ell} {k \choose k-\ell}}{{k + k' \choose 2 \ell + k' - k}}.  \eoff

This coalescent probability is inversely proportional to the population size of the fitness class, $n_{k-\ell}$, and the effective selection coefficient within that class, $s_{k-\ell}$, modified by the combinatoric coefficient $A_{\ell}^{k,k'}$.  As we will see, this has a clear intuitive interpretation.  Fitness class $k-\ell$ has size $n_{k-\ell}$, so the coalescence probability per real generation is $\frac{1}{n_{k-\ell}}$.  We will see that each lineage spends of order $s_{k-\ell}$ generations in that class, so the total coalescence probability in this class has the form $\frac{1}{n_{k-\ell}} \frac{1}{s_{k-\ell}}$.  This is multiplied by $A_{\ell}^{k, k'}/2$, which we will show describes the probability that the two individuals are in class $k-\ell$ at the same time.  In other words, the probability coalescence occurs in a class equals the inverse population size of the class times the number of generations lineages spend together in that class.  In the following sections of this paper we derive Eq. \ref{coalprob} in the two alternative ways mentioned in the Introduction: by explicitly considering the lineage frequency distribution and by following the path summation method of \citet{hudsonkaplan94}, \citet{gordocharlesworth02}, and \citet{bartonetheridge04}.

\subsection{Calculating statistics describing sequence variation}
Our approach of treating mutation events as timesteps, and computing coalescence probabilities at each timestep, allows us to make a precise mapping to coalescence theory in which certain quantities have a different meaning than in the traditional theory.  In this framework, we can calculate a simple analytic expression for the probability two lineages sampled from particular fitness classes will coalesce in any other fitness class.  These fitness-class coalescence probabilities allow us to explicitly calculate the structure of genealogies in this ``mutation time.''  We can then compute the distribution of any statistic describing expected sequence variation by averaging over the fitness classes our original individuals come from.  For a statistic $x$ that depends on genealogies between two individuals, for example, we write expressions of the form \eon P(x) = \sum H(k, k') \mathrm{Prob}[k, k' \textrm{ coalesce in } k - \ell] P(x|k, k', \ell), \label{avoverdistn} \eoff where $H(k, k')$ describes the probability two individuals sampled at random from the population come from classes $k$ and $k'$ respectively.

From the form of these expressions and our simple result for the coalescence probabilities, we can immediately see the main effect of selection on the structure of genealogies.  The discussion following \eq{coalprob} implies that the effect of negative selection is similar to that of an effective population size that changes as time recedes into the distant past --- i.e. some $N_e(t)$.  This intuition has been suggested by earlier work (see e.g. \citet{seger10}).  As we will see, our analysis describes the precise form of $N_e(t)$:  it follows the distribution $n_{k-\ell}$ as $\ell$ increases further to the past, modified by the coefficient $A_{\ell}^{k,k'}$. We will also see that this picture of time-varying population size has limits:  different pairs of individuals have a different $N_e(t)$.  As is clear from \eq{avoverdistn}, these different histories are averaged according to the distribution $H(k, k')$.  While it is the average $N_e(t)$ between pairs that determines the distribution of pairwise statistics, this lack of a single $N_e(t)$ describing all individuals means that statistical power may exist in larger samples to distinguish negative selection from neutral population expansion.  We explore these general conclusions of our analysis in detail in the Discussion.

Note that in the standard neutral coalescent, one first calculates the distribution of coalescence times and then imagines mutations occurring as a Poisson process throughout the coalescent tree, with rates proportional to branch lengths.  In our fitness-class coalescent, by contrast, the coalescence times \emph{are} the mutations.  To avoid confusion, from here on we will refer to the effective ``generations'' in our model as ``steps,'' and refer to the fitness-class coalescent ``times'' as the ``steptimes.''  We will reserve the word ``time'' to refer to the actual coalescent time, measured in actual generations.

After determining a fitness-class coalescent tree, we can invert our mapping to determine the structure of genealogies in real time.  We will do this by calculating how the steptime in our fitness-class coalescent model translates into an actual time in generations.  This will allow us to relate the distribution of branch lengths in steptimes to an actual coalescent tree in generations.  We can then treat neutral mutations as is usually done in the standard coalescent: as a Poisson process with probabilities proportional to branch lengths.

Our fitness-time coalescent requires a number of approximations which limit its applicability.  Most importantly, we neglect Muller's ratchet, and more generally ignore the effects of fluctuations in the size of each fitness class.  We have considered these approximations in \citet{allelebased}, and return to consider them in more detail in the Discussion.  We find that within a broad and biologically relevant parameter regime they lead to systematic but small corrections to our results.  Despite these limitations, our approach also has several advantages relative to previous work.  The fitness-time coalescent approach makes many otherwise difficult analytic calculations tractable, allows us to compute the diversity at the selected sites in addition to linked neutral sites, and may offer a useful basis for practical methods of coalescent simulation and inference.

\section{Model}

We now turn to the details of our model, which is identical to the model we studied in \citet{allelebased}.  We imagine a finite haploid population of constant size $N$. Each individual has a genome composed of a large number of sites.  Each site is assumed to begin in some ancestral state, and can mutate with some constant rate.  Each mutation is assumed to be either neutral or to confer some fitness disadvantage $s$ (where by convention $s > 0$).  We work within an infinite-sites approximation, where the probability that two mutations at the same site segregate simultaneously within the population is negligible.

We assume that there is no epistasis for fitness, so each deleterious mutation contributes multiplicatively to the fitness of each individual.  We assume that all deleterious mutations carry the same fitness cost $s$, and that $s \ll 1$, so that the fitness of an individual with $k$ deleterious mutations is approximately $w_k = 1 - s k$.

The dynamics of competing individuals are assumed to follow the diffusion limit of the standard Wright-Fisher model. In each generation an individual acquires a new deleterious mutation, somewhere in its genome, with probability $\ud$. Thus, $\thetad/2 \defeq N \ud$ is the per-genome scaled deleterious mutation rate. Similarly, neutral mutations occur at a rate $\un$ per individual per generation, and we define $\thetan/2 \defeq N \un$. Whenever a mutation arises, it is assumed to arise at site for which there are no other segregating polymorphisms in the population (the infinite-sites assumption).  We focus exclusively on the case of perfect linkage, where we imagine that all the sites we are considering are in an asexual genome or within a short enough distance in a sexual genome that recombination can be entirely neglected. Although our model is defined for haploids, this assumption means that our analysis also applies to diploid populations provided that there is no dominance (i.e. being homozygous for the deleterious mutation carries twice the fitness cost as being heterozygous).  In this case, our model is equivalent to that considered by \citet{hudsonkaplan94}.

For the bulk of this paper, we will assume that Muller's ratchet can be neglected.  While this assumption presented minimal problems in the context of the allele-based analysis in \citet{allelebased}, it is more problematic here.  Thus we will return to the question of the importance of Muller's ratchet in more detail in the Discussion.

We believe that our model is the simplest possible null model based on a concrete picture of mutations at individual sites that can describe the effects of a large number of linked negatively selected sites on patterns of genetic variation.  In \citet{allelebased} we discuss in more detail its relationship with other models which have been introduced in earlier related work.

\section{Allelic Diversity in the Deleterious Mutation-Selection Balance}

Our analysis aims to develop a fitness-class coalescent theory that involves tracing the ancestry of individuals as they change in fitness by acquiring deleterious mutations.  In order to do this, we need to first understand the distribution of fitnesses within the population and the structure of lineage diversity amongst individuals within a given fitness class.  We have analyzed these topics in detail in \citet{allelebased}.  Here we briefly summarize the results relevant for our subsequent coalescent analysis.

\begin{figure}
\includegraphics[scale=1]{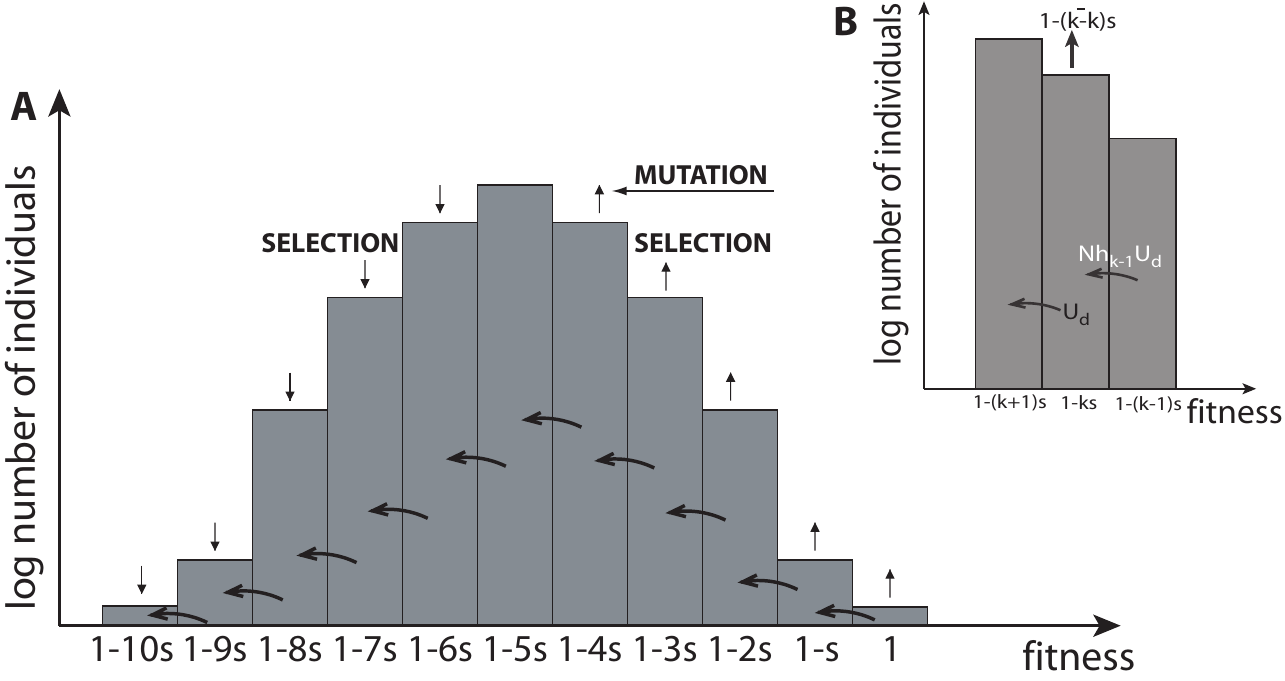}
\caption{The distribution of the fraction of the population in each fitness class.  \textbf{(a)} The distribution of the number of individuals as a function of fitness, where the most beneficial class is arbitrarily defined to have fitness $1$, and each deleterious mutation introduces a fitness disadvantage of $s$.  Mutations move individuals to less-fit classes, and selection balances this by favoring the classes more fit than average. The shape of the depicted steady state distribution is a result of this mutation--selection balance. The inset \textbf{(b)} shows the processes which lead to this balance within a given fitness class; this is explored in more detail in \citet{allelebased}.}
\label{fig1} \end{figure}

\begin{figure}
\includegraphics[width=6.5in]{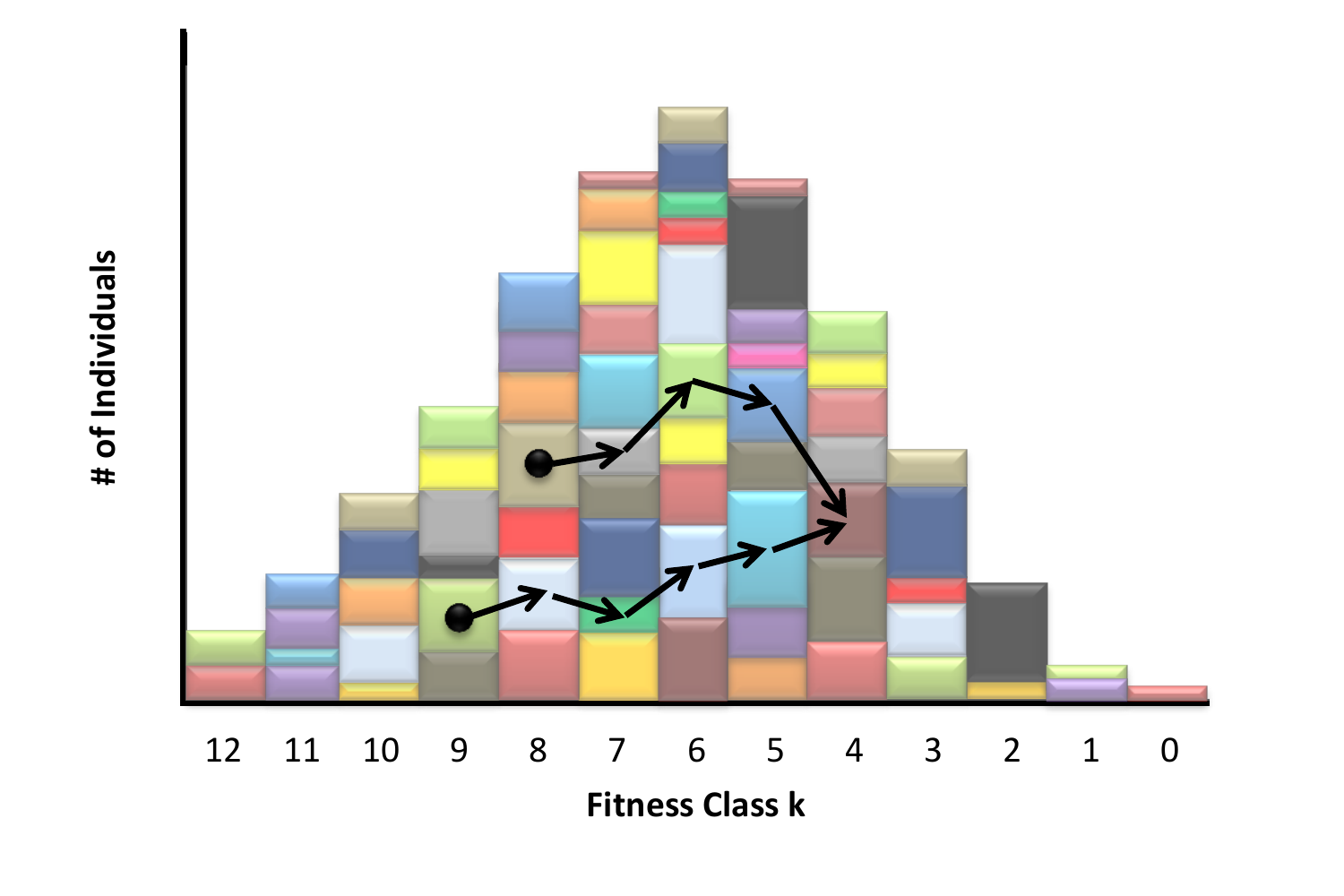}
\caption{Each fitness class in the population is composed of many lineages, each of which was created by a single mutation and is (in our infinite-sites model) genetically unique.  In \citet{allelebased} we described the distribution of lineage frequencies within each fitness class.  Shown is a schematic cartoon in which each lineage is depicted in a different color. The arrows denote an example of the fitness-class coalescence process for two individuals sampled from classes 8 and 9.  These individuals came from different lineages, and these lineages were created by mutations from different lineages within the next most-fit class (as shown by the arrows).  The arrows trace the ancestry of the two individuals back through the different lineages that successively founded each other, until they finally coalesce in the class third from right.  }
\label{fig2}
\end{figure}

In our model all deleterious mutations have the same fitness cost $s$, and so we can classify individuals based on their Hamming class, $k$, relative to the wildtype (which by definition has $k=0$).  That is, individuals in class $k$ have $k$ deleterious mutations more than the most-fit individuals in the population.  Note that not all individuals in class $k$ have the same set of $k$ deleterious mutations. Furthermore, $k$ refers only to the number of \emph{deleterious} mutations an individual has; individuals with the same $k$ can have different numbers of neutral mutations.  We normalize fitness such that by definition all individuals in class $k = 0$ have fitness 1.  Individuals in class $k$ then have fitness $1 - ks$ (\fig{fig1}).

We showed in \citet{allelebased} that the balance between mutation and selection leads to a steady state in which the fraction of the population in fitness class $k$, which we call $h_k$, is given by a Poisson distribution with mean $\ud/s$, \eon \label{hk} h_k = \frac{ e^{-\ud/s}}{k!} \left( \frac{\ud}{s} \right)^k. \eoff  This is consistent with the earlier work by \citet{haigh78}, and means that the average fitness in the population is $1 - \ud$, and that $\bar k = \frac{\ud}{s}$.

We will later need to understand the distributions of timings, $\probtimes{k-1}{k}(t)$, at which an individual mutates from class $k-1$ to class $k$.  We can calculate this by noting that the probability that an individual in class $k$ arose from a mutation in an individual in class $k-1$ rather than a reproduction event from an individual in class $k$ is \eon \frac{N \ud h_{k-1}}{N h_k (1-\ud) + N \ud h_{k-1}}. \eoff Substituting in the steady state values for the $h_k$, this becomes \eon \frac{1}{1 + \frac{1}{k} \left( \frac{1}{s} - \frac{\ud}{s} \right)} \approx \frac{1}{1 + \frac{1}{sk}} \approx sk \eoff  This means that we have \eon \probtimes{k-1}{k}(t) = s k e^{-s k t}. \label{timejumpone} \eoff  Note that this calculation is identical to the equivalent distribution of mutation timings computed by \citet{gordocharlesworth02} following the approach of \citet{hudsonkaplan94}.

We now consider the lineage structure within the mutation-selection balance. Consider a fitness class $k$, which has an overall frequency $h_k$ (\fig{fig1}b).  The frequency $h_k$ is maintained by a stochastic process in which the class is constantly receiving new individuals from class $k-1$ due to mutations.  In our infinite-alleles approximation, each such mutation creates a lineage which is an allele that is unique within the population.  Each lineage fluctuates in frequency for a while before eventually dying out, perhaps after acquiring additional mutations that found new lineages in fitness class $k+1$.  At any given moment, there is some frequency distribution of lineages in each class $k$ (see \fig{fig2}).  While the identity of these lineages changes over time, there is a probability distribution that at any moment there is a given frequency distribution of lineages.  In steady state, this probability distribution does not change with time.

In \citet{allelebased}, we calculated this steady state probability distribution of the frequency distribution of lineages.  For our purposes here, it is most useful to consider these results in the absence of neutral mutations; we will consider the diversity at neutral sites separately below.  In the absence of neutral mutations, we noted that new lineages are founded in class $k$ at a rate $\theta_k/2$, where \eon \theta_k = 2 N h_{k-1} \ud. \eoff  These individuals are then removed from class $k$ at a per capita rate \eon s_k \equiv -\ud - s (k - \bar k). \eoff  We refer to $s_k$ as the \emph{effective selection coefficient} against an allele in class $k$, because it is the rate at which any particular lineage in class $k$ loses individuals, and we defined \eon \gamma_k = N s_k. \eoff  Our model then reduces to the situation studied by the Poisson Random Field model of \citet{sawyerhartl92} and \citet{hartlsawyer94}.  Thus the frequency distribution of lineages (alleles) in fitness class $k$ follows a Poisson Random Field (PRF) with effective parameters $\theta_k$ and $\gamma_k$.  That is, the number of distinct lineages in class $k$ with a frequency between $a$ and $b$ (relative to the total size of the fitness class $N h_k$) is Poisson distributed with mean \eon \int_a^b f_k(x) dx, \eoff where \eon f_k(x) = \frac{\theta_k}{x (1-x)} \frac{1- e^{-2 \gamma_k (1-x)}}{1 - e^{-2 \gamma_k}}. \label{fk} \eoff  This is equivalent to saying that the probability that there exists a lineage in class $k$ with frequency (in the entire population) between $x$ and $x+dx$ is $f_k(x) dx$, for infinitesimal $dx$.

Note that this analysis involves various implicit approximations, and the results are valid within a specific parameter regime.  We describe these approximations and limitations in detail in \citet{allelebased}.  Most importantly, our approach neglects the fact that although each fitness class will have an average size $h_k$, in a finite population there will be fluctuations around this $h_k$.  Furthermore, our PRF analysis neglects the fact that there is a correlation between the size of a lineage and the size of a fitness class conditional on that lineage existing.  We analyzed these approximations in Appendix B of \citet{allelebased}, and described in detail the parameter regimes in which they are valid.  Note that all of the results we describe below include the corrections for correlations detailed in that Appendix.  In the Discussion, we return to discuss in more detail a key aspect of this approximation --- that we neglect the effects of Muller's ratchet --- which is particularly relevant for the present work.

Most importantly for our subsequent analysis, note that our Poisson Random Field result implies that on average the sum of all the frequencies of all the alleles in fitness class $k$ is simply \eon h_k = \int_0^1 x f_k(x) dx, \eoff which implies that the frequency of the fitness class within the total population is $h_k$, and that the probability that two individuals chosen at the same time at random from fitness class $k$ both come from the same lineage is \eon \int_0^1 \frac{x^2}{h_k^2} f_k(x) dx. \eoff

\section{The Fitness-class Coalescent Probabilities}

We are now in a position to calculate the degree of relatedness between two individuals sampled from the population.  Our goal is to understand the probability distribution of the fitness-class coalescence steptimes for two individuals chosen at random from the population.  We begin by calculating the coalescence probability in each step.  For now we neglect neutral mutations entirely and focus on formulating the fitness-class coalescent framework; we defer the calculations of neutral diversity to a later section.  In this section we focus on the PRF-based method for calculating coalescent probabilities; we present an alternative derivation based more directly on the method of \citet{hudsonkaplan94} in the next section.

First, imagine that by chance we pick two individuals from the same fitness class $k$. This class has a total frequency $h_k$ as given in \eq{hk}, and within the class there is a probability $f_{k}(x)$ as given in \eq{fk} that there exists a lineage with frequency $x$.  Thus there is probability \eon \pck{k}{k} = \int_0^1 x^2 f_k(x) \label{qtwoeqn} \eoff that these two chosen individuals come from the same lineage (note this expression contains the same implicit approximations as our calculation of $Q_2$ in \citet{allelebased}) If so, they are genetically identical and the coalescence steptime is $0$. If not, we want to calculate the probability they coalesce in class $k-1$, $\pck{k}{k-1}$.  If the lineage of individual $A$ in class $k$ was founded by a mutation from class $k-1$ a time $t_1$ ago, and the lineage of individual $B$ in class $k$ was founded by a mutation a time $t_2$ ago, the probability the two individuals came from a common lineage in class $k-1$ is \eon \pck{k}{k-1} = \int dx dy dt_1 dt_2 \probtimes{k-1}{k,k}(t_1,t_2) \frac{x f_{k - 1}(x)}{h_{k-1}} \frac{y \g{k-1}(y \rightarrow x, |t_2-t_1|)}{h_{k-1}}.   \label{pckeqn} \eoff  Here $\probtimes{k-1}{k,k}(t_1, t_2)$ is the joint distribution of $t_1$ and $t_2$, and $\g{k-1}(y \rightarrow x, |t_2-t_1|)$ is the probability a lineage in class $k-1$ changes in frequency from $x$ to $y$ in time $|t_2 - t_1|$ (where $y$ could be $0$, corresponding to a lineage that has already mutated back to class $k-2$ by the time the second individual mutates to class $k-1$).  We return to the forms of these functions below.

Note that all of these expressions assume that the distribution $h_k$ is constant in time.  This is the same assumption we used in calculating $f_k(x)$.  As we showed in \citet{allelebased}, this is a good approximation in class $k$ provided that $N h_k s k \gg 1$.  As in \citet{allelebased}, we only require in practice that this condition hold in the classes in the bulk of the fitness distribution; it can fail near the tails of the distribution without affecting our results because by definition only a very small fraction of the population are found in these tails.  Note however that for the purposes of the present paper, certain additional complications can arise from fluctuations in $h_k$ in the high-fitness tail of the distribution, leading to Muller's ratchet.  We neglect these ratchet effects here, but return to address them in the Discussion.  These formulas also assume that the probability a single lineage represents a substantial fraction of the size of a fitness class can be neglected.  We discussed a correction for this effect in Appendix B of \citet{allelebased}, and all of the results described below include this correction.

If the two individuals coalesced in this first step, the coalescent steptime is $1$.  If not (which occurs with probability $1 - \pck{k}{k-1}$), we have to consider the probability they coalesce at the next step (i.e. in the mutations that took them from class $k-2$ to $k-1$).  This probability is \eon \pck{k}{k-2}=\int dx dy dt_1 dt_2 \probtimes{k-2}{k,k}(t_1,t_2) \frac{x f_{k-2}(x)}{h_{k-2}}  \frac{y \g{k-2}(y \rightarrow x, |t_2-t_1|)}{h_{k-2}} \eoff  Here $t_1$ is the time the ancestor of individual $A$ in class $k$ mutated from class $k-2$ to $k-1$, and analogously for $t_2$; $\probtimes{k-2}{k,k}(t_1, t_2)$ is the joint distribution of these times, and $f_{k-2}(x)$ and $\g{k-2}$ are defined as above.  If the two individuals did not coalesce in this step, we can continue in the same vein and calculate $\pck{k}{k-3}$, and so on.

So far we have imaged that both individuals that we originally selected from the population came from the same class $k$. This will not generally be true.  Rather, when we pick two individuals at random, they will come from classes $k$ and $k'$ with probability \eon \hdistn (k, k') = \left\{ \begin{array}{ll} 2 h_k h_{k'} & \quad \textrm{if } k \neq k' \\ h_k^2 & \quad \textrm{if } k = k' \end{array} \right. \label{heqn} \eoff  For convenience we choose $k \leq k'$.  We define $\pckk{k}{k'}{k-\ell}$ to be the probability that two individuals from classes $k$ and $k'$ coalesce in class $k - \ell$.  Note that $\pckk{k}{k'}{k-\ell} = 0$ for $\ell < 0$.  For $\ell \geq 0$ we have \eon \pckk{k}{k'}{k - \ell}=\int dx dy dt_1 dt_2 \probtimes{k-\ell}{k, k'}(t_1, t_2) \frac{x f_{k-\ell}(x)}{h_{k-\ell}}  \frac{y \g{k-\ell}(y \rightarrow x, |t_2-t_1|)}{h_{k-\ell}}. \label{pckkeqn} \eoff  Of course the fact that $k' > k$ means that typically $t_1$ will be larger than $t_2$, and have a broader distribution.

From the set of coalescence probabilities \eq{pckkeqn}, we can calculate the probability distribution of coalescence steptimes between two individuals.  We describe these steptimes by the distribution of classes in which coalescence occurs; given that we pick two individuals from classes $k$ and $k'$ (with $k < k'$ by convention) the probability that they coalesce in class $k-\ell$ is simply \eon \pst{k}{k'}{\ell} = \pckk{k}{k'}{k-\ell} \prod_{j=0}^{\ell-1} \left[ 1 - \pckk{k}{k'}{k-j} \right]. \eoff  Note that this expression contains a subtle approximation:  if two lineages coalesce in class $k-\ell$ they were more likely to have coexisted in class $k-\ell+1$ and hence slightly more likely to have coalesced there than we have accounted for.  We neglect this effect here; it is closely related to the nonconditional approximation discussed in more detail below and in Appendix A.  We also note that, assuming that the probability that three lineages coalesce in a given step is negligible, we can in principle calculate the distribution of coalescent tree shapes and branch lengths in steptimes for a sample of any number of individuals.

\subsection{Computing the Coalescence Probabilities}
We now have a formal structure describing the structure of coalescent genealogies in the presence of negative selection. It remains, however, to evaluate the coalescent probabilities in each step, and to use these probabilities to calculate the probability distribution of genealogies.

We begin by noting that the coalescent probabilities all depend on the transition probability for the change in the frequency of a lineage from $x$ to $y$ in a time $|t_1 - t_2|$ in class $k-\ell$, $\g{k-\ell}(y \rightarrow x, |t_2-t_1|)$.  This transition probability was calculated by \citet{kimura55c} and can be expressed as an infinite sum of Gegenbauer polynomials.  Fortunately, it always appears in the context of an integral \eon I_G = \int y \g{k-\ell}(y \rightarrow x, |t_2-t_1|) dy, \eoff which is simply the average of $y$ over $\g{k-\ell}$.  Hence this integral is given by the deterministic result for the change in the frequency of the lineage, \eon I_G = x e^{-s (k-\ell) |t_2-t_1|}. \label{getridofg} \eoff This simple expression for $I_G$ makes our approach analytically tractable.

We now begin by evaluating the probability that two individuals chosen from fitness class $k$ coalesce in class $k-1$. Applying \eq{getridofg} to \eq{pckeqn}, we have \eon \label{onestep} \pck{k}{k-1} = \int dx dt_1 dt_2 \probtimes{k-1}{k,k}(t_1,t_2) \frac{x}{(h_{k-1})^2} f_{k - 1}(x)  x e^{-s(k-1) |t_1-t_2|}. \eoff  Since the two individuals mutated independently from class $k-1$, we have $\probtimes{k-1}{k,k}(t_1,t_2) = \probtimes{k-1}{k}(t_1) \probtimes{k-1}{k}(t_2)$, where $\probtimes{k-1}{k}(t)$ is given by \eq{timejumpone}.  This gives \eon \pck{k}{k-1} = \int dx \frac{x^2 f_{k-1}(x)}{h_{k-1}^2} \int dt_1 dt_2 (s k)^2 \exp \left[ -sk (t_1 + t_2) - s(k-1)|t_1 - t_2| \right]. \eoff We can do the time integral by ordering $t_1$ and $t_2$, and find it gives $\frac{k}{(2k-1)}$.  The $dx$ integral is more complex; we discussed integrals of this form in Appendices A and B of \citet{allelebased} and found that \eon \int_0^1 dx x^2 f_{k-\ell}(x) \equiv I_x^{k-\ell} = \frac{1}{1 + 2 N h_{k-\ell} s (k-\ell)}. \eoff  Plugging in this result, we have \eon \pck{k}{k-1} = \frac{1}{1 + 2 N h_{k-1} s (k-1)} \frac{k}{2k-1}. \label{onestepresult} \eoff

We now wish to calculate the probability two individuals both chosen from fitness class $k$ coalesce in an arbitrary class $k - \ell$.  First consider the probability of coalescence in class $k-2$.  This is given by \eaon \pck{k}{k-2} & = & \int \probtimes{k-2}{k,k} (t_1, t_2) \frac{x^2 f_{k-2}(x)}{h_{k-2}^2} \exp \left[ -s (k-2) |t_1-t_2| \right] dt_1 dt_2 dx \\ & & = I_x^{k-2} \int \probtimes{k-2}{k,k}(t_1, t_2) \exp \left[ -s (k-2) |t_1-t_2| \right] dt_1 dt_2. \label{secondstep} \eaoff

The time $t_1$ is now the sum of the time for one individual to have mutated from class $k-2$ to class $k-1$ plus the time for it to have mutated from class $k-1$ to class $k$, and analogously for $t_2$.  However, in order for the two lineages to coalesce in class $k-2$, they must \emph{not} have coalesced in class $k-1$.  We refer to the probability distribution of the times when these individuals mutated from class $k-1$ to class $k$ conditional on them not having coalesced in class $k-1$ as $\probtimes{k-1}{k,k} (t_1, t_2 | nc)$.  We discuss this full calculation in Appendix A.  Here we make use of a simpler approximation:  since the coalescence probability in each step will turn out to be small, conditioning on not coalescing in class $k-1$ does not shift the distribution of mutation timings much.  To be precise, $\probtimes{k-1}{k,k} (t_1, t_2 | nc)$ differs from $\probtimes{k-1}{k} (t_1) \probtimes{k-1}{k}(t_2)$ only by a factor proportional to $\pck{k}{k-1}$.  In what follows, we will therefore neglect the complications associated with the probability distributions of the mutant timings conditional on non-coalescence, and use the simpler distributions of unconditional timings.  Note that by a similar token we have also implicitly neglected the fact that coalescence did not occur in class $k$ in computing the distribution of mutation timing relevant for computing the probability of coalescence in class $k-1$.  We refer to this as the non-conditional approximation, and discuss its validity further in Appendix A.

In the non-conditional approximation, the probability that two individuals both chosen from fitness class $k$ coalesce in an arbitrary class $k - \ell$ is \eon \label{noncon1} \pck{k}{k-\ell} = \int \probtimes{k-\ell}{k,k}(t_1,t_2) \frac{x^2 f_{k-\ell}(x)}{h_{k-\ell}^2} e^{-s(k-\ell)|t_1-t_2|} dt_1 dt_2 dx, \label{twentynine} \eoff where in our approximation $\probtimes{k-\ell}{k,k}(t_1,t_2)$ is the unconditional distribution of the times at which the two individuals sampled in class $k$ originally moved from class $k-\ell$ to class $k-\ell+1$ by acquiring a deleterious mutation.  Since $t_1$ and $t_2$ are independent in the non-conditional approximation, we have $\probtimes{k-\ell}{k,k}(t_1,t_2) = \probtimes{k-\ell}{k}(t_1) \probtimes{k-\ell}{k}(t_2)$.  We calculate these distributions of mutant timings $\probtimes{k-\ell}{k}(t)$ in Appendix B.  Plugging these in, and evaluating the integrals as described in Appendix C, we find \eon \pck{k}{k-\ell} = \frac{1}{1 + 2 N h_{k-\ell} s (k-\ell)} \frac{ {k \choose \ell}^2}{{2k \choose 2 \ell}},  \label{pcknoncon} \eoff  where ${a \choose b} \equiv \frac{a!}{b!(a-b)!}$.

This is our final result for the coalescence probability in class $k-\ell$ of two individuals chosen from the same class $k$. Note that the dependence on the parameters of the evolutionary process is entirely contained in the factor $\frac{1}{1+ 2 N h_{k-\ell} s (k-\ell)}$.  Thus the result \eq{pcknoncon} is simply \eon \pck{k}{k-\ell} = \frac{1}{1 + 2 N h_{k-\ell} s (k-\ell)} A^k_{\ell}, \eoff where $A^k_{\ell}$ is a numerical coefficient which depends on $k$ and $\ell$ but not on the population parameters.

This general form for the coalescence probabilities makes intuitive sense.  $N h_{k-\ell}$ is the population size of class $k-\ell$, and $\frac{1}{s(k-\ell)}$ is the average number of generations that an individual spends in class $k-\ell$ before mutating away.  Since the per-generation coalescent probability in a population of size $n$ is proportional to $\frac{1}{n}$, it makes sense that the coalescent probability in class $k-\ell$ is approximately proportional to one over the population size of this class times the number of generations individuals spend in this class.  The additional $1$ in the denominator captures the fact that the individuals might mutate away from the class before coalescing there (which reduces the average time they spend in the class together).  The numerical factor multiplying this basic scaling, $A_{\ell}^k$ comes from the integrals over the probability distribution of mutant timings (i.e. the $dt_1$ and $dt_2$ integrals).  It reflects the probability that the ancestors of the two individuals we are considering were both in class $k-\ell$ at the same time, since they could not otherwise coalesce there.  The factor of $\frac{1}{2 s (k-\ell)}$ is the average amount of time that the two individuals spend together in class $k-\ell$ given that they are ever in that class at the same time.

From this result, we can also form an intuitive picture of the shape of genealogies in the presence of negative selection. We have just seen that coalescence probability per actual generation depends on the parameters as $\frac{1}{N h_{k-\ell}}$, where the relevant value of $\ell$ increases as we go back in time.  Thus the structure of genealogies in the presence of negative selection is similar to having a variable population size as we go back in time.  The precise nature of this variable population size is encoded in the fitness distribution $h_{k-\ell}$.  For example, if we imagine sampling two individuals from the same below-average fitness class, the probability distribution of their genealogies is like having a population size that initially increases and then decreases as we look backwards in time.  Of course, this analogy only goes so far.  Most importantly, the coalescent steptimes are related to the statistics describing genetic diversity in a different way from how normal coalescent times are usually related to these statistics.  Further, in general we will not happen to sample two individuals in the same fitness class, a complication we now turn to.

\subsection{General coalescence probabilities in the non-conditional approximation}
Thus far we have focused on the coalescence probabilities starting from a sample of two individuals from the same fitness class $k$.  However, when we sample two individuals from the population at random, it is likely that they come from different fitness classes.  In general, the probability that two individuals sampled at random from the population come from classes $k$ and $k'$ respectively is $\hdistn (k, k')$, as defined in \eq{heqn}.

Given that we sample two individuals from classes $k$ and $k'$, where by convention we choose $k' > k$, the coalescence probability in the non-conditional approximation is \eon \pckk{k}{k'}{k-\ell} = \int \probtimes{k-\ell}{k}(t_1) \probtimes{k-\ell}{k'}(t_2) \frac{x^2}{h_{k-\ell}^2} f_{k-\ell}(x) e^{-s(k-\ell)|t_1-t_2|} dx dt_1 dt_2. \label{thirtysix} \eoff  We substitute in our expressions for $\probtimes{k-\ell}{k}(t)$ and evaluate the integrals in Appendix C; we find \eon \pckk{k}{k'}{k-\ell} = \frac{1}{1 + 2 N h_{k-\ell} s(k-\ell)} \aklm{k}{k'}{\ell}, \label{pckknoncon} \eoff where \eon \aklm{k}{k'}{\ell} = \frac{{k' \choose k-\ell} {k \choose k-\ell}}{{k + k' \choose 2 \ell + k' - k}}. \label{aeqn} \eoff

\eq{pckknoncon} is the complete solution for coalescent probabilities in the non-conditional approximation.  As in the previous subsection, the parameter dependence is simple and the probability of coalescence in a given fitness class is proportional to the inverse population size of the fitness class and the time an average individual spends in that fitness class.  This is multiplied by a $k$, $k'$, and $\ell$-dependent numerical factor which decreases as $k'-k$ increases, reflecting the fact that the larger $k'-k$ is, the less likely the ancestors of the two sampled individuals are to have been in a given fitness class at the same time.  The dependence of $\aklm{k}{k'}{\ell}$ on $\ell$ is more complex, but reflects the probability that the ancestors of the two individuals we are considering were in class $k-\ell$ at the same time.

\begin{figure}
\includegraphics[width=6.5in]{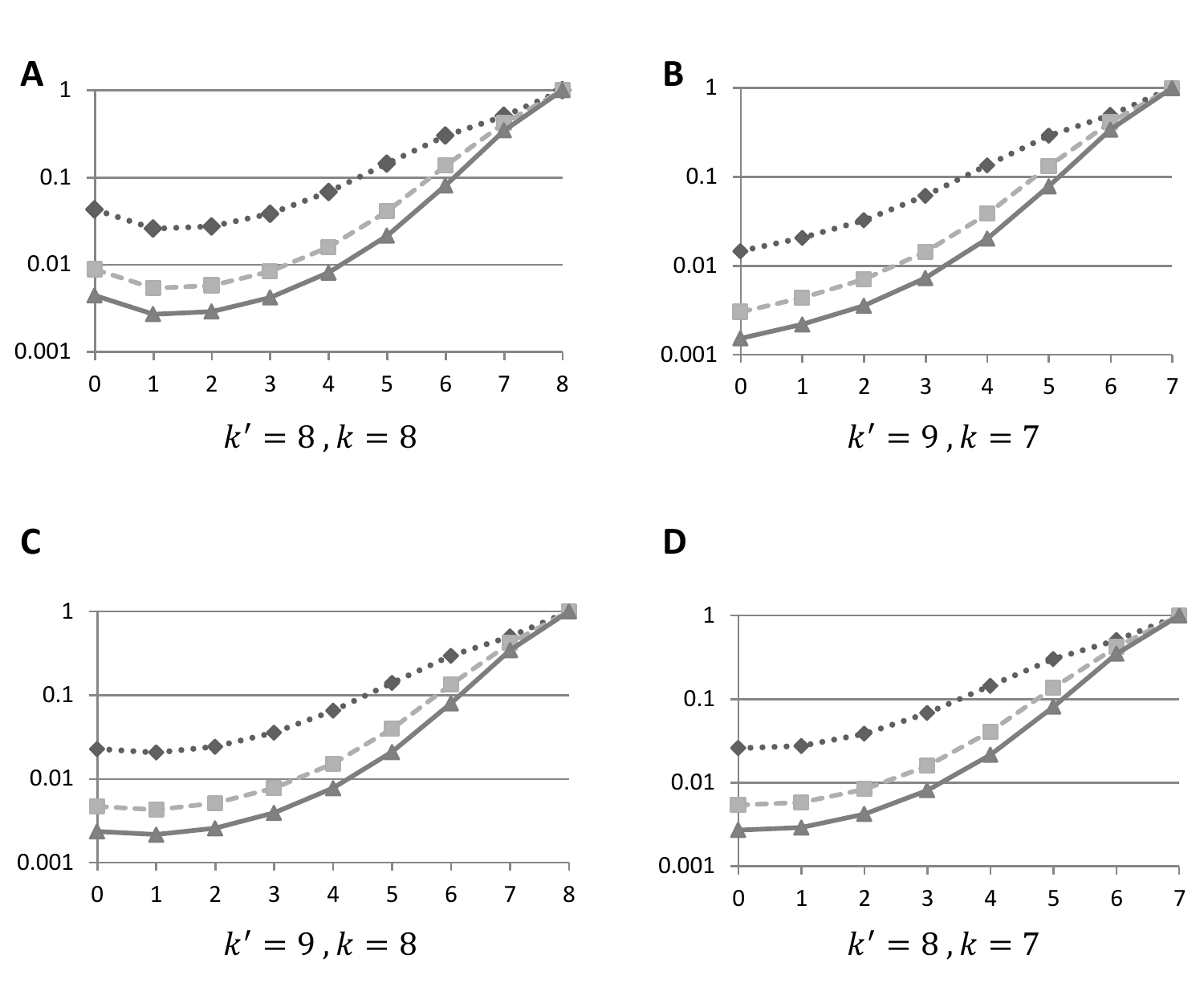}
\caption{Examples of the coalescence probabilities $\pckk{k}{k'}{\ell}$ for two individuals sampled from fitness classes $k$ and $k'$ to coalesce in class $k-\ell$, shown as a function of $\ell$.  Here $\ud/s = 8$, $s = 10^{-3}$, and results are shown for $Ns = 10$ (dotted lines), $Ns = 50$ (dashed lines), and $Ns = 100$ (solid lines).  }
\label{fig5}
\end{figure}

In \fig{fig5} we show examples of coalescence probabilities calculated from our theoretical framework within the non-conditional approximation for different population parameters. We see that the probability of coalescence steadily increases for longer steptimes (classes with larger fitness), and decreases with increasing selection coefficients and population size.

\section{A sum of ancestral paths approach}

Our analysis thus far has focused on using the lineage structure within each fitness class to determine the coalescence probabilities. \citet{hudsonkaplan94} proposed a somewhat different way to look at the same problem: they considered a sample of individuals and, without explicitly describing lineage structure, computed the relative probabilities that the next event to occur backwards in time would involve a mutation or coalescent event.  For example, if two individuals are in the same fitness class, the next event could be either coalescence within that class or a mutation event.  The rates at which these events occur determines their relative probabilities.  In this manner, \citet{hudsonkaplan94} were able to generate a recursion relation for the mean time to a common ancestor, their Eq. (12).  \citet{gordocharlesworth02} used this equation as the basis for a coalescent simulation.  A similar logic was used earlier in by \citet{kaplanhudson88} to develop analogous diffusion equations for the transition probabilities between states; \citet{bartonetheridge04} developed this approach to compute the effect of selection on genealogies in a two-locus system.

Recursion relations of the \citet{hudsonkaplan94} form can be solved numerically, and have been used to generate data describing coalescent statistics, but have not yet led to an analytic description of the structure of genealogies in the presence of negative selection at many linked sites.  We now demonstrate that these numerical methods are equivalent to our lineage-based formalism above, by showing that the \citet{hudsonkaplan94} approach can be used to derive identical analytical formulas for the coalescent probabilities.  We refer to this as a ``sum of ancestral paths'' approach, because it relies on summing over all possible paths of individual ancestry through the fitness distribution.  The equivalence of this approach to our lineage-structure calculations means that our analytical results in this paper match earlier numerical and simulation results based on the \citet{hudsonkaplan94} formulation.

In order to calculate the coalescence probabilities for a sample of two individuals, we consider the set of all possible ancestral paths these individuals may have followed. Each path is represented by an ordered set of events, backwards in time. These events may either be deleterious mutation events, which move one of the ancestral lineages to the previous fitness class, or coalescence events, which merge the two ancestral lineages. In the absence of back mutations, the ancestral lineages may only move toward higher-fitness classes, such that movement through the distribution is irreversible. As a consequence, in order for two individuals to coalesce in class $k-\ell$, each ancestral lineage must undergo a series of deleterious mutation events, bringing them from their initial classes to class $k-\ell$. The lineages must then coalesce before any additional deleterious mutations occur.

For example, in order for two individuals sampled from class $k$ to coalesce in class $k-1$, the first event, backwards in time, must be a deleterious mutation. This mutation can occur in either individual. After this event, one of the ancestral lineages is still in class $k$, while the other is in class $k-1$. The second event, backwards in time, must be a deleterious mutation event in the ancestral lineage that remains in class $k$. Both ancestral lineages are now in class $k-1$. Finally, the third event must be a coalescent event. Note that there are a total of two paths, since either individual may have been the first to mutate.

In general, in order for two individuals sampled from classes $k'$ and $k$ to coalesce in class $k-\ell$, the first $k'-k+2\ell$ events must consist of $k'-k+\ell$ deleterious mutation events in the ancestral lineage that began in $k'$, and $\ell$ deleterious mutation events in the ancestral lineage that began in $k$. The final event must then be a coalescent event. Note that there are a total of $\binom{k'-k+2\ell}{l}$ possible paths, reflecting the number of ways to order the mutation events in one lineage with those in the other.  To calculate the coalescence probability, we sum the probabilities of each path that results in this particular coalescence event.

\subsection{The probabilities of each event}
The probability of each path is the product of the probability of each event in the path. In order to determine the probability of each event, we first consider the rates. As in our lineage structure approach, we neglect neutral mutations for now; we will consider their effects in a later section below. We saw above that the distribution of times since a deleterious mutation occurred in an individual in class $k$ is $\probtimes{k}{k-1}(t) = s k e^{-s k t}$.  If the two individuals are in different classes, they are not able to coalesce. Therefore, the probability of each event is simply: \eaon P(\textrm{1st Event is Del. Mut. in k}|k,k')& = &\frac{s k}{s k + s k'} \\ P(\textrm{1st Event is Del. Mut. in k'}|k,k')&=&\frac{s k'}{s k +s k'} . \eaoff If the two individuals are in the same class, the next event may either be a coalescent event or a deleterious mutation. Within each class, coalescence is a neutral process that occurs with rate $1/Nh_k$. Therefore, we have that: \eaon P(\textrm{1st Event is Coal.}|k,k) & = & \frac{1/(N h_k)}{s k + s k + 1/(N h_k)} = \frac{1}{1+2 N h_k s k} = I_x^k \\ P(\textrm{1st Event is Del. Mut.}|k,k) & = & \frac{2s k}{s k + s k + 1/(N h_k)} = \frac{2 N h_k s k}{1+2 N h_k s k} = 1-I_x^k . \eaoff  These probabilities are analogous to those used by \citet{gordocharlesworth02}, and similar expressions can be derived as a simple extension of the analysis of \citet{bartonetheridge04}.

\subsection{The sum over possible ancestral paths}
Using these probabilities, we now calculate the probability of coalescence in a given class. First, consider sampling two individuals from the same fitness class $k$. In order for these two individuals to coalesce in class $k$, the first event must be a coalescent event. Thus we have:  \eon \pckk{k}{k}{k} = I_x^k, \eoff equivalent to our earlier lineage-based result.  In order for these individuals to coalesce in class $k-1$, the first event must be a deleterious mutation event. Since both individuals' ancestral lineages are currently in class $k$, the probability the first event is a deleterious mutation event is $1-I_x^k$. After this event, there is now one ancestral lineage in class $k-1$, and one in class $k$. The next event must be a deleterious mutation in the latter, which occurs with probability $\frac{k}{2k-1}$. Finally, the third event must be a coalescent event. This implies \eon \pst{k}{k}{1} =  (1-I_x^{k})I_x^{k-1} \frac{k}{2k-1} . \eoff  Note that this logic has given us an expression for the probability that the coalescent steptime is $1$, $\pst{k}{k}{1}$, and not the probability of coalescence in this class given that coalescence has not yet occurred, $\pckk{k}{k}{k-\ell}$, because we have already included the probability that the coalescence event does not happen in class $\ell$.

We can continue to extend this logic to subsequent fitness classes.  For example, for coalescence to occur in class $k-2$, there are six possible paths. We can label them as AABBc, BBAAc, ABABc, ABBAc, BABAc, and BAABc, where A corresponds to a mutation in the first individuals' ancestral lineage, B corresponds to a mutation in the second individuals' ancestral lineage, and c corresponds to a coalescent event. We can calculate the probability of each path.  For example, \eon P(AABBc)=\left(\frac{1 - I_x^k}{2}\right) \left( \frac{k - 1}{2k - 1} \right) \left(\frac{k}{2k - 2} \right) \left(\frac{k - 1}{2k - 3} \right) I_x^{k-2}. \eoff The probability of path BBAAc is identical, since it has the same probabilities at each step. However, the remaining four paths have a different probability, because the ancestral lineages exist together in the $k-1$ class at the same time. This distorts the probability of mutations at that step, since coalescence could also have occurred.  For paths of this type, we have \eon P(ABABc)=\left( \frac{1 - I_x^k}{2} \right) \left( \frac{k}{2k - 1} \right) \left( \frac{1-I_x^{k - 1}}{2}\right) \left( \frac{k - 1}{2k - 3} \right) I_x^{k-2} . \eoff We add up each path to find \eaon \pst{k}{k}{2} & = & I_x^{k - 2}\frac{ k(k - 1)}{4(2k - 1)(2k - 3)}  \left(2 \left(1 - I_x^k \right) + 4 \left( 1-I_x^k \right) \left( 1-I_x^{k - 1} \right) \right) \\ & = & I_x^{k-2}\frac{ 3k(k - 1)}{2(2k - 1)(2k - 3)}  \left(1-I_x^k-\frac{2}{3}I_x^{k-1}+\frac{2}{3}I_x^kI_x^{k-1}\right). \eaoff

It is informative to consider the form of this result. The $I_x^{k-2}$ factor is the probability that the two ancestral lineages coalesce in class $k-2$, given that they existed in class $k-2$ at the same time.  The remaining factors represent the probability that the two ancestral lineages existed at the same time in class $k-2$. This consists of a leading order term $\frac{ k(k - 1)}{4(2k - 1)(2k - 3)}$ (identical to our earlier result for $A^{k}_{\ell=2}$), multiplied by a correction due to the distortion in paths from the possibility of coalescence in previous steps.

We can continue on to consider the probability of coalescence in class $k-3$. There are now a total of $\binom{6}{3}$ possible paths. These can be split into four types, depending upon whether the two ancestral lineages coexisted in both classes $k-1$ and $k-2$ (e.g. ABABABc), in class $k-1$ only (e.g. ABAABBc), in class $k-2$ only (e.g. AABBABc), or in neither (e.g. AAABBBc). The probability of each type of path is identical, except for a distortion factor $(1-I_x^{k-i})$ for each class $k-i$ in which the two ancestral lineages were together at the same time. The probabilities can be calculated as before, and summed to yield $\pst{k}{k}{3}$. Using similar logic, we can extend this approach to the situation where two individuals are sampled from different classes, $k'$ and $k$.

In Appendix D, we describe the details of carrying out this summation over all possible paths to determine the coalescent probabilities.  We find \eaon \pst{k}{k'}{\ell} & = & I_x^{k - \ell}\frac{\binom{k'}{k - \ell}\binom{k}{k - \ell}}{\binom{k' + k}{k' - k + 2\ell}} \left[ 1 - \sum_{i = 0}^{\ell - 1} \frac{\binom{k' - k + 2 i}{i}\binom{2\ell - 2i}{\ell - i}}{\binom{k' - k + 2 \ell}{\ell}}I_x^{k - i} + \right. \\ & & \left. \sum_{i = 0}^{\ell - 2} \sum_{j > i}^{\ell - 1} \frac{\binom{k' - k + 2i}{i}\binom{2j - 2i}{j - i}\binom{2\ell - 2j}{\ell - j}}{\binom{k' - k + 2\ell}{\ell}}I_x^{k - i}I_x^{k - j}-\ldots \right], \label{sumpathscoalprob} \eaoff where as always we have assumed $k \leq k'$ by convention.  The form of this solution is intuitive. The factor $I_x^{k-\ell}$ is the probability of coalescence in class $k-\ell$, given that the two ancestral lineages existed in this class at the same time. The remaining factors reflect the probability that the two lineages are together in class $k-\ell$ at some point. This consists of a leading order term, which is identical to the $\aklm{k}{k'}{\ell}$ calculated previously, times a correction. The correction represents the distortion in the paths due to the possibility that coalescence could have occurred at previous steps. There are a total of $l+1$ terms in the correction, each of which is known and calculable.

Fortunately, provided $2 N h_k s k \gg 1$, we can neglect the higher-order terms in \eq{sumpathscoalprob}.  This is equivalent to calculating the probability of coalescence in a given class, without considering the possibility that coalescence events could have occurred in previous classes.  Thus it converts our expression for $\pst{k}{k'}{\ell}$ into an expression for $\pckk{k}{k'}{k-\ell}$.  Neglecting these terms also implicitly makes the non-conditional approximation, as we did in the PRF method, because it assumes that the fact that coalescence did not occur in previous classes does not distort the likelihood of taking particular paths. Making this approximation, we find \eon \pckk{k}{k'}{k-\ell} = \frac{1}{1 + 2 N h_{k-\ell} s(k-\ell)} \aklm{k}{k'}{\ell}, \eoff which exactly matches our expression for the coalescence probabilities in the non-conditional approximation in our PRF approach, \eq{pckknoncon}.

The condition $2 N h_k s k \gg 1$ is the condition we are already assuming in treating the frequencies of each class, $h_k$ as constant.  Thus the results from the PRF method and the sum of ancestral paths are exactly equivalent in the regime where they are valid.  We discuss the correspondence between approximations in the sum of ancestral paths method as compared to the PRF method in more detail in Appendix D.

\section{The Structure of Genealogies and Statistics of Genetic Diversity}

We can now use the coalescence probabilities described above to calculate the structure of genealogies in the presence of negative selection.  We can then use these genealogies to calculate various statistics describing the genetic diversity within the population.  We know the coalescent probabilities in each step of our fitness-class coalescent process, so in principle we can calculate the probability of any genealogy relating an arbitrary number of individuals using methods analogous to those used in standard neutral coalescent theory.  This would then allow us to calculate the distribution of any statistic describing the genetic diversity among these individuals, again using methods analogous to neutral coalescent theory.

Here we will focus on the simplest genealogical relationship:  the distribution of the time to the most recent common ancestor of two individuals, which demonstrates the main ideas in the simplest context.  This allows us to calculate the distribution of the per-site heterozygosity $\pi$.  This is the only statistic relevant to a sample of two individuals.  In larger samples, provided the total number of individuals sampled is not too large, the coalescent probabilities between any pair of sampled individuals are independent to those between any other pair.  Thus the distribution of per-site heterozygosity $\pi$ we expect in such a sample is equivalent to the distribution of $\pi$ we calculate here.

In our fitness-class coalescent framework, it is natural to consider diversity at the negatively selected sites separately from diversity at linked neutral sites.  We focus first on the distribution of coalescent steptimes and $\pi_d$, the per-site heterozygosity at negatively selected sites alone, ignoring neutral mutations.  We will then turn to the connection between steptimes and actual times in generations, which will enable us to calculate the distribution of neutral diversity, including the per-site heterozygosity at neutral sites $\pi_n$.  In analyzing data, we will of course typically not know \emph{a priori} which sites are neutral and which are negatively selected.  In such a situation, we merely add up the expected diversity at neutral sites and negatively selected sites, so that the total expected per-site heterozygosity is $\pi = \pi_d + \pi_n$.

\subsection{Distribution of steptimes and $\pi_d$}
We begin by imagining that we sample two individuals at random from the \emph{same} fitness class $k$.  By construction, the number of negatively selected sites at which they will be polymorphic is twice their coalescent steptime, $\pi_d = 2 \ell$.  We therefore have \eon \ppid (\pi_d = 2 \ell) = \pst{k}{k}{\ell}, \eoff where $\ppid (\pi_d = 2 \ell)$ is the probability $\pi_d = 2 \ell$.

More generally, if two individuals sampled from classes $k$ and $k'$ coalesce in class $k-\ell$, we have $\pi_d = 2 \ell + k' - k$.  This means we have \eon \ppid (\pi_d = 2 \ell + k' - k | k, k') = \pst{k}{k'}{\ell} . \eoff  We can average this over the distributions of $k$ and $k'$ to find the distribution of $\pi_d$ amongst individuals sampled at random from the population.  We find \eon \ppid (\pi_d) = \sum_{\ell} \sum_{k=0}^\infty \hdistn (k,k' = k + \pi_d - 2\ell) \pst{k}{k' = k+ \pi_d - 2 \ell}{\ell}, \eoff where the first sum runs from $\ell = 0$ to the largest integer less than or equal to the smaller of $k$ or $\pi_d/2$.  Note that in practice we only have to evaluate the sum over $k$ from $0$ to a multiple of $\ud/s$, since $\hdistn (k, k')$ will be negligible for larger $k$.

These results for the distributions of genealogy lengths and of $\pi_d$ involve several sums.  However, all the terms in these sums are straightforward and the numerical evaluations of their values are simple and fast.  In \fig{fig6} we show a representative example of the predicted distribution of the per-site heterozygosity at negatively selected sites, $\rho(\pi_d)$, compared to simulation results.  We explore the significance of the shape of the distribution $\rho(\pi_d)$, how this distribution depends on the parameter values, and the source of the small but systematic deviations between the theoretical predictions and the simulation results in the Discussion.

\subsection{The relationship between steptimes and time in generations}
So far we have focused on the genealogies measured in steptimes, which allowed us to calculate the distribution of heterozygosity among negatively selected sites.  We would now like to relate the steptimes to actual times in generations. To do this, we consider the probability that a coalescence event occurred at time $t$, given two individuals sampled from classes $k$ and $k'$ that coalesced in class $k - \ell$, $\probt (t|k,k',\ell)$.  This can be divided into two parts:  the time since the ancestors of these two individuals were both in class $k-\ell$ together, which has a distribution $\probt_1 (t|k,k',\ell)$, and the time to coalescence once in this class, $\probt_2 (t|k,k',\ell)$.

We compute these distributions in Appendix E, and find \eon \probt_1 (t|k, k', \ell) = s \pi_d e^{-s(k' + k) t}(e^{st} - 1)^{\pi_d - 1} \binom{k'+k}{\pi_d}, \label{realtimescond3} \eoff where we have made use of the fact that $\pi_d = k' - k + 2 \ell$, and \eon \probt_2(t|k',k, \ell) = \left( 2s(k - \ell) + \frac{1}{Nh_{k-\ell}} \right) e^{-(s (k - \ell) + \frac{1}{Nh_{k-l}} ) t}. \eoff The total real time since coalescence is the sum of these two times, so we have \eon \psi(t|k', k,\ell) = \psi_1(t|k',k,\ell) \otimes \psi_2(t|k', k, \ell) . \eoff We compute this convolution in Appendix E, and find \eon \psi(t|k',k,\ell) = \sum_{i = 0}^{n-1} s \pi_d (-1)^{\pi_d -i-1} \binom{\pi_d - 1}{i} \binom{k' + k}{\pi_d} \frac{B}{A - B} \left( e^{-sBt} - e^{-sAt} \right) , \label{realtimescond} \eoff where we have defined $A \equiv k' + k - i$ and $B \equiv k - \ell + \frac{1}{Nsh_{k-\ell}}$.

Note that, making the usual approximation $N h_{k-\ell} s (k-\ell) \gg 1$, this expression can be simplified; we find \eon \probt (t|k',k,\ell) = s(\pi_d + 1) e^{-s(k'+k)t} (e^{st} - 1)^{\pi_d} \binom{k' + k}{\pi_d + 1} .  \eoff  However, it is important to note that while this approximation may be valid in the bulk of the distribution, it will always fail when coalescence occurs in the zero-class, where $s(k-\ell)=0$.  In this case, we must use the more complex expression \eq{realtimescond} or (in the case when the coalescence time within the $0$-class can be neglected compared to the time taken to descend from the $0$-class) the expression \eq{realtimescond3}.

Averaging over the possible values of $k$, $k'$, and $\ell$, we find the overall distribution of actual coalescent time between two randomly chosen individuals, \eon \probt (t)= \sum_{k = 0}^\infty \sum_{m=0}^\infty \sum_{\ell = 0}^k \probt (t|k,k',\ell) \pst{k}{k+m}{\ell} \hdistn (k,k+m), \label{timetotal} \eoff where the distributions $\hdistn (k, k+m)$, $\pst{k}{k+m}{\ell}$, and $\probt (t|k,k',\ell)$ are as given above.  However, as we will see below, in calculating neutral diversity we will typically find it easier to work directly with $\probt (t|k, k', \ell)$ rather than this unconditional distribution for $\probt (t)$.

\subsection{The neutral heterozygosity $\pi_n$}
From the distributions of real times to a common ancestor described above, we can calculate the distribution of $\pi_n$, the neutral heterozygosity.  Since the neutral mutations occur as a Poisson process with rate $\un$, and there are a total of $2 t$ generations in which these mutations can occur, $\pi_n$ follows a Poisson distribution with mean $U_n t$, where $t$ is drawn from the distribution of coalescence times, \eq{timetotal}.  We have \eon \ppin (\pi_n) = \int_0^\infty \frac{\left[ 2 \un t \right]^{\pi_n}}{\pi_n!} e^{-2 \un t} \probt (t) dt. \eoff  In \fig{fig7}, we compare this distribution of neutral heterozygosity (as modified by the corrections described in Appendix A) to direct simulations.  We find good general agreement to the shape of the distribution, though there are slight systematic errors (presumably due to effects of Muller's ratchet, which we explore further in the Discussion).  Note that, like our results for the diversity at negatively selected sites, these results differ dramatically from the exponential distribution a neutral model or effective population size approximation would predict; we describe these comparisons further in the Discussion.

We note that to calculate the distribution of total heterozygosity $\pi = \pi_n + \pi_d$, we must account for the fact that $\pi_d$ and $\pi_n$ are not independent:  large $\pi_d$ means a large coalescent steptime and hence makes a large $\pi_n$ more likely.  The distribution of $\pi_d$ is independent of $\pi_n$, and is given by $\ppid (\pi_d)$ above.  Above we found $\probt (t|k, k', \ell)$, which implies that \eon \ppin (\pi_n| k, k', \ell) = \int_0^\infty \frac{\left[ 2 \un t \right]^{\pi_n}}{\pi_n!} e^{-2 \un t} \probt (t | k, k' \ell) dt . \label{pineqn} \eoff  We can compute this integral; we find \eon \ppin (\pi_n | k' ,k, \ell) = \sum_{i = 0}^{\pi_d -1} \pi_d (-1)^{\pi_d -i - 1} \binom{\pi_d - 1}{i} \binom{k' + k}{\pi_d }\frac{B}{A - B} \left( \frac{( \frac{2 U_n}{s})^{\pi_n}}{( \frac{2 U_n}{s} + B)^{\pi_n + 1}} - \frac{( \frac{2U_n}{s})^{\pi_n}}{( \frac{2 U_n}{s} + A)^{\pi_n + 1}} \right) , \eoff
where we have defind \eon A = k' + k - i, \qquad B = 2k - 2 \ell + \frac{1}{Nsh_{k-l}} . \eoff Since $\pi_d = 2 \ell + k - k'$, this implies \eon \ppin (\pi_n | \pi_d) = \sum_{\pi_d = 2 \ell + k - k'} \rho(\pi_n | k, k', \ell). \eoff  The distribution of $\pi$ is then given by \eon \rho(\pi) = \sum_{\pi_n + \pi_d = \pi} \ppid (\pi_d) \ppid(\pi_n | \pi_d). \eoff  This is no more difficult to calculate than $\ppin (\pi_n)$, since it involves analogous sums.  However, while the distribution of $\pi$ is clearly important in analyzing sequence data, in this paper we focus on the distributions of $\pi_n$ and $\pi_d$ separately, which provides a more complete picture of the source of all aspects of the genetic variation.

\subsection{The mean pairwise heterozygosity}
Above we have calculated the distribution of heterozygosity for both neutral and deleterious mutations.  It is straightforward to average these results to calculate the mean pairwise heterozygosity for both neutral and deleterious mutations.  In \fig{fig8} and \fig{fig9} we show how this mean heterozygosity depends on population size, mutation rate, and selection strength, for neutral and deleterious mutations respectively.  We see that the dependence of $\ev{\pi_d}$ on the population size is fairly weak.  While it increases roughly linearly with $N$ in the weak selection regime, this quickly saturates and for $Ns$ substantially greater than $1$ the mean heterozygosity becomes almost independent of population size.  The dependence on $\ud/s$, by contrast, is much stronger.  The dependence of $\ev{\pi_n}$ on the parameters is also interesting:  this depends weakly on the parameters for small $N$ or $\ud/s$, but for larger $N$ becomes roughly linear.  These results make intuitive sense, particularly in light of the ``mutation-time'' approximation that we introduce in the Discussion, where we discuss these figures in more detail.

\subsection{Neutral heterozygosity from a sum of ancestral paths}
An alternative way to compute neutral heterozygosity is to further extend the sum of ancestral paths approach which we used above to provide an alternative derivation of the coalescence probabilities.  In this formulation, we do not make any connection to real times.  This means we cannot use it directly to calculate the distributions of the times to most recent common ancestors of a sample.  However, this approach does provide an alternative way to compute the distribution of neutral heterozygosity, $\ppin (\pi_n)$.  We carry out this computation in Appendix G, and show that it leads to results identical to our analysis above.

\subsection{Statistics in larger samples}\label{S3}
The distributions of $\pi_n$ and $\pi_d$ described above are very different from the distributions of heterozygosity expected in the absence of selection.  We could certainly measure the distribution of pairwise heterozygosity from a sample of many individuals from a population, and use this to infer the action of selection.  However, it may also be useful to understand the expected distribution of other statistics describing the variation in larger samples.  The relationship between these different statistics will typically be different than expected in the neutral case, making them useful in constructing other statistical tests for selection.

One statistic often used to describe variation in larger samples is the total number of segregating sites among a sample of $n$ individuals, $S_n$.  Here we describe how our framework allows us to calculate the distribution of $S_3$; similar methods can be used to calculate the distribution of $S_n$ for larger $n$.  One common test for neutrality, Tajima's $D$, is based on a comparison between the observed values of $\pi$ and $S_n$; our results for $S_3$ could in principle be used to show how this statistic should be expected to behave in the presence of purifying selection.  As we will see, it is unwieldy to calculate closed form expressions for these quantities in our framework, so here we merely lay out a prescription for calculating $S_3$.

We first consider the distribution of $S_3^d$, the number of segregating negatively selected sites among three randomly sampled individuals.  In order to calculate the probability a sample has a particular $S_3^d$, we imagine picking three individuals at random from the population and calculate the probability of the coalescence events that lead to that $S_3^d$. We illustrate such a situation where three individuals are sampled from classes $k$, $k'$, and $k''$ in \fig{fig3}.  Two of these three lineages coalesced in class $k_1$. We call the steptime at which two of the three lineages coalesced $\tau_3$ (see \fig{fig3}).  We next need to calculate the distribution of $\tau_2$, the total steptime to common ancestry of the three individuals.  This time of course cannot be smaller than $\tau_3$.  Given values of $\tau_3$ and $\tau_2$, it is clear from \fig{fig3} that the total number of segregating negatively selected sites is $S_3^d = 2 \tau_2 + \tau_3$.

Calculating the joint distribution of $\tau_2$ and $\tau_3$ is tedious, because we must sum over all possible orderings of the coalescence events, but it can be computed using either our lineage structure method or the sum of ancestral paths approach.  The basic result is analogous to our results for the coalescence steptime between a pair of individuals: coalescence probabilities within a given class are proportional to the inverse size of that class times the number of real generations the ancestors of given individuals typically spend in that class, times a factor that reflects the time that the ancestors of sampled individuals are present in each class at the same time.

The number of segregating sites $S_3^d$ is given by \eon S_3^d = \tau_3 + 2 \tau_2 - (k'' - k) - (k'' - k'). \eoff  Thus using the distributions of $\tau_3$ and $\tau_2$, and averaging over the distributions of $k$, $k'$, and $k''$, we can calculate the full distribution of $S_3^d$.  Given a particular value of $S_3^d$, there is a relationship between the steptimes and actual times (analogous to \eq{realtimescond}), which we could use to find the distribution of the total number of segregating neutral sites $S_3^n$.  More complex statistics involving even larger samples can be computed using similar methods.

However, while this analysis provides a prescription for calculating the distribution of $S_3^d$ and $S_3^n$, it is clear that the full distributions are opaque. In the Discussion we provide a simple approximation for $S_n$ in a specific parameter regime we refer to as the ``mutation-time'' regime, but the complexities of the general calculation are tangential to the ideas behind our framework, so we do not pursue them further here.  However, these issues will be important to explore in future work aiming to use this framework for data analysis, and our approach here can be used as the basis for genealogical simulations.  Further, since our methods allow us to quickly compute the probability of a given genealogical history and to draw a particular genealogy from the appropriate distribution, they may provide a useful basis for importance sampling or MCMC methods to infer selection pressures from data.

\section{Numerical Simulations of the Genetic Diversity}\label{Simulations}

We compare the predictions of our fitness-class coalescence analysis to Monte Carlo simulations of the Wright-Fisher model. In our simulations, we consider a
population of constant size $N$ and we keep track of the frequencies of all
genotypes over successive, discrete generations.  In each generation, $N$
individuals are sampled with replacement from the preceding generation, according
to the standard Wright-Fisher multinomial sampling procedure \citep{ewensbook} in
which the chance of sampling an individual is determined by its fitness relative
to the population mean fitness.

In our simulations, each genotype is characterized by the set of sites at which it harbors deleterious mutations and the set of sites at which it harbors neutral
mutations. In each generation, a Poisson number of deleterious mutations are
introduced, with mean $N \ud$, and a Poisson number of neutral mutations are
introduced, with mean $N \un$; each new mutation is ascribed to a novel site,
indexed by a random number. The mutations are distributed randomly and
independently among the individuals in the population (so that a single individual might receive multiple mutations in a given generation). The simulations record the  time (in generations) at which each distinct genotype was first introduced.

Starting from a monomorphic population, all simulations were run for at least $\frac{1}{s} \log(\ud/s)$ or $N$ generations (whichever was larger), to ensure relaxation both to the steady-state mutation-selection equilibrium and to the PRF equilibrium of allelic frequencies within each fitness class.  The final state of the population --- i.e. the frequencies of all surviving genotypes --- was recorded at the last generation.  In order to produce the empirical distributions of $\pi_d$, and $\pi_n$ shown in \fig{fig6} and \fig{fig7}, we averaged across at least 300 independent populations for each parameter set.

Our simulations allow for random fluctuations in the frequencies of each fitness class, and for Muller's ratchet.  In most of the parameter regimes we explored, the ratchet proceeded during the simulation, so that the least loaded class at the end of each simulation typically contained anywhere from no deleterious mutations (typical for $\ud/s = 2$) to more than a dozen (typical for $\ud/s = 4$).  We see that despite these effects, our theory agrees well with the simulations, although there are small systematic errors that are signatures of the effects of the ratchet.  Generally speaking these errors increase as we increase $\ud/s$, but become less severe for larger $N$ or $s$.  We consider these effects of Muller's ratchet in more detail in the Discussion.

\section{Discussion}

In recent years, both experimental studies and sequence data have pointed to the general importance of selective forces among many linked variants in microbial and viral populations, and on short distance scales in the genomes of sexual organisms \citep{hahn08}.  Our analysis provides a framework for understanding how one particular type of selection --- pervasive purifying (i.e. negative) selection against deleterious mutations --- affects the structure of genetic variation at the negatively selected sites themselves and at linked neutral loci.  This type of selection is presumably widespread in many populations, in which there is a selective pressure to maintain existing genotypes and mutations away from these genotypes at a variety of loci are deleterious.

A variety of earlier work has addressed aspects of this problem, as described in the Introduction.  The key insight of our approach is that instead of following the true ancestral process, we develop a \emph{fitness-class} genealogical approach which focuses on how individuals ``move'' through the fitness distribution.  Here each mutation plays the role of a reproductive event that moves individuals through the fitness distribution, and each fitness class is a ``generation'' in which coalescence can occur with some probability. We calculate this probability using a simple approximation based on the PRF model of \citet{sawyerhartl92}, rather than by considering the actual reproductive process within that class.  By extending formulas originally computed by \citet{hudsonkaplan88} and \citet{bartonetheridge04}, we showed that these coalescent probabilities can also be computed using a summation of ancestral paths based on the structured coalescent described by \citet{hudsonkaplan94}.  Hence the conclusions from our analysis also describe the simulations of \citet{gordocharlesworth02} and are consistent with all other results based on this structured coalescent approach.  Our work is also closely related to recent work in continuous-fitness model by \citet{ofallonadler10}, which uses a similar framework to analyze the weak-selection regime but not the $Ns \gg 1$ situation we study here.

Our approach leads to simple expressions for the coalescent probability at each step in
our fitness-class genealogical process.  This makes it a complete effective coalescent
theory:  using these probabilities, we can calculate the probability that a sample of
individuals has any particular ancestral relationship.  Our coalescent probabilities are
different from those in the standard Kingman coalescent \citep{kingman82a}, so the
structure of genealogies has a different form.

Of course, since our process is an effective rather than an actual coalescent, the
relationship between a fitness-class genealogy and the expected statistics of genetic
variation given that genealogy is different than in the standard neutral coalescent.
Given a particular genealogy measured in steptimes, the numbers of deleterious mutations
\emph{are} the coalescent times, and to calculate the statistics of neutral variation we
have to make use of the relationship between steptimes and actual coalescence times.  This
contrasts with the Kingman coalescent, where numbers of neutral mutations are typically
Poisson-distributed variables with means proportional to coalescence times
\citep{wakeleybook}.  However, we can account for these differences by starting with the
distribution of fitness-class genealogies and then converting these genealogies into
actual coalescence times.

In this paper, we have used this fitness-class approach to calculate simple statistics
describing genetic variation, in particular the distribution of pairwise heterozygosity.
This leads to analytic expressions for the quantities of interest, although these
expressions involve sums which are most easily calculated numerically.  These are easy to
compute, and do not become harder to evaluate in larger populations, and hence are more
efficient to evaluate than either simulations or calculations within the ancestral
selection graph.

\subsection{An Intuitive Picture of the Structure of Genealogies}
The most important aspect of our analysis is not the specific results for heterozygosity, which match the conclusions of earlier simulations.  Rather, the fitness-class coalescent approach allows us to draw several important general conclusions about how negative selection distorts the structure of genealogies.  For two individuals drawn from particular fitness classes, the effect of negative selection is similar to that of an effective population size that changes as time recedes into the past, as has been suggested by earlier work.  However, this is not a population size that decreases in a simple way into the past.  Our analysis shows the exact form of this time dependent population size.  Further, it is clear from our analysis that this is not the only effect of negative selection on genealogies.  There are two key complications.  First, the statistics of genetic variation (particularly at the deleterious sites themselves) depend on the structure of genealogies differently in our fitness-class coalescent than in the standard neutral coalescent.  Second, different pairs of individuals have a different time-varying effective population size.  This means that genetic diversity cannot be represented by a single time-varying effective $N_e(t)$ for the whole population, which means that it may be possible to develop statistical tests to distinguish negative selection from population size.  All of these general intuitive conclusions about the structure of genealogies in our fitness-class coalescent are illustrated in \fig{newfig}.

We now pause to make this intuitive picture of the shape of typical genealogies more precise.  In general the probability that two individuals will coalesce within class $k$ has the form $P_c \approx \frac{A}{2} \frac{1}{n_k s k}$, where $n_k$ is the population size of that class, $s k$ is the effective selection pressure against individuals within that class, and $A$ is a constant that depends on which classes the lineages began in, but not on any of the population parameters.  We have seen that each lineage spends on average $\frac{1}{s k}$ generations in class $k$.  Thus we can think of each individual as seeing a historical effective population size as shown in \fig{newfig}c: it starts in some class $k$ with size $n_k$ and spends $\frac{1}{sk}$ generations in that class before moving to class $k-1$, and so on.

If we sample two individuals, however, they will not always be in the same class at the same time.  This effect reduces the coalescence probabilities in each class, as captured by the factor $A/2$.  This factor is the average fraction of the $\frac{1}{sk}$ generations each lineage spends in class $k$ that the two lineages spend there together.  Alternatively, we can think of this factor as consisting of two parts:  $A$ is the probability that the two lineages are ever in the same class at the same time, and $\frac{1}{2 s k}$ is the average amount of time that they coexist in the class if they coexist at all (they each spend on average $\frac{1}{s k}$ generations there, but on average overlap for only half this time if they overlap at all). While the two lineages are in the class at the same time, the per-generation coalescent probability is $\frac{1}{n_k}$.

This logic implies that genealogies in the presence of purifying selection look like neutral genealogies with a specific type of historical population size dependence.  Imagine for example we picked two individuals from the same fitness class $k$.  They each spend on average $\frac{1}{s k}$ generations in class $k$, and during that time they have a probability $ \frac{A}{2} \frac{1}{n_k}$ per (real) generation of coalescing (this probability includes the fact that on average they are both in the class simultaneously for only a fraction of the mean time each spends there).  So roughly speaking, they have an effective population size of $N_e \sim 2 n_k / \aklm{k}{k}{\ell=0}$ for the first $\frac{1}{ s k}$ generations.  If they fail to coalesce, they then move to class $k-1$, where they spend $\frac{1}{s (k-1)}$ generations and have a probability $\frac{A}{2} \frac{1}{n_{k-1}}$ per generation of coalescing, and hence an effective population size $N_e \sim 2 n_{k-1}/ \aklm{k}{k}{\ell=1}$ for this time.  If they again fail to coalesce, they move to class $k-2$, and so on.

So far, this picture of a time-dependent population size is rather crude, but we can make it more precise.  Specifically, we can write the coalescence probability between two individuals sampled from class $k$ and $k'$ as a function of time in generations as \eon \probt (t|k, k') = \sum_{\ell=0}^k \pst{k}{k'}{\ell} \probt (t|k, k', \ell). \eoff  We can then define the time-dependent effective population size between these individuals, $N_e(t)$, as the inverse probability of coalescence at time $t$ given that coalescence has not yet occurred, \eon \frac{1}{N_e(t)} = \frac{\probt (t|k, k')}{1 - \int_0^t \probt (t'|k, k') dt'}. \eoff  In other words, the $N_e(t)$ is defined as usual as the inverse of the probability that the two individuals will coalesce at time $t$ given that they have not yet done so.

We illustrate this precise time-dependent population size $N_e(t)$ in \fig{newfig}d.  We see that for two individuals sampled from the same fitness class, $N_e(t)$ typically increases into the recent past and then decreases into the more distant path.  This reflects the fact that the two individuals are becoming less likely to be in the same fitness class in the recent past, but that as time recedes into the distant past they are likely to be in the highly fit classes which have smaller $n_k$.  For two individuals sampled from classes near but not identical to each other, $N_e(t)$ starts high and then drops before exhibiting a pattern similar to that among individuals sampled from the same class.  This reflects the fact that it takes at least a short time before the two individuals have any chance of being in the same class.  Finally, for two individuals sampled from more distant classes, $N_e(t)$ simply declines into the past, both because longer ago they were more likely to be in the same class and more likely to be in the small classes near the high-fitness tail.

Averaging over the whole population, \fig{newfig}d shows the precise time-dependent population size $N_e(t)$ for two randomly sampled individuals.  This average $N_e(t)$ initially stays roughly constant as time recedes into the past before decreasing thereafter.  For these two randomly sampled individuals, selection is indistinguishable from this particular historically varying population size (although this particular type of variation in population size is presumably rather unusual). The distribution of coalescence times between this pair of individuals looks the same as neutral coalescent histories with this specific population size history.  The deleterious mutation rates and selection pressures only matter in that they determine the form of this population size history.

However, a key difference from a neutral population of time-varying size is that, as is clear in \fig{newfig}d, pairs of individuals do not typically come from the same fitness class. Rather, they come at random from different parts of the fitness distribution, and those that come from different places have ancestries characterized by different historically varying population sizes.  The total distribution of ancestry is the sum of all of these.  In other words, the genetic variation within the population is like that in a population where some individuals had one type of historical population size history, while others had another. If we restrict ourselves to pairwise statistics such as $\pi$, the average $N_e(t)$ across pairs of individuals will accurately describe the genetic diversity.  However, when we consider appropriately defined statistics in larger samples, the fact that there is no single $N_e(t)$ for the whole population could be important.  It remains an interesting question for future work to explore how to exploit this fact to develop statistical tests to distinguish the effects of purifying selection from that of a historically varying effective population size.

\subsection{Approximations underlying our approach}
Our analysis relies on three key approximations.  First, both our lineage-structure and our sum of ancestral paths methods assume that we can neglect fluctuations in the total frequency $h_k$ of each class.  Related to this approximation, we have also implicitly assumed that the probability a lineage in class $k$ reaches a frequency close to $h_k$ can be neglected.  In \citet{allelebased}, we analyzed these approximations in detail and showed that they will hold in class $k$ whenever $N h_k s k \gg 1$.  In practice, this condition will often break down in the high and low-fitness tails of the fitness distribution. Fortunately, provided it holds in the bulk of the distribution in which most individuals will be sampled (which will typically be true provided $N s \gg 1$), our approach will still be a good approximation.

Our second key approximation is the non-conditional approximation, which we discuss in more detail in Appendix A. This approximation is also made in a more subtle way in the summing over ancestral paths method, as described in Appendix D, though we note that in computing the distribution of $\pi_d$ it is possible to avoid this approximation in this method.

Our final and most important approximation is that we assume that Muller's ratchet can be neglected.  We can think of this as the most extreme aspect of our approximation neglecting fluctuations in the sizes of each fitness class.  This approximation can sometimes be problematic; we discuss it in detail below.

Although we have focused primarily on situations when selection is weak compared to total deleterious mutation rates, our approach is also valid regardless of whether $s$ is strong or weak compared to $\ud$.  However, when selection is sufficiently strong ($Ns \gg 1$ and $\ud/s < 1$), then an effective population size approximation accurately describes the patterns of genetic variation, as we describe below.  Thus our methods are primarily useful for situations where selection is weak compared to mutation rates.

\subsection{Relationship with an effective population size approximation}
\citet{charlesworth93} considered how selection against many linked deleterious mutations affects linked neutral diversity in a model identical to ours.  These authors found that when selection is sufficiently strong, the shape of genealogies and hence the statistics of variation at linked neutral sites is identical to the neutral case, with a reduced effective population size. We refer to this as the effective population size (EPS) approximation.

The idea behind the EPS approximation is that when selection is strong, deleterious mutations are quickly eliminated from the population by selection.  Thus if we sample individuals from the population, they must have very recently descended from individuals within the class of individuals which had no deleterious mutations (the $0$-class).  The EPS approximation assumes that the time for this to happen can be neglected, and that individuals never coalesce before it does.  These individuals then coalesce within the $0$-class as a neutral process with effective population size equal to the size of that $0$-class, which is $N e^{-\ud/s}$.  Thus the genetic diversity within the population is identical to that in a neutral population of reduced size $N_e = N e^{-\ud/s}$.

The EPS approximation is valid provided that the neutral coalescence time within the $0$-class, $t_{neut}$, is large compared to the time it takes for a typical individual to have descended from the $0$-class, $t_{desc}$.  We know $t_{neut} \sim N e^{-\ud/s}$, and since a typical individual comes from fitness class $k \sim \ud/s$, we have that $t_{desc} \sim \sum_{j=1}^{\ud/s} \frac{1}{js} \sim \frac{1}{s} \ln \left( \frac{\ud}{s} \right)$.  This means that the EPS approximation will be valid provided \eon N s e^{-\ud/s} \gg \ln
\left( \frac{\ud}{s} \right) . \eoff  Because of the exponential term on the left hand side of this expression, it is clear that the EPS approximation is a strong-selection, weak-mutation limit.  It will tend to be valid provided that $Ns > 1$ and $\ud < s$, but whenever $\ud$ becomes much larger than $s$, it will typically break down even in enormous populations.

Our analysis describes the effects of background selection beyond the EPS approximation.  We do not assume that the coalescence time through the fitness distribution is small compared to the coalescence times within the $0$-class, or that coalescence cannot occur among individuals carrying deleterious mutations.  It is precisely these two effects that lead to distortions away from the neutral expectations, making it impossible to describe genealogies using neutral theory with a revised effective population size.  Although our analysis is a generalization of the EPS approximation, it is not inconsistent with it.  However, we have focused primarily on situations where the EPS approximation breaks down, and coalescence times through the fitness distribution are large compared to those in the $0$-class, because this is the situation where our approach is most useful.

Note also that in many situations it may be the case that there are many linked weakly selected mutations \emph{and} many linked strongly selected mutations.  In such circumstances, the process we consider and the EPS approximation can act simultaneously, each for different classes of mutations. Imagine we had one class of mutations with fitness cost $s_1$ which occur with mutation rate $U_1$, where $U_1 < s_1$ and $N s_1 \gg 1$ so that the EPS approximation applies.  At the same time, imagine another class of mutations with fitness cost $s_2$ which occur with mutation rate $U_2$, where $U_2 \gg s_2$ so that the EPS approximation breaks down for these mutations.  In this case, the genetic diversity we expect to see will be characteristic of our fitness-class coalescent theory (with $\ud = U_2$ and $s = s_2$), but with a reduced effective population size $N_e = N e^{-U_1/s_1}$. In other words, the strongly selected mutations reduce the effective population size because all individuals are very recently descended from an individual that had no large-effect mutations, but the coalescence time through the distribution of weakly selected mutations cannot be neglected.

\subsection{A ``Mutation-time" Approximation}
We have seen that our analysis accounts for two effects missing from the EPS approximation: coalescence events outside the $0$-class, and the time it takes for individuals to have descended from the $0$-class.  Whenever $\ud/s$ and $N$ are both sufficiently large, the former effect can be neglected while the latter is still important, because the number of lineages in each fitness class becomes large and hence coalescence events are very unlikely to occur outside of the $0$-class.  This leads to an approximation which we can think of as a generalization of the EPS approximation.  Rather than considering primarily the diversity generated within the most-fit background, we focus instead on the diversity that accumulates while lineages move between different less-fit backgrounds.  Hence we term this approach a ``mutation-time approximation'' (MTA) for short.  In this approximation, we assume that all individuals coalesce within the $0$-class, as with the EPS approximation.  However, unlike the EPS approximation, we consider the time it took for individuals to descend from the $0$-class in addition to the coalescence time within the $0$-class.  This approximation is valid for large $N$ (when even $N h_1$ is enormous compared to $\frac{1}{s}$) so that coalescence always occurs in the $0$-class.

In this mutation-time approximation our results become much simpler and provide a useful intuitive picture of the structure of genealogies and genetic variation.  Consider the deleterious heterozygosity $\pi_d$ of two individuals sampled from fitness classes $k$ and $k'$.  In this approximation, these two individuals always coalesce in the $0$-class so we always have $\pi_d = k + k'$.  Since two individuals are sampled from classes $k$ and $k'$ with probability $\hdistn (k, k')$, the distribution of $\pi_d$ in the population as a whole is extremely simple:  we have \eon \ppid (\pi_d) = \sum_{k = \pi_d - k'} \hdistn (k, k') = e^{-2 \ud/s} \frac{1}{\pi_d!} \left(\frac{2 \ud}{s} \right)^{\pi_d} . \eoff

\begin{figure}
\includegraphics[width=4.0in]{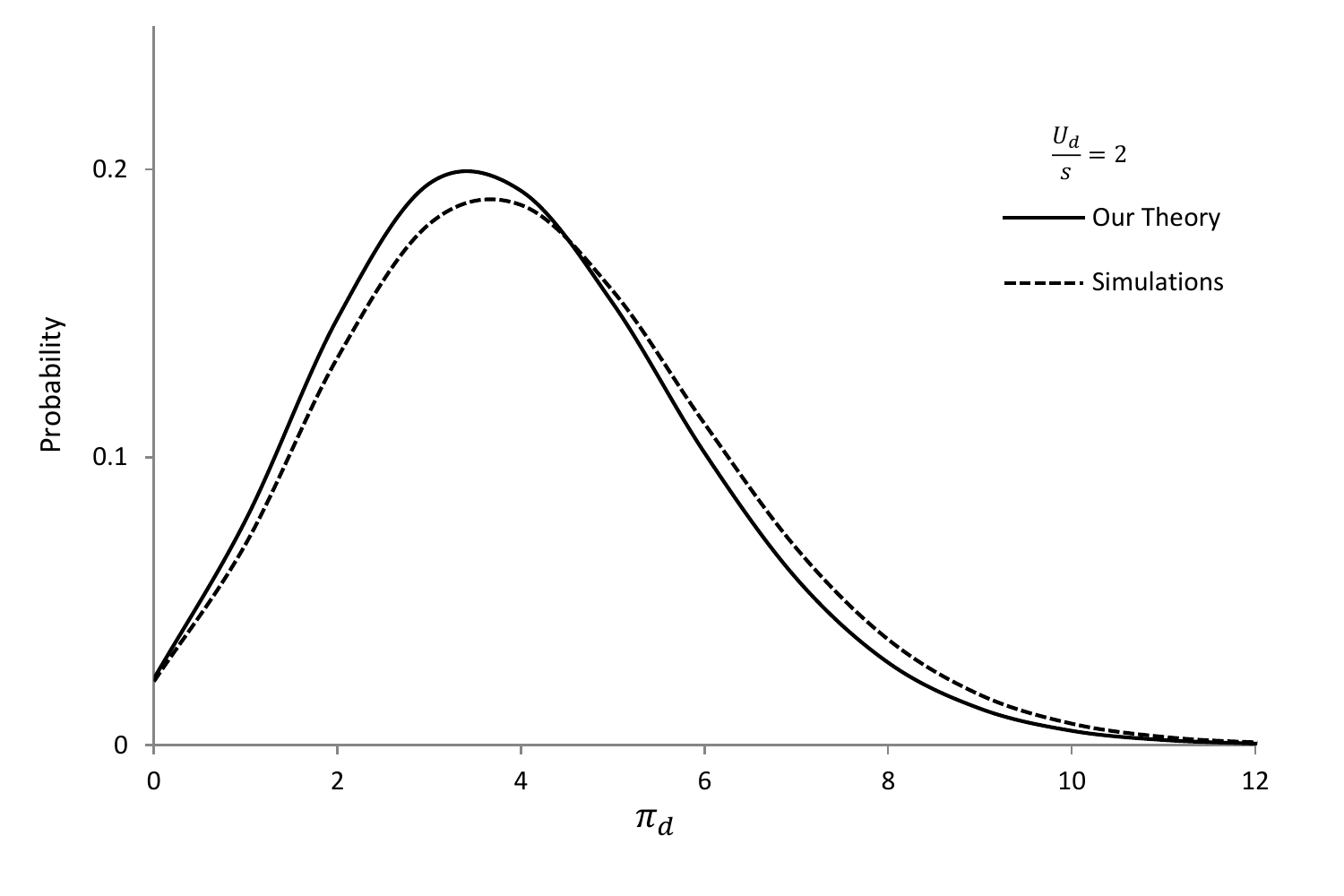}
\includegraphics[width=4.0in]{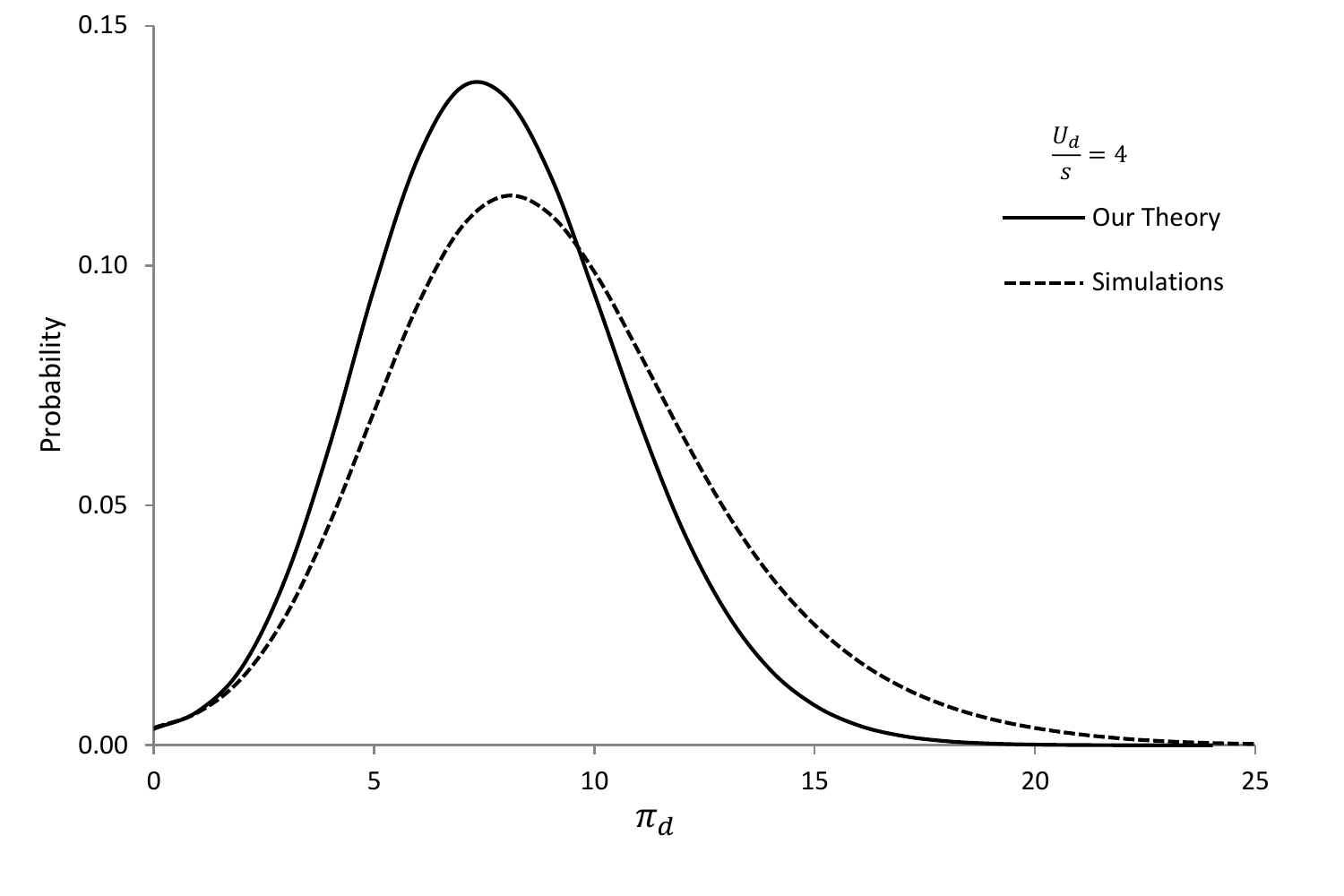}
\caption{Characteristic examples of the distribution of $\pi_d$.  Here $N = 5 \times 10^4$, $s = 10^{-3}$ and in \textbf{(a)} $\ud/s = 2$, while in \textbf{(b)} $\ud/s = 4$.  Theoretical predictions are shown as a solid line, simulation results as a dashed line.  The fit to simulations is good, but we tend to slightly underestimate the coalescence times, and this tendency is worse for larger $\ud/s$.  This is due to Muller's ratchet, which becomes more problematic as we increase $\ud/s$.  This systematic underestimate becomes less severe (for all values of $\ud/s$) as $N$ increases, as expected, but comprehensive simulations for much larger $N$ are computationally prohibitive.  }
\label{fig6}
\end{figure}

This simple approximation makes it clear why the distribution of $\pi_d$ looks the way it does, and explains how it varies with $\ud/s$ and with $N$, both in this mutation-time approximation and more generally.  For large $N$, when coalescence outside the $0$-class can be neglected, two individuals from class $k$ and $k'$ have $\pi_d = k + k'$.  Thus the distribution of $\pi_d$ has roughly the same shape as the distribution of fitness within the population.  The mean $\pi_d$ is $2 \ud/s$, since the average individual comes from class $k = \ud/s$.  Smaller and larger $\pi_d$ are less likely; the distribution of fitness in the population has variance equal to the mean, so the variance of the distribution of $\pi_d$ is also roughly equal to its mean.  As $N$ gets smaller, there is sometimes coalescence outside of the $0$-class.  This reduces $\pi_d$ given $k$ and $k'$.  Hence as we reduce $N$, the distribution of $\pi_d$ shifts somewhat leftwards, with a peak somewhat below $2 \ud/s$, and has slightly more variance since there is a less definite correspondence between $k, k'$, and $\pi_d$.  Since $\pi_n$ is determined by $\pi_d$, this also explains why the distribution of $\pi_n$ has the peaked form we observe, and how it depends on $\ud/s$ and $N$ (note that for $\pi_n$ the coalescence time within the $0$-class, which increases linearly with $N$, must also be included).  All of these intuitive expectations are reflected in our results, as shown in \fig{fig6}, \fig{fig7}, \fig{fig8}, and \fig{fig9}.  Note for example that in \fig{fig6}, the peak of $\pi_d$ is slightly below $2 \ud/s$ (reflecting the finite population size) and has variance about equal to its mean; we have verified that as $N$ increases the shape of the distribution remains roughly the same, but the mean increases towards $2 \ud/s$ and the variance decreases slightly.

More complex statistics of sequence variation are similarly straightforward to calculate in the mutation-time approximation. When considering larger samples, the genetic diversity is determined by the fitness classes these individuals come from, which is always simple since the probability a given individual is sampled from fitness class $k$ is just the Poisson-distributed $h_k$.  This approximation may therefore prove useful in developing simple and intuitive expressions for various statistics.  For example, we can use this approximation to calculate a simple expression for the distribution of the total number of segregating negatively selected sites in a sample of size $n$, $S_n^d$, which as we have seen above is otherwise rather involved.  We have \eon \rho(S_n^d = x) = \sum_{k_1, k_2, \ldots k_n} h_{k_1} h_{k_2} \ldots h_{k_n}, \eoff where the sum is over sets of the $k_i$ that sum to $x$.  We find \eon \rho(S_n^d = x) = e^{-n \ud/s} \frac{1}{x!} \left( \frac{n \ud}{s} \right)^x . \eoff  This is a distribution which is peaked around a mean value of $\frac{n \ud}{s}$, for the same reasons the distribution of $\pi_d$ looks as it does.  We note however that as we increase the sample size $n$ the population size $N$ must be even larger for this MTA approximation to hold.

We can also calculate the distributions of actual coalescence times and hence the distributions of statistics describing neutral diversity in the mutation-time approximation.  Consider the distribution of the real coalescence time between two individuals chosen from classes $k$ and $k'$. In the mutation-time approximation where the coalescence time within the $0$-class can be neglected, the actual coalescence time is as given in \eq{realtimescond3}, \eon \probt_1 (t|k, k') = s (k+k') e^{-s(k+k')t} \left( e^{st} - 1 \right)^{k + k' - 1} .  \eoff  Averaging over the values of $k$ and $k'$, we have \eon \probt_1 (t) = \sum_{k=0}^{k'} \sum_{k'=0}^\infty \hdistn (k, k') \probt(t | k, k'). \eoff  The distribution of coalescence times once within the $0$-class is, as before, $\probt_2 (t) = \frac{1}{N h_0} e^{-t/(N h_0)}$.  From this distribution of real coalescence times, we can find the distribution of neutral heterozygosity $\pi_n$ in the usual way, \eon \ppin (\pi_n) = \int_0^\infty \frac{\left[ 2 U_n t \right]^{\pi_n}}{\pi_n!} e^{-2 U_n t} \probt (t) dt. \eoff

We can immediately see that the average coalescence time in this MTA approximation is $t \approx \sum_0^{\ud/s} \frac{1}{si} + N h_0 \approx \frac{1}{s} \log \left( \ud / s \right) + N h_0$.  We therefore expect that the neutral heterozygosity will on average be \eon \ev{\pi_n} \sim \frac{2 \un}{s} \log \left( \frac{2 \ud}{s} \right) + 2 N h_0 \un. \eoff  The first term in this expression comes from the time to descend through the fitness distribution, while the second term comes from the time to coalesce within the $0$-class.  If this latter term is large compared to the former, the EPS approximation applies.  In the opposite case where the time to descend through the distribution dominates, we can see from the MTA approximation that, as with $\pi_d$, the shape of this distribution of $\pi_n$ is primarily determined by the shape of $\hdistn (k, k')$.  In this case, we find that the peak in $h_k$ at $k = \ud/s$ leads to the peak in the distribution of real times and hence the peak in the distribution of $\pi_n$.  The width of the distribution of $\pi_n$ is somewhat wider, however, since even given individuals coming from fitness classes near the mean, there is a broad distribution of possible real times, and a broad distribution of $\pi_n$ even given a particular real time.

This average heterozygosity would correspond to an effective population size of \eon N_e \sim \frac{1}{s} \log \left( \frac{2 \ud}{s} \right) + N h_0, \eoff but as we have seen this effective population size cannot correctly describe the full distribution of $\pi_n$ nor its relationship to other statistics describing the genetic diversity.  For smaller values of $N$ where the mutation-time approximation breaks down, the average $\pi_n$ would be somewhat lower than the MTA predicts, and its distribution somewhat broader.

\subsection{Muller's Ratchet}
We have neglected Muller's ratchet throughout our analysis, and assumed that the fitness distribution $h_k$ is fixed.  Yet Muller's ratchet will certainly occur, and in some circumstances could have a significant impact on genetic diversity \citep{gordocharlesworth02, seger10}.  Thus this is a potentially important omission from our theory.  In this section we discuss some of the complications associated with Muller's ratchet that are important to keep in mind when considering our approach.  We discuss the parameter regimes where neglecting Muller's ratchet should be reasonable, and those where it is likely to cause more serious problems.  We provide rough estimates of how large we expect these problems to be, and suggest a few possible ways in which future work might incorporate Muller's ratchet into our general framework.

Muller's ratchet causes several related problems within our theoretical framework.  First, it causes the values of $h_k$ to change with time, and means they may not always follow a Poisson distribution.  This changes the distribution of lineage frequencies within each class, and hence changes the coalescence probabilities.  After a ``click'' of the ratchet, the whole distribution $h_k$ shifts in a complicated way, eventually reaching a new state where it is shifted left (so the class that was originally at frequency $h_k$ is now at frequency $h_{k-1}$, and so on).  In a similarly complex way, the PRF distribution of lineage frequencies in class $k$ shifts from $f_k$ to $f_{k-1}$, and so on.  This naturally changes the coalescence probabilities in each class.  Fortunately, since the coalescence probabilities in class $k$ are generally very similar to those in classes $k+1$ or $k-1$, this effect is unlikely to lead to major inaccuracies provided the ratchet does not click many times within a coalescent time.  This is true except when we start considering coalescence in classes close to the $0$-class, where the $k$-dependence becomes significant.  This can be thought of as an additional problem associated with Muller's ratchet, and is associated with the fact that the ratchet shifts the whole fitness distribution.  This effect is easiest to see with an example:  imagine we sample two individuals within the $k$-class, and that these individuals did not coalesce before their ancestors were both in the $0$-class.  At the time (in the past) when these individuals' ancestors were in the $0$-class, this current $0$-class might have been the $1$-class or $2$-class (or higher).  Thus these two individuals within the $0$-class might not coalesce until, for example, their ancestors were in what is currently the ``$-2$''-class.  This clearly means that we might in fact have $\pi_d > 2k$, which our analysis assumes is impossible.  In fact, we observe precisely this effect in simulations, and it is the reason why we commonly observe systematic deviations where the simulated values of $\pi_d$ are larger than our theory predicts.

From this discussion it is clear that the key factor in determining whether Muller's ratchet can reasonably be neglected is how many times the ratchet ``clicks'' in a coalescence time.  We have seen above that an average individual coalesces through the fitness distribution in a time at most of order $\frac{1}{s} \ln \left( \ud/s \right)$ generations.  Once within the $0$-class, coalescence times are of order $N e^{-\ud/s}$.  We must compare these times to the time it takes for the ratchet to ``click.'' The rate of the ratchet is a complex issue that has been analyzed by \citet{gordocharlesworth00a}, \citet{gordocharlesworth00b}, and \citet{kimstephan02} in the regime where $N e^{-\ud/s} > 1$ and by \citet{gessler95} in the regime where $N e^{-\ud/s} < 1$.  No general analytic expressions exist which are valid across all parameter regimes.  However, provided the ratchet does not typically move a substantial fraction of the width of the fitness distribution in the coalescence time of two random individuals, it will be a small correction to $\pi_d$, and neglecting it is a reasonable first approximation.  In practice we find in our simulations that for the parameter regimes we consider, the ratchet causes $\pi_d$ to be at most of order $2$ larger than our theoretical predictions, corresponding roughly to a single click of the ratchet during a typical coalescence time.

The discussion above suggests a way to incorporate Muller's ratchet within our theoretical framework, albeit in an ad-hoc way.  The ratchet shifts the distribution $h_k$ underneath the fitness-class coalescent process.  The details of this shift are complicated, but on average every click of the ratchet shifts the distribution one step to the left.  We can define $k_{min}$ to be the number of deleterious mutations (relative to the optimal genotype) in the most-fit individual at any given time.  For the case where $N e^{-\ud/s} > 1$, the rest of the distribution will be approximately a Poisson distribution, but with $h_k$ replaced by $h_{k-k_{\min}}$.  Muller's ratchet can then be thought of as a process by which $k_{min}$ increases over time.  This increase is a random process, but has some average rate, leading to an average $k_{min}(t)$.  As we look backwards in time during the fitness-class coalescent process, the value of $k_{min}$ is decreasing due to Muller's ratchet.  This suggests a simple approximation:  we replace the actual value of $k$ with an ``effective'' value of $k$ that accounts for the fact that $k_{min}$ decreases as we look backwards in time.  For each step through the fitness distribution, we imagine that $k_{min}$ has decreased by the appropriate amount, and hence the effective value of $k$ in the new fitness class is decreased by less than $1$ compared to the old fitness class.  When $N e^{-\ud/s} < 1$ the ratchet is an almost deterministic process, so a similar approximation may prove useful, but in this case the distribution $h_k$ is on average shifted from the Poisson form \citep{gessler95}.  To incorporate the ratchet into our analysis in this situation, we first must recalculate the relevant coalescence probabilities given the expected average form of $h_k$, and then carry out the above program.  These and other methods to account for Muller's ratchet remain an interesting topic for future work.

Despite the potential relevance of Muller's ratchet in practical situations, we note that it does not affect our results in the standard coalescent limit.  As is apparent from our general expressions for the coalescence probabilities, the structure of our fitness-class coalescent theory does not depend on all three parameters $N$, $\ud$, and $s$ independently.  Rather, it depends only on the combinations $N \ud$ and $Ns$.  Thus our theory makes sense in the standard limit where $N \ud$ and $Ns$ are held constant while we take $N \to \infinity$.  In this limit, Muller's ratchet does not occur.  Whether this means we can neglect the ratchet for large but finite $N$ depends on the convergence properties of the coalescent limit. This is a difficult limit to explore with simulations, because it requires large population sizes.  However, we have used simulations to verify in a few cases that, as expected, increasing $N$ while keeping $N \ud$ and $Ns$ constant does not change the predicted structure of genealogies but decreases some of the systematic differences between theoretical predictions and the simulations which are suggestive of the effect of the ratchet.  Note that while this ratchet-free limit does not change the structure of genealogies in our fitness-class coalescent, the distribution of real coalescent times does change, since all real timescales are proportional to $s$.  Thus, as might be expected, we must also take $N \un$ constant as $N \to \infinity$ if we wish neutral diversity to also remain unaffected in this limit.

Note that this ratchet-free limit, while fairly standard in coalescent theory, is somewhat different from the mutation-time approximation we discussed above.  Of course, we can easily imagine a population which is large enough that the mutation-time approximation applies, and \emph{then} take the standard coalescent limit.

\subsection{Conclusion}
Our fitness-class coalescent approach provides a framework in which we can compute distributions of genealogical structures in situations where many linked negatively selected sites distort patterns of genetic variation.   We have used this framework to calculate the distributions of a few simple statistics describing sequence variation.  It remains for future work to use this fitness-class coalescent approach to compute a wide array of statistics to better understand the details of how purifying selection on many linked sites distorts patterns of genetic variation.  The eventual goal will be to use our results to help interpret the increasing amounts of sequence data which seem to point to the importance of negative selection on many linked sites.

\section*{Acknowledgments}

We thank Daniel Fisher and John Wakeley for many useful discussions, which inspired our fitness-class coalescent approach.  MMD acknowledges support from the James S. McDonnell Foundation.  AMW thanks the Princeton Center for Theoretical Science at Princeton University, where she was a fellow during some of her work on this paper. LEN is supported by the Department of Defense through the National Defense Science and Engineering Graduate Fellowship Program, and also acknowledges support from an NSF graduate research fellowship.  JBP acknowledges support from the James S. McDonnell Foundation, the Alfred P. Sloan Foundation, the David and Lucille Packard Foundation, the Burroughs Wellcome Fund, Defense Advanced Research Projects Agency (HR0011-05-1-0057), and the US National Institute of Allergy and Infectious Diseases (2U54AI057168).  Many of the computations in this paper were run on the Odyssey cluster supported by the FAS Sciences Division Research Computing Group at Harvard University.

\newpage

\section*{Appendix A:  The full conditional calculation}

In the main text, we focused primarily on the non-conditional approximation to the coalescence probabilities, which led to our simple expression for the coalescenct probabilities, \eq{pckknoncon}.  We saw in the main text that this non-conditional approximation can be relaxed by keeping the higher order terms in \eq{sumpathscoalprob}.  In this Appendix, we show how this approximation can be relaxed in our lineage-structure framework by carrying out the full conditional calculation for some of the simplest possible cases.  We use this to understand the structure of the conditional results and discuss the validity of the non-conditional approximation.  We note that the full conditional result can also be obtained from the sum of ancestral paths approach, as described in Appendix D, and the validity of the non-conditional approximation can be directly assessed with that approach.

We begin by considering the full conditional result for the probability that two individuals both sampled from class $k$ coalesce in class $k-2$.  In the main text we found that this coalescence probability is \begin{eqnarray} \pck{k}{k-2} & = & \int \probtimes{k-2}{k,k} (t_1, t_2) \frac{x^2}{(h_{k-2})^2} f_{k-2}(x) \exp \left[ -s (k-2) |t_1-t_2| \right] dt_1 dt_2 dx \\ & & = I_x^{k-2} \int \probtimes{k-2}{k,k}(t_1, t_2) \exp \left[ -s (k-2) |t_1-t_2| \right] dt_1 dt_2. \end{eqnarray}

In order to evaluate this integral, we need to determine the probability distribution of mutant timings $\probtimes{k-2}{k,k}(t_1,t_2)$.  The time $t_1$ is now the sum of the time for one individual to have mutated from class $k-2$ to class $k-1$ plus the time for it to have mutated from class $k-1$ to class $k$, and analogously for $t_2$.  However, in order for the two lineages to coalesce in class $k-2$, they must \emph{not} have coalesced in class $k-1$.  To illustrate the main point, we neglect the distortion in the mutant timings due to the fact that individuals did not coalesce in class $k$ and focus only on the distortions due to the fact that coalescence did not occur in class $k-1$; if desired, the former distortion can also be included using analogous methods.  We refer to the probability distribution of the times when these individuals mutated from class $k-1$ to class $k$ conditional on them not having coalesced in class $k-1$ as $\probtimes{k-1}{k,k} (t_1, t_2 | nc)$.  The distribution of the times for these individuals to then have mutated from class $k-2$ to class $k-1$ is then given by \eon \probtimes{k-2}{1step} = [s(k-1)]^2 e^{-s(k-1)(t_1 + t_2)}, \eoff as in the first step.  Thus the distribution of $t_1$ and $t_2$ is given by \eon \probtimes{k-2}{k,k} (t_1, t_2) = \probtimes{k-1}{k,k} (t_1,t_2|nc) \star \probtimes{k-2}{1step}(t_1,t_2), \label{convolution} \eoff where $\star$ indicates a convolution.  Note that much of the time when the individuals did coalesce in class $k-1$, they did so because $t_1$ happened to be close to $t_2$ (since this increases the chance the two individuals mutated from the same lineage).  Thus in $\probtimes{k-1}{k,k} (t_1,t_2|nc)$, $t_1$ and $t_2$ are on average further apart than in $\probtimes{k-1}{k,k} (t_1,t_2)$, and $t_1$ and $t_2$ are no longer independent random variables.

We now need to calculate $\probtimes{k-1}{k,k} (t_1,t_2|nc)$.  We have \eon \label{conditional} \probtimes{k-1}{k,k} (t_1,t_2|nc) = \frac{\probtimes{k-1}{k,k}(t_1,t_2) - \probtimes{k-1}{k,k}(t_1,t_2|c) \pck{k}{k-1}}{1 - \pck{k}{k-1}}, \eoff where $\probtimes{k-1}{k,k} (t_1,t_2|c)$ is the distribution of timings of mutations from class $k-1$ to $k$ given that the lineages \emph{do} coalesce in class $k-1$.  Applying the general probability identity $P(t_1,t_2|c) = \frac{1}{P(c)} P(c|t_1,t_2) P(t_1,t_2)$, and reading off the coalescence probability given $t_1$ and $t_2$ from \eq{onestep}, we find that \eon \probtimes{k-1}{k,k} (t_1,t_2|c) = \frac{I_x^{k-1}}{\pck{k}{k-1}} \probtimes{k-1}{k,k} (t_1,t_2) e^{-s(k-1)|t_1-t_2|}. \label{didcoalesce} \eoff  We therefore find \eon \probtimes{k-1}{k,k}(t_1,t_2|nc) = \frac{1}{1-\pck{k}{k-1}} \left[ (sk)^2 e^{-s k (t_1+t_2)} - I_x^{k-1} (sk)^2 e^{-2 k (t_1+t_2)} e^{-s(k-1)|t_1-t_2|} \right]. \eoff

Plugging this into our convolution formula for $\probtimes{k-2}{k,k}(t_1,t_2)$ and evaluating the integrals by separating out the possible time orderings, we find \eon \probtimes{k-2}{k,k}(t_1,t_2) = \frac{k^2 \left[ s (k-1) \right]^2}{1 - \pck{k}{k-1}} e^{-s(k-1)(t_1+t_2)} \left[ \left( 1 - e^{-s t_1} \right) \left( 1 - e^{-2 t_2} \right) - \frac{I_x^{k-1}}{k-2} B \right], \label{twostepcond} \eoff where we have defined \eaon B & = & \frac{1}{(k-2)} \left[ 1 - e^{-2 s \min (t_1,t_2)} - \frac{2}{k} \left( 1 - e^{-s k \min (t_1,t_2)} \right) \right. \nonumber \\ & & \left. + \frac{1}{k} \left( 1 - e^{-2 k |t_1 - t_2|} \right) \left( e^{-2 s \min (t_1,t_2)} - e^{-s k \min (t_1,t_2)} \right) \right]. \eaoff  We can now use this expression in \eq{secondstep} to calculate the coalescence probability $\pck{k}{k-2}$.  Since the result is tedious and does not further illuminate the structure of the full conditional calculation, we do not do so explicitly here, but the integrals are straightforward to evaluate with the methods we have used above.

To motivate the validity of the non-conditional approximation, we need to consider the full calculation going back one additional step.  Thus we consider the probability that two individuals both sampled from class $k$ coalesce in class $k-3$, $\pck{k}{k-3}$.  This will be given by \eon \pck{k}{k-3} = \int \probtimes{k-3}{k,k}(t_1,t_2) \frac{x^2}{h_{k-3}^2} f_{k-3}(x) e^{-s(k-3)|t_1-t_2|} dt_1 dt_2 dx, \eoff where here $\probtimes{k-3}{k,k}(t_1,t_2)$ is the distribution of the time at which the ancestors of the two sampled individuals originally mutated from class $k-3$ to class $k-2$, conditional on them not coalescing in classes $k-2$ or $k-1$.

We can calculate $\probtimes{k-3}{k,k}(t_1,t_2)$ in the same way we calculated $\probtimes{k-2}{k,k}(t_1,t_2)$.  Explicitly, \eon \probtimes{k-3}{k,k}(t_1,t_2)= \probtimes{k-2}{k,k}(t_1, t_2|nc) \star \probtimes{k-3}{1step} (t_1, t_2) , \label{threeinit}\eoff where analogously to the expression in the previous step \eon \probtimes{k-2}{k,k}(t_1, t_2|nc)=\frac{1}{1-\pck{k}{k-2}}\left[\probtimes{k-2}{k,k} (t_1, t_2)- \probtimes{k-2}{k,k}(t_1, t_2|c)\pck{k}{k-2}\right] . \eoff We note that $\probtimes{k-2}{k,k} (t_1, t_2)$ is the expression in \eq{twostepcond} we calculated above. As before, we have \eon \probtimes{k-2}{k,k}(t_1, t_2|c) \pck{k}{k-2}=I_x^{k-2} \probtimes{k-2}{k,k}(t_1, t_2) e^{-s(k-2) |t_1-t_2|} , \eoff hence we can write \eon \probtimes{k-2}{k,k}(t_1, t_2|nc)=\frac{\probtimes{k-2}{k,k} (t_1, t_2)}{1-\pck{k}{k-2}}\left[1- I_x^{k-2}e^{-s(k-2) |t_1-t_2|} \right] . \eoff Plugging the above expression back into Eq. \ref{threeinit}, we obtain \eaon  \probtimes{k-3}{k,k}(t_1,t_2)&=& \frac{s^2 (k-1)^2 k^2 s^2 (k-2)^2}{(1-\pck{k}{k-1})(1-\pck{k}{k-2})} e^{-s(k-2) (t_1+t_2)} \int_0^{t_2} \int_0^{t_1} e^{s(k-2)(y+z)} e^{s(k-1)(y+z)}  \nonumber \\ && \times \left[ 1-I_x^{k-2} e^{-s(k-z)|y-z|}\right] \left[ (1-e^{-sy})(1-e^{-sz})-\frac{I_x^{k-1}}{k-2} B\right]. \eaoff

We could evaluate the integrals in the above expression for $\probtimes{k-3}{k,k}(t_1,t_2)$ in the same way that we did in our calculation for $\probtimes{k-2}{k,k}(t_1,t_2)$.  We would then substitute this result for $\probtimes{k-3}{k,k}(t_1,t_2)$ into an analogous calculation of $\probtimes{k-4}{k,k}(t_1,t_2)$, and so on.  In this way we can build up the full conditional results.  The most useful way to go about this is to separate the results into powers of $I_x$, which is a small parameter related to the coalescent probability in each step.  We see from the expression for $\probtimes{k-3}{k,k}(t_1,t_2)$ that there is a term in $(I_x)^0$, which is exactly the non-conditional approximation.  There are two terms involving $(I_x)^1$, and a single term involving $(I_x)^2$.  In general, in the expression for $\probtimes{k-\ell}{k,k}(t_1,t_2)$, we will have one $(I_x)^0$ term (which equals the result in the non-conditional approximation) plus $\ell$ terms proportional to $I_x$, ${2 \choose \ell}$ terms proportional to $(I_x)^2$, and so on.  Fortunately, the dependence on the population parameters is entirely contained within these powers of $I_x$.  That is, the coefficients of these various powers of $I_x$ depend \emph{only} on $k$ and $\ell$, and not at all on the population parameters $N$, $s$, and $\ud$.  Thus we could simply calculate a table of coefficients once, and then would be able to understand all the distributions of mutant timings (and from this all the coalescent probabilities).

In practice, it is easier to make these full conditional calculations within the sum of ancestral paths approach.  As we have seen in the main text, that approach leads naturally to a power series in $I_x$ of exactly the form described above, in which the leading order term is the non-conditional approximation and the additional terms represent the conditional corrections.  This calculation shows that provided $I_x \ll 1$, which is true provided our usual condition that $N h_k s k \gg 1$ holds, these higher order terms are all small, and our non-conditional approximation is valid.

These full conditional results are, however, very complex and unilluminating.  Therefore we focus here on understanding the general structure of these results, and on showing why the non-conditional approximation is good description of the distribution of mutation timings.  We can see that at each step back through the fitness distribution, the probability distribution of times shifts from the non-conditional results by a factor which is roughly proportional to the coalescence probability at that step.  That is, in general we have \eon \probtimes{k-\ell}{k,k}(t_1,t_2) = \frac{1}{1-\pck{k}{k-\ell}} \left[ \probtimes{k-\ell}{k,k}(t_1,t_2) - \pck{k}{k-\ell} \probtimes{k-2}{k,k}(t_1,t_2|c) \right]. \eoff  The first term in square brackets reflects the fact that the probability distribution at a given step conditional on non-coalescence at that step is almost equal to the unconditional probability distribution at that step.  The second term represents the correction:  note that it is proportional to the coalescence probability in that step, $\pck{k}{k-\ell}$.  The nature of the correction can be seen by plugging in the distribution of times conditional on coalescence, giving \eon \probtimes{k-\ell}{k,k}(t_1,t_2) = \frac{\probtimes{k-\ell}{k,k}(t_1,t_2)}{1 - \pck{k}{k-\ell}} \left[1 - I_x^{k-\ell} e^{-s(k-\ell) | t_1 - t_2|} \right]. \eoff  We see that the correction acts to reduce the probability that $|t_1 - t_2|$ is small --- that is, it makes it more likely that $t_1$ and $t_2$ are further apart, because this is more likely to be the case given that coalescence did not occur.

Since at each step the shift in the distribution of mutant timings is proportional to the coalescence probability, and the coalescence probability at each step is small, it seems clear that the non-conditional approximation where we simply ignore this shift in mutant timings is reasonable.  However there is one potential caveat we must consider:  although the shift in the distribution of mutation timings due to conditioning on non-coalescence is small \emph{in each step}, we typically take many steps before the lineages coalesce.  In fact, since the shift in mutation timings is proportional to the coalescence probability, and we typically go back a number of steps of order one over the coalescence probability, in principle the shifts in mutation timings could add up to a substantial shift.

Fortunately, there are three factors which prevent this from happening.  First, the shift in mutation timings at each step is always to reduce the probability of times $t_1$ and $t_2$ where $|t_1-t_2| \lesssim \frac{1}{(k-\ell)s}$.  Since at each step $\ell$ is increasing, and the range of separations between mutation timings at which coalescence can happen is also increasing, the shifts in mutation timings from many steps ago are not a huge factor in determining coalescence probabilities in a particular step.  That is, though the shifts in mutation timings add up over many steps, the shifts most relevant to the coalescent probability in a given step do not.  Second, the coalescence probabilities at each step are different.  This reduces the chance that we take enough steps to shift the overall mutation timings substantially by the time we coalesce.  Finally, and most importantly, we will see that the there is a substantial probability that the ancestors of the two individuals sampled do not coalesce until they are in the most-fit class.  This means that the total sum of coalescence probabilities (and hence the total possible weight in the shift of mutation timings) remains small even in the worst case where the two lineages do not coalesce for the maximum possible number of steps.  The non-conditional approximation will always be good in the regime where this is true.  All of these heuristic conclusions are reflected in the fact that the full conditional result we calculate in the sum of ancestral paths approach is equal to the non-conditional result plus corrections that are small provided $I_x \gg 1$.

\section*{Appendix B:  The non-conditional distributions of mutant timings}

Within the non-conditional approximation we need to calculate the distribution of mutant timings, as used in \eq{twentynine} and \eq{thirtysix}. Specifically, we need to calculate \eon \probtimes{k-\ell}{k}(t) = \probtimes{k-1}{k}(t) \star \probtimes{k-2}{k-1}(t)  \star \probtimes{k-3}{k-2}(t) \star \ldots \star  \probtimes{k-\ell}{k-\ell+1}(t)  \label{convexplicit},  \eoff where $\star$ refers to a convolution and \eon \probtimes{k-\ell}{k-\ell+1}(t)=s(k-\ell+1)e^{-s(k-\ell+1)t} , \label{realonegeneral} \eoff as motivated in \eq{timejumpone}.  In general, the convolution of $n$ exponential distributions with parameters $\lambda_1 \ldots \lambda_n$ is given by \eon \sum_{i=0}^{n - 1} \lambda_i e^{-\lambda_i t} \prod_{j = 0,\neq i}^{n-1} \frac{\lambda_j}{\lambda_j - \lambda_i} . \eoff  Applying this identity with $\lambda_i = s(k-i)$, we find \eon \probtimes{k-\ell}{k}(t) = \sum_{i=0}^{\ell - 1} s e^{-s(k - i) t} \left( \frac{ \displaystyle \prod_{j = 0}^{\ell - 1} k - j}{ \displaystyle \prod_{j = 0,\neq i}^{\ell-1} i - j} \right) \eoff  We can simplify this expression by noting that \eon \prod_{j=0}^{\ell - 1}(k-j) = \frac{k!}{(k - \ell)!} , \eoff and similarly that \eon \prod_{j = 0, \neq i}^{\ell - 1}(i - j) = i!(\ell - 1 - i)!(-1)^{\ell -1-i}.  \eoff This means we have \eon \probtimes{k-\ell}{k}(t) = \sum_{i=0}^{\ell - 1} s \ell e^{-s(k - i) t}(-1)^{\ell - i - 1} \binom{\ell - 1}{i} \binom{k}{k - \ell} . \eoff

We can evaluate this sum by recognizing the binomial expansion formula \eon (1+x)^n = \sum_{i=0}^n x^i {n \choose i}, \eoff where we identify $x = -e^{st}$.  We find \eon \probtimes{k-\ell}{k} (t) = s \ell {k \choose \ell} e^{-skt} \left( e^{st} -1 \right)^{\ell-1}. \label{thesecond} \eoff  More generally, we have \eon \probtimes{b}{a}(t) = s (a-b) {a \choose b} e^{-s a t} \left( e^{st} - 1 \right)^{a-b-1}. \label{thethird} \eoff

\section*{Appendix C:  General coalescence probabilities in the non-conditional approximation}

The probability of coalescence for two individuals originally in two different classes $k$ and $k'$, as defined in \eq{thirtysix} can be rewritten as \eon \pckk{k}{k'}{k'-\ell} = \frac{1}{1 + 2N h_{k-\ell} s(k-\ell)} \left[ I_1 +I_2\right], \eoff where we have defined \eaon I_1 & = & \int_0^{\infty} \probtimes{k-\ell}{k'}(t_1) e^{-s(k - \ell)t_1} \int_0^{t_1} \probtimes{k-\ell}{k} (t_2) e^{s(k-\ell)t_2} dt_2 dt_1 \\ I_2&=&\int_0^{\infty} \probtimes{k-\ell}{k} (t_2) e^{-s(k-\ell)t_2} \int_0^{t_2} \probtimes{k-\ell}{k'} (t_1) e^{s(k-\ell) t_1} dt_1 dt_2 . \eaoff

Note that both $I_1$ and $I_2$ involve integrals of the form \eon I_a = \int_0^t \probtimes{b}{a} (t') e^{s b t'} dt'. \eoff  Plugging in the results for the non-conditional distributions of mutant timings, \eq{thethird}, and making use of the binomial expansion formula for $(1+x)^n$ noted in Appendix B, we find this integral becomes \eaon I_a & = & s(a-b) {a \choose b} \int_0^t e^{s (b-a) t'} \left( e^{st'} -1 \right)^{a-b-1} dt' \\ & = & s (a - b) {a \choose b} \sum_{i=0}^{a-b-1} (-1)^{a-b-1+i} {a-b-1 \choose i} \int_0^t e^{s(b-a+i)t'} dt' \\ & = & (a-b) {a \choose b} (-1)^{a-b} \sum_{i=0}^{a-b-1} \frac{(-1)^i}{a-b} {a-b \choose i} \left( e^{s(b-a+i)t} -1 \right) \\ & = & {a \choose b} (-1)^{a-b} \sum_{i=0}^{a-b} (-1)^i {a-b \choose i} \left( e^{s(b-a+i)t} - 1 \right) \\ & = & {a \choose b} (-1)^{a-b} e^{s(b-a)t} \sum_{i=0}^{a-b} \left( -e^{st} \right)^i {a-b \choose i} \\ & = & {a \choose b} e^{s(b-a)t} \left( e^{st} - 1 \right)^{a-b}. \eaoff

We now substitute this result for $I_a$ into our expressions for $I_1$ and $I_2$.  We note that both have terms of the form \eon I_b = \int_0^\infty \probtimes{b}{a} (t) {c \choose b} e^{-sct} \left(e^{st}-1 \right)^{c-b} dt. \eoff  Using similar manipulations to those above, we find \eaon I_b & = & (a - b) {a \choose b} {c \choose b} \int_0^\infty e^{-s (a + c) t} \left( e^{st} - 1 \right)^{a+c-2b-1} dt \\ & = & s (a-b) {a \choose b} {c \choose b} (-1)^{a+c-1} \sum_{i=0}^{a+c-2b-1} {a + c - 2b -1 \choose i} (-1)^i \int_0^\infty e^{-s(a+c-i)t} dt \\ & = & (a - b) {a \choose b} {c \choose b} (-1)^{a+c-1} \sum_{i=0}^{a+c-2b-1} (-1)^i {a + c - 2 b - 1 \choose i} \frac{1}{a+c-i} . \eaoff  Using the partial fraction decomposition \eon \frac{1}{{n+x \choose n}} = \sum_{i=1}^n (-1)^{i-1} {n \choose i} \frac{i}{x + i}, \eoff we find \eon I_b = \frac{\frac{a - b}{a + c - 2b} {a \choose b} {c \choose b} (-1)^{a+c}}{{-2b -1 \choose a + c - 2b}} = \frac{\frac{a - b}{a+c-2b} {a \choose b}{c \choose b} (-1)^{2b}}{{a+c \choose a+c-2b}}. \eoff

We can now use this result for $I_b$ to determine $I_1$ and $I_2$, and hence compute $\pckk{k}{k'}{k'-\ell}$.  We find \eon \pckk{k}{k'}{k'-\ell} = \frac{1}{1 + 2 N h_{k-\ell} s (k - \ell)} \frac{{k' \choose k-\ell}{k \choose k- \ell}}{{k + k' \choose 2 \ell + k' -k}}. \eoff  As we noted in the main text, this is just \eon \pckk{k}{k'}{k-\ell} = \frac{1}{1 + 2 N h_{k-\ell} s(k-\ell)} \aklm{k}{k'}{\ell}, \eoff with $\aklm{k}{k'}{\ell}$ as defined in \eq{aeqn}.  Note that when $k = k'$, this result simplifies to $\pckk{k}{k}{k-\ell}$ as defined in the main text, as expected.

\section{Appendix D: Computing sums of ancestral paths}

In this appendix, we describe the calculation of $\pst{k}{k'}{\ell}$ using the sum of ancestral paths approach.

\subsection{Calculation of $\pst{k}{k}{3}$}
We begin by considering a simpler specific case, where $k = k'$ and $\ell=3$. There are a total of $\binom{6}{3} = 20$ possible ancestral paths by which two individuals sampled from class $k$ can coalesce in class $k-3$. These can be separated into four types, according to whether the two ancestral lineages were ever together in classes $k-1$ or $k-2$. We can list all paths of each type, using the notation that A is a mutation event in the first lineage, and B is a mutation event in the second lineage. We have
\[\underbrace{ \left( \begin{array}{c} ABABAB \\ABABBA \\ABBAAB \\ABBABA \\BAABAB \\BAABBA \\BABAAB \\BABABA \end{array} \right)}_{\binom{2}{1} \binom{2}{1} \binom{2}{1} = 8 \textrm{ ways}} \qquad \underbrace{ \left( \begin{array}{c} ABAABB \\ABBBAA \\BAAABB \\ BABBAA \end{array} \right)}_{\binom{2}{1} \left(\binom{4}{2} - \binom{2}{1} \binom{2}{1} \right) = 4 \textrm{ ways}} \qquad \underbrace{ \left( \begin{array}{c} AABBAB \\AABBBA \\BBAAAB \\BBAABA \end{array} \right)}_{\binom{2}{1} \left(\binom{4}{2} - \binom{2}{1} \binom{2}{1} \right) = 4 \textrm{ ways}} \qquad \underbrace{ \left( \begin{array}{c} AAABBB \\AABABB \\BBBAAA \\BBABAA \end{array} \right)}_{\binom{6}{3} - \textrm{others} = 4 \textrm{ways}} . \]

The probabilities of all paths of a particular type are identical.  We can calculate the probability of each of the four types of paths using the same logic as outlined in the main text.  We find \eaon P(AAABBBc) & = & I_x^{k-3}\frac{k(k-1)(k-2)}{8(2k-1)(2k-3)(2k-5)} \left( 1-I_x^k \right), \\ P(AABBABc) & = & I_x^{k-3}\frac{k(k-1)(k-2)}{8(2k-1)(2k-3)(2k-5)} \left( 1-I_x^k \right) \left( 1-I_x^{k-1} \right),  \\ P(ABAABBc) & = & I_x^{k-3} \frac{k(k - 1)(k - 2)}{8(2k - 1)(2k-3)(2k-5)} \left( 1-I_x^k \right) \left( 1-I_x^{k-2} \right), \\ P(ABABABc) & = & I_x^{k-3}\frac{k(k-1)(k-2)}{8(2k-1)(2k-3)(2k-5)} \left( 1-I_x^k \right) \left( 1-I_x^{k-1} \right) \left( 1-I_x^{k-2} \right) . \eaoff  Summing over all the possible paths, we find \eon \pst{k}{k}{3} = I_{k-3} \frac{ \binom{k}{k - 3} \binom{k}{k-3}}{ \binom{2k}{6}} \left[ 1- \frac{ \binom{2}{1} \binom{4}{2}}{ \binom{6}{3}} I_{k-1}-\frac{ \binom{2}{1} \binom{4}{2}}{ \binom{6}{3}}I_{k-2} + \frac{ \binom{2}{1} \binom{2}{1} \binom{2}{1}}{ \binom{6}{3}} I_{k-1} I_{k-2} \right] . \eoff

We now pause to consider the form of the probabilities of each type of ancestral path.  These probabilities differ only by factors of $(1-I_x^{k-i})$.  One such factor arises each time the two ancestral lineages are together in class $k-i$. In other words, we can rewrite the probability of each path as the probability of an undistorted path (defined to be a path in which the contributions due to the possibility of coalescence in previous classes are neglected), times a correction for each class in which the two lineages are together: \eaon P(AAABBBc) & = & P(\textrm{Undistorted Path}) \left(1-I_x^k\right) \\ P(AABBABc) & = & P(\textrm{Undistorted Path}) \left(1-I_x^k\right)\left(1-I_x^{k-1}\right) \\ P(ABAABBc) & = & P(\textrm{Undistorted Path}) \left(1-I_x^k\right)\left(1-I_x^{k-2}\right) \\ P(ABABABc) & = & P(\textrm{Undistorted Path}) \left(1 - I_x^k \right) \left(1 - I_x^{k-1} \right) \left(1 - I_x^{k - 2} \right) . \eaoff By definition, the ``undistorted path'' probability is the probability neglecting the contributions due to the possibility of coalescence in previous steps, and is therefore the same for all paths.  We have \eaon P(\textrm{Undistorted Path}) & = & \frac{k (k-1) (k-2) k (k-1) (k-2)}{2k (2k-1) (2k-2) (2k-3) (2k-4) (2k-5)} I_x^{k - \ell} \\ & = & \frac{ \frac{k!}{(k - 3)!}\frac{k!}{(k - 3)!}}{\frac{2k!}{(2k - 6)!}} I_x^{k-\ell} . \eaoff Using these results, we can write $\pst{k}{k}{3}$ as \eaon \pst{k}{k}{3} & = & \left[\textrm{\# of Paths} \right] P(\textrm{Undistorted Path}) \left[ F_k (1 - I_x^k) + F_{k, k-1} (1 - I_x^{k}) (1 - I_x^{k - 1}) \right. \nonumber \\ & & \left. + F_{k, k-2} (1 - I_x^{k}) (1 - I_x^{k - 2}) + F_{k, k-1, k-2} (1 - I_x^{k})(1 - I_x^{k - 1}) (1 - I_x^{k-2}) \right] , \eaoff where we have defined $F_{\{a\}}$ to be the fraction of paths that are together in the set of classes $\{a\}$ (and are not together in any other class).

\subsection{Calculation of $\pst{k'}{k}{\ell}$}
We now use this approach to calculate the coalescence probability in the general case.  The probability of any particular ancestral path from $k$ and $k'$ to $k-\ell$ is the product of the individual probabilities of each mutational step that makes up this path.  Each such individual probability consists of three parts:  a numerator, which depends only on the current class of the lineage that mutates, divided by a denominator, which depends only on the sum of the current set of classes for both lineages, times a correction factor of $(1 - I_x^{k-i})$ if the two lineages are in the same class at that step.

Although in each ancestral path the mutations will occur in a different order, all paths will ultimately consist of the same set of mutations ($k'\rightarrow k'-1 \rightarrow \ldots \rightarrow k-\ell$ and $k\rightarrow k-1 \rightarrow \ldots \rightarrow k-\ell$). Therefore, regardless of the path taken, the product of the numerators from each step will be identical. Similarly, the sum of the current set of classes will begin at $k'+k$, and decrement by one each time a deleterious mutation occurs, until both lineages are in the final class ($k'+k\rightarrow k'+k-1 \rightarrow\ldots \rightarrow 2k-2\ell$). Therefore, regardless of the path taken, the product of the denominators from each step will also be identical. Therefore, the paths will differ only by the correction factor $(1-I_x^{k-i})$ for each class in which the two ancestral lineages are together.  This means that, analogous to the case of $\pst{k}{k}{3}$ we described above, the probability of each path is the probability of an ``undistorted path'' times the appropriate correction factor.  The probability of the undistorted path is \eon P(\textrm{Undistorted Path}) = \frac{k' (k' - 1) \ldots( k- \ell+ 1) k (k - 1) \ldots ( k - \ell + 1)}{(k' + k)(k' + k - 1) \ldots (2k - 2 \ell + 1)} I_x^{k - \ell} . \eoff

We can now sum up all possible paths to obtain \eaon \pst{k'}{k}{\ell} & = & \left[\textrm{\# of Paths} \right] P(\textrm{Undistorted Path}) \left[ F_{\emptyset} + \sum_{i=0}^\ell F_{k-i} (1 - I_x^{k-i}) \nonumber \right. \\ & & + \sum_{i=0}^{\ell-1} \sum_{j > i}^\ell F_{k-i, k-j} (1 - I_x^{k-i}) (1 - I_x^{k - j}) \\ & & \left. + \sum_{i=0}^{\ell-2} \sum_{j>i}^{\ell-1} \sum_{m > j}^{\ell} F_{k-i, k-j, k-m} (1 - I_x^{k-i}) (1 - I_x^{k - j}) (1 - I_x^{k - m}) + \ldots  \right] , \nonumber \eaoff where as before $F_{\{a\}}$ is the fraction of paths that are together in the set of classes $\{a\}$ (and are not together in any other class).  Note that there are a total of $\ell+1$ terms in this equation, representing the possibility that the two lineages can be together in anywhere from $0$ to $\ell$ of the classes.  We can rearrange these terms to write \eaon \pst{k'}{k}{\ell} & = & \left[\textrm{\# of Paths}\right] P(\textrm{Undistorted Path}) \left[ 1 - \sum_{i=0}^\ell G_{k-i} I_x^{k-i} \nonumber \right. \\ & & + \sum_{i=0}^{\ell-1} \sum_{j > i}^\ell G_{k-i, k-j} I_x^{k-i} I_x^{k - j} \\ & & \left. - \sum_{i = 0}^{\ell - 2} \sum_{j>i}^{\ell - 1} \sum_{m > j}^{\ell} G_{k-i, k-j, k-m} I_x^{k - i} I_x^{k - j} I_x^{k - m} + \ldots  \right] , \nonumber \eaoff where we have defined $G_{\{a\}}$ to be the fraction of paths that are together in \emph{at least} the set of classes $\{a\}$.

We can evaluate each of these factors of $G$.  For example, the fraction of paths that are together in class $k-i$ equals the number of ways for the two lineages to descend from classes $k'$ and $k$ to be together in class $k-i$, $\binom{k'-k+2i}{i}$, times the number of ways for the two lineages to descend from class $k-i$ to be together in class $k-\ell$, $\binom{2i-2\ell}{i-\ell}$, divided by the total number of ways for the two lineages to descend from classes $k'$ and $k$ to be together in $k-\ell$, $\binom{k'-k+2\ell}{\ell}$.  Using this logic, we find \eaon \pst{k'}{k}{\ell} & = & \left[\textrm{\# of Paths} \right] P(\textrm{Undistorted Path}) \\ & & \times \left[ 1 - \sum_{i = 0}^{\ell - 1}
\frac{\binom{k'-k+2 i}{i}\binom{2\ell - 2i}{\ell - i}}{\binom{k'-k+2
\ell}{\ell}}I_x^{k - i} + \sum_{i = 0}^{\ell - 2} \sum_{j >
i}^{\ell - 1} \frac{\binom{k'-k+2i}{i}\binom{2j - 2i}{j - i}\binom{2\ell -
2j}{\ell - j}}{\binom{k'-k+2\ell}{\ell}}I_x^{k - i}I_x^{k - j} \ldots \right] \nonumber . \eaoff

The total number of paths is $\binom{k'-k+2\ell}{\ell}$, so we finally find that the full probability of coalescence in class $k-\ell$ is \eaon
\pst{k}{k'}{\ell} & = & I_x^{k - \ell}\frac{\binom{k'}{k - \ell}\binom{k}{k -
\ell}}{\binom{k' + k}{k' - k + 2\ell}} \left[ 1 - \sum_{i = 0}^{\ell - 1}
\frac{\binom{k' - k + 2 i}{i}\binom{2\ell - 2i}{\ell - i}}{\binom{k' - k + 2
\ell}{\ell}}I_x^{k - i} + \right. \nonumber \\ & & \left. \sum_{i = 0}^{\ell - 2} \sum_{j >
i}^{\ell - 1} \frac{\binom{k' - k + 2i}{i}\binom{2j - 2i}{j - i}\binom{2\ell -
2j}{\ell - j}}{\binom{k' - k + 2\ell}{\ell}}I_x^{k - i}I_x^{k - j}-\ldots \right] .  \eaoff
This is \eq{sumpathscoalprob} from the main text.  Note that it equals our non-conditional result for $\pckk{k}{k'}{\ell}$ times a correction factor.  There are a total of $\ell+1$ terms in this correction factor.  This full correction factor can be arbitrarily complex for large $\ell$, so we do not write out a general form here.  However, it is straightforward to calculate for any values of $k$, $k'$, and $\ell$; a Mathematica script to do so is available on request.

\section*{Appendix E: The correspondence between steptimes and real times}

In this Appendix, we calculate the correspondence between steptimes and the actual times measured in generations.  Our goal is to calculate the probability distribution of real coalescence times, $\probt (t|k,k',\ell)$, given that individuals were initially in classes $k$ and $k'$ and coalesced in class $k-\ell$.

To begin, we neglect the coalescence time within class $k-\ell$, and consider the time at which an ancestor of one of the two sampled individuals first mutated from class $k-\ell$ to class $k - \ell + 1$, $\probt_1 (t|k, k', \ell)$.  We first calculate the joint distribution of the times at which both ancestors mutated out of the class, $\probtimescond{k-\ell}{k, k'}(t_1, t_2)$.  Conditional on coalescence in class $k - \ell$, $\probtimescond{k-\ell}{k, k'}(t_1, t_2)$, is given by the probability of $t_1$ and $t_2$ and coalescence divided by the total probability of coalescence.  That is, \eon \probtimescond{}{} (t_1,t_2) = \frac{P(coal|t_1, t_2) P(t_1, t_2)}{P(coal)}. \eoff  Substituting in the relevant expressions from the main text, this gives \eon \probtimescond{k-\ell}{k, k'} (t_1,t_2) = \frac{1}{\aklm{k}{k'}{\ell}} \probtimes{k - \ell}{k, k'}(t_1,t_2) e^{-s(k-\ell)|t_1-t_2|}. \eoff

The time at which the first ancestor mutated out of class $k-\ell$ is the longer of the two times $t_1$ and $t_2$, \eon \probt (t|k,k',\ell)  =  \left[ \int_0^{t} \probtimescond{k-\ell}{k, k'} (t_1,t) dt_1 + \int_0^{t} \probtimescond{k-\ell}{k, k'}(t,t_2) dt_2 \right] . \eoff  Substituting in our expression for $\probtimescond{k-\ell}{k, k'}(t_1,t_2)$ and carrying out the integrals as in Appendix C, we find \eon \probt_1 (t|k, k', \ell) = s \pi_d e^{-s(k' + k) t}(e^{st} - 1)^{\pi_d - 1}\binom{k'+k}{\pi_d},  \eoff where we have used $\pi_d = k' - k + 2 \ell$.

We can alternatively calculate $\probt_1 (t|k, k', \ell)$ using our sum of ancestral paths approach.  As before, we imagine two individuals sampled from classes $k$ and $k'$ and condition on them coalescing in class $k-\ell$.  Consider a case where $k \neq k'$.  Then the first event in the history of these two individuals must be a deleterious mutation.  Since these mutations happen at rate $sk$ and $sk'$ in each lineage, the distribution of times since this mutation occurred in one of the two ancestral lineages is \eon P(t)=s(k + k')e^{-s(k + k') t}. \eoff   With probability $\frac{k'}{k+k'}$, this mutation is in the lineage sampled from class $k'$, in which case the two lineages are now in classes $k$ and $k'-1$. Alternatively, the mutaion occurred in the lineage sampled from $k$ and the lineages are in classes $k-1$ and $k'$.

We can now consider the time to the next event backwards in time.  If the two lineages are in the same class (but not yet in class $k-\ell$), the distribution of times to the next deleterious mutation event is somewhat shorter, because we are conditioning on coalescence not occuring.  However, provided that $2 s k_1 \gg \frac{1}{N h_k}$ (the condition we are already making elsewhere), this shortening of the time will be a small correction and neglecting it is a good approximtion.

Making this approximation, the rate at which the next deleterious mutation event occurs when the two lineages are in classes $k_1$ and $k_2$ is just $s (k_1 + k_2)$.  Regardless of the order in which these mutations happen between the two lineages, this sum is simply decreased by $s$ at each step.  This will continue until the both ancestral lineages are in class $k-\ell$. Therefore, the distribution of times until the original mutation out of class $k-\ell$ is given by: \eon \psi_1(t|k' ,k, \ell) = s(k' + k)e^{-s(k' + k)t} \otimes s(k '+k - 1)e^{-s(k' +k-1)t} \otimes \ldots \otimes s(2k -2 \ell+1)e^{-s(2k - 2 \ell+ 1)t} . \eoff
This can be written as \eon \psi_1(t|k',k,\ell) = \lambda_0 e^{-\lambda_0t} \otimes \lambda_{1} e^{-\lambda_1 t} \otimes \ldots \otimes \lambda_{k' - k + 2 \ell- 1} e^{- \lambda_{k' - k + 2 \ell-1}t} ,  \eoff where we have defined: \eon \lambda_i = s(k'+k-i) . \eoff  We can compute this convolution as in Appendix B (compare to \eq{convexplicit} for $\probtimes{2k - 2\ell}{k + k'}(t)$).  We find \eon \probt_1 (t|k, k', \ell) = s \pi_d e^{-s(k' + k) t}(e^{st} - 1)^{\pi_d - 1}\binom{k'+k}{\pi_d}, \eoff identical to the result of our lineage structure calculation above.

\subsection{Distribution of Coalescence Times}
To calculate the correspondence between steptimes and real times, we now need to add the time it takes two individuals two coalesce in class $k-\ell$, $\probt_2(t|k, k', \ell)$, to the time it took them both to get to that class, $\probt_1(t|k, k', k-\ell)$.  The rate of coalescence once in class $k-\ell$ is $\frac{1}{N h_{k-\ell}}$, so we have \eon \psi_2(t|k',k,\ell)=\left(2s(k-\ell)+1/Nh_{k-\ell}\right)e^{-(s(k-\ell)+1/Nh_{k-l})t}. \eoff
Putting this together, the full distribution of times since coalescence is
\eon \psi(t|k', k,\ell) = \psi_1(t|k',k,\ell) \otimes \psi_2(t|k', k, \ell) . \eoff
Carrying out this convolution (and expanding the binomial factor $(e^{st}-1)^{\pi_d -1}$ in $\probt_1$), we find \eon \psi(t|k',k,\ell) = \sum_{i = 0}^{n-1} s \pi_d (-1)^{\pi_d -i-1} \binom{\pi_d - 1}{i} \binom{k' + k}{\pi_d} \frac{B}{A - B} \left( e^{-sBt} - e^{-sAt} \right) , \eoff where we have defined $A \equiv k' + k - i$ and $B \equiv k - \ell + \frac{1}{Nsh_{k-\ell}}$.

\section*{Appendix F:  An alternative approach to neutral diversity}

Instead of calculating the distribution of neutral heterozygosity by first computing the distribution of real times, we could alternatively incorporate them directly into the sum of ancestral paths framework.  This completely bypasses the correspondence with real coalescence times.  To do this, we characterize ancestral paths not only by the ordering of deleterious mutation and coalescence events, but also by the ordering of neutral mutations.  This means that if we sample two individuals $A$ and $B$, there are five types of events that can happen in their ancestral paths:  a deleterious mutation (DM) in $A$ or in $B$, a neutral mutation (NM) in either $A$ or in $B$, and or a coalescence (C) event (if $A$ and $B$ are currently in the same class).

We now imagine that we sample two individuals from classes $k$ and $k'$, and that they coalesce in class $k-\ell$.  Our goal is to calculate the probability distribution of $\pi_n$ given $k$, $k'$, and $\ell$, $\ppin (\pi_n|k, k', \ell)$.  We will find it helpful to divide the five types of events that can occur into two classes:  neutral mutations on the one hand, and deleterious mutations or coalescence (which we call ``steps'') on the other.  We begin by computing the probability that a given number of NMs occur before the next DM or C events (i.e. the number of neutral mutations that occur at this ``step'').  We have \eon P(\textrm{a NMs, then DM in $k'$ or $k'$} | k' , k) = \left( \frac{ \frac{2 U_n}{s}}{k' + k+ \frac{2 U_n}{s}} \right)^a \frac{k + k'}{k' + k + \frac{2 U_n}{s}} , \eoff where we have made our usual assumption that $N h_k s k \gg 1$, allowing us to neglect the rates of coalescence events (when $k = k'$) in writing this expressions.

This probability only depends on the sum of the current classes the individulas are in.  At each subsequent step, regardless of the path taken, this sum of the classes will decrease by one.  Therefore, the probability that $a_i$ neutral mutations occur at step $i$ is independent of the path taken.  This observation allows us to calculate the probability that a given total number of neutral mutations have occurred since coalescence. We first calculate the probability that a given number of neutral mutations have occurred since the first deleterious mutation out of the $k-\ell$ class. We will add in the additional neutral mutations once in the $k-\ell$ class at the end.

In order for $\pi_n$ neutral mutations to have occurred since the first deleterious mutation out of class $k-\ell$, we require that $a_0$ mutations occurred at the first step, $a_1$ mutations occurred at the second step, and so on, such that $a_0 + a_1 + \ldots + a_{k' - k + 2\ell - 1} = \pi_n$. This gives \eon \ppin (\pi_n = X|k', k, \ell) = \frac{ \frac{(k' + k)!}{(2k - 2 \ell)!}}{ \frac{( \frac{2 U_n}{s} + k' + k)!}{(\frac{2 U_n}{s} + 2k-2\ell)!}} \sum_{| \vec{a}| = X} \left( \frac{2 U_n/s}{2 U_n/s +k +k'} \right)^{a_0} \ldots \left( \frac{2 U_n/s}{2 U_n/s + 2k - 2l + 1} \right)^{a_{k' - k + 2l - 1}}. \eoff  We can  define $x \equiv 2U_n/s+k+k'$, recognize $\pi_d = k' -k + 2 \ell$, and relabel the $a_i$ as \eon a_{0} \rightarrow X-b_0, \quad a_{1} \rightarrow b_0-b_1, \quad \ldots \quad a_{\pi_d-2} \rightarrow b_{\pi_d-3} - b_{\pi_d - 2}, \quad a_{\pi_d-1} \rightarrow b_{\pi_d-2}.  \eoff  This gives \eaon \ppid (\Pi_n = X|k',k, \ell) & = & \frac{ \binom{k' + k}{\pi_d}}{ \binom{ \frac{2U_n}{s} +k' + k}{\pi_d}} \left( \frac{2 U_n}{s} \right)^X \left( \frac{1}{x} \right)^X \sum_{b_0 = 0}^X \left( \frac{x}{x - 1} \right)^{b_0} \\ & & \sum_{b_1 = 0}^{b_0} \left( \frac{x - 1}{x - 2} \right)^{b_1} \ldots \sum_{b_{\pi_d- 2}=0}^{b_{\pi_d -3}} \left( \frac{x- \pi_d+2}{x -\pi_d +1} \right)^{b_{\pi_d - 2}} \nonumber . \eaoff

To simnplify this expression, it is helpful to define a function f such that: \eaon \mathbf{f} \left(A, B \right) & \equiv & \left( \frac{1}{x} \right)^X \sum_{b_0 = 0}^X \left( \frac{x}{x-1} \right)^{b_0} \\ & & \sum_{b_1=0}^{b_0} \left(\frac{x-1}{x-2} \right)^{b_1} \ldots \sum_{b_{A-1}=0}^X \left( \frac{x-A+1}{x-A} \right)^{b_0} \sum_{b_{A} =0}^{b_{A-1}} \left( \frac{x - A}{x - B} \right)^{b_{A}} \nonumber \eaoff In other words, $\mathbf{f}\left(A,B\right)$ is a set of $A$ nested sums, each of the same form, except for the final sum, which can have a different denominator. Using this definition, we have \eon P(\Pi_n = X|k',k,\ell) = \frac{ \binom{k' + k}{\pi_d}}{ \binom{ \frac{2 U_n}{s} + k' + k}{\pi_d}} \left( \frac{2U_n}{s} \right)^X \mathbf{f} \left( \pi_d-2,\pi_d - 1 \right) . \eoff The virtue of this definition is that this sum can be solved recursively. We have \eon\sum_{b_{A} = 0}^{b_{A - 1}} \left(\frac{x - A}{x-B}\right)^{b_{A}} = \frac{x- B}{A-B} - \frac{x-A}{A -B} \left( \frac{x-A}{x-B} \right)^{b_{A-1}} . \eoff Therefore we have \eon \mathbf{f} \left(A, B \right) = \frac{x - A}{B - A} \mathbf{f} \left(A-1, B \right) - \frac{x - B}{B - A} \mathbf{f} \left(A-1, A \right) .  \eoff Repeatedly inserting this result yields:
\eaon \mathbf{f}\left(A, A+1 \right) & \rightarrow & \frac{(x - A)(x - A-1)}{1} \left(\frac{ \mathbf{f} \left( A-1, A+1 \right)}{x - A-1}- \frac{ \mathbf{f} \left( A-1, A \right)}{x - A} \right) \nonumber \\ \mathbf{f}\left(A, A+1\right) & \rightarrow & \frac{(x - A+1) (x - A) (x - A - 1)}{2} \left[ \frac{ \mathbf{f} \left( A-2, A+1 \right)}{x - A - 1} - \frac{ 2 \mathbf{f} \left(A - 2, A \right)}{x - A} + \frac{ \mathbf{f} \left( A - 2, A - 1 \right)}{x - A + 1} \right] \nonumber \\ & \vdots & \nonumber \\ \mathbf{f} \left( A, A + 1 \right) & \rightarrow & (m + 1) \binom{x - A - 1 + m}{m + 1} \sum_{i = 0}^m \frac{(-1)^{i + m}}{x - A - 1 + i} \binom{m}{i} \mathbf{f} \left(A - m, A + 1 - i \right) . \eaoff  Note that $\mathbf{f}(-1,B)=1/B^X$, since there are no more sums to compute. Thus, for $m=A+1$ we have
\eon \mathbf{f} \left(A, A + 1 \right) = (A+2) \binom{x}{A + 2} \sum_{i=0}^{A+1} \frac{(-1)^{i + A + 1}}{(x - A- 1 + i)^{X + 1}} \binom{A+1}{i} .  \eoff Relabeling the sum and taking $A = \pi_d-2$, we have \eon \mathbf{f} \left(\pi_d-2, \pi_d - 1 \right) = \pi_d \binom{x}{\pi_d} \sum_{i = 0}^{\pi_d - 1} \frac{(-1)^{i}}{(x -i)^{X + 1}} \binom{\pi_d - 1}{i}.  \eoff

We can now substitute these results into our expression for $\pi_n$, to find
\eon \ppin_1 (\Pi_n=X|k',k,\ell) = \pi_d\binom{k'+k}{\pi_d} \left(\frac{2U_n}{s} \right)^X\sum_{i=0}^{\pi_d-1}\frac{(-1)^{i}}{(2U_n/s+k+k'-i)^{X+1}}\binom{\pi_d-1}{i}\eoff
Note, however, that this is only the distribution of neutral mutations since the first deleterious mutation out of class $k-l$. It is also possible for neutral mutations to occur prior to the coalescence event. Adding in this factor, we find \eaon \ppin (\Pi_n=X|k',k,\ell) & =& \pi_d\binom{k'+k}{\pi_d} \sum_{i=0}^{\pi_d-1}(-1)^{i} \binom{\pi_d-1}{i} \\ & & \times \sum_{X=0}^{\pi_n} \frac{\left(2U_n/s \right)^X}{(2U_n/s + k+ k'-i)^{X +1}} \left( \frac{2 N_{k-l} U_n}{1 + 2N_{k-l} U_n+ 2N_{k-l} s (k-l)} \right)^{\pi_n-X} . \nonumber  \eaoff Rearranging this expression gives \eon \ppin (\pi_n | k' ,k, \ell) = \sum_{i = 0}^{\pi_d -1} \pi_d (-1)^{\pi_d -i - 1} \binom{\pi_d - 1}{i} \binom{k' + k}{\pi_d }\frac{B}{A - B} \left( \frac{( \frac{2 U_n}{s})^{\pi_n}}{( \frac{2 U_n}{s} + B)^{\pi_n + 1}} - \frac{( \frac{2U_n}{s})^{\pi_n}}{( \frac{2 U_n}{s} + A)^{\pi_n + 1}} \right) , \eoff where we have defind \eon A = k' + k - i, \qquad B = 2k - 2 \ell + \frac{1}{Nsh_{k-l}} , \eoff  identical to our earlier result.

\clearpage

\newpage

\bibliographystyle{genetics}
\bibliography{negselcoallib}

\begin{thebibliography}{37}

\bibitem[{\sc Barton {\rm and} Etheridge}(2004)]{bartonetheridge04}
{\sc Barton, N.~H. {\rm and} A.~M. Etheridge}, 2004 The effect of selection on
  genealogies. Genetics {\bf 166}: 1115--1131.

\bibitem[{\sc Charlesworth}(1994)]{charlesworth94}
{\sc Charlesworth, B.}, 1994 The effect of background selection against
  deleterious mutations on weakly selected, linked variants. Genetical Research
  {\bf 63}: 213--227.

\bibitem[{\sc Charlesworth {\em et~al.\/}}(1993){\sc Charlesworth, Morgan, {\rm
  and} Charlesworth}]{charlesworth93}
{\sc Charlesworth, B., M.~T. Morgan, {\rm and} D.~Charlesworth}, 1993 The
  effect of deleterious mutations on neutral molecular variation. Genetics {\bf
  134}: 1289--1303.

\bibitem[{\sc Charlesworth {\em et~al.\/}}(1995){\sc Charlesworth,
  Charlesworth, {\rm and} Morgan}]{charlesworth95}
{\sc Charlesworth, D., B.~Charlesworth, {\rm and} M.~T. Morgan}, 1995 The
  pattern of neutral molecular variation under the background selection model.
  Genetics {\bf 141}: 1619--1632.

\bibitem[{\sc Comeron {\rm and} Kreitman}(2002)]{comeronkreitman02}
{\sc Comeron, J.~M. {\rm and} M.~Kreitman}, 2002 Population, evolutionary and
  genomic consequences of interference selection. Genetics {\bf 161}: 389--410.

\bibitem[{\sc Comeron {\em et~al.\/}}(2008){\sc Comeron, Williford, {\rm and}
  Kliman}]{comeron08}
{\sc Comeron, J.~M., A.~Williford, {\rm and} R.~M. Kliman}, 2008 The
  hill-robertson effect: Evolutionary consequences of weak selection and
  linkage in finite populations. Heredity {\bf 100}: 19--31.

\bibitem[{\sc Desai {\em et~al.\/}}(2010){\sc Desai, Nicolaisen, Walczak, {\rm
  and} Plotkin}]{allelebased}
{\sc Desai, M.~M., L.~E. Nicolaisen, A.~M. Walczak, {\rm and} J.~B. Plotkin},
  2010 The structure of allelic diversity in the presence of purifying
  selection. Genetics {\bf xxx}.

\bibitem[{\sc Etheridge {\rm and} Griffiths}(2009)]{EtheridgePMID19341750}
{\sc Etheridge, A.~M. {\rm and} R.~C. Griffiths}, 2009 A coalescent dual
  process in a moran model with genic selection. Theoretical Population Biology
  {\bf 75}: 320--330.

\bibitem[{\sc Etheridge {\em et~al.\/}}(2010){\sc Etheridge, Griffiths, {\rm
  and} Taylor}]{EtheridgePMID20685218}
{\sc Etheridge, A.~M., R.~C. Griffiths, {\rm and} J.~E. Taylor}, 2010 A
  coalescent dual process in a moran model with genic selection, and the lambda
  coalescent limit. Theoretical Population Biology {\bf 78}: 77--92.

\bibitem[{\sc Ewens}(2004)]{ewensbook}
{\sc Ewens, W.~J.}, 2004 {\em Mathematical Population Genetics: I. Theoretical
  Introduction\/}. Springer, New York, NY.

\bibitem[{\sc Gessler}(1995)]{gessler95}
{\sc Gessler, D. D.~G.}, 1995 The constraints of finite size in asexual
  populations and the rate of the ratchet. Genetical Research {\bf 66}:
  241--253.

\bibitem[{\sc Gordo {\rm and} Charlesworth}(2000{a})]{gordocharlesworth00a}
{\sc Gordo, I. {\rm and} B.~Charlesworth}, 2000{a} The degeneration of asexual
  haploid populations and the speed of muller's ratchet. Genetics {\bf 154}:
  1379--1387.

\bibitem[{\sc Gordo {\rm and} Charlesworth}(2000{b})]{gordocharlesworth00b}
{\sc Gordo, I. {\rm and} B.~Charlesworth}, 2000{b} On the speed of muller's
  ratchet. Genetics {\bf 156}: 2137--2140.

\bibitem[{\sc Gordo {\em et~al.\/}}(2002){\sc Gordo, Navarro, {\rm and}
  Charlesworth}]{gordocharlesworth02}
{\sc Gordo, I., A.~Navarro, {\rm and} B.~Charlesworth}, 2002 Muller's ratchet
  and the pattern of variation at a neutral locus. Genetics {\bf 161}:
  835--848.

\bibitem[{\sc Hahn}(2008)]{hahn08}
{\sc Hahn, M.~W.}, 2008 Toward a selection theory of molecular evolution.
  Evolution {\bf 62}: 255--265.

\bibitem[{\sc Haigh}(1978)]{haigh78}
{\sc Haigh, J.}, 1978 The accumulation of deleterious genes in a
  population-muller's ratchet. Theoretical Population Biology {\bf 14}:
  251--267.

\bibitem[{\sc Hartl {\rm and} Sawyer}(1994)]{hartlsawyer94}
{\sc Hartl, D.~L. {\rm and} S.~A. Sawyer}, 1994 Selection intensity for codon
  bias. Genetics {\bf 138}: 227--234.

\bibitem[{\sc Hermisson {\em et~al.\/}}(2002){\sc Hermisson, Redner, Wagner,
  {\rm and} Baake}]{hermisson02}
{\sc Hermisson, J., O.~Redner, H.~Wagner, {\rm and} E.~Baake}, 2002
  Mutation-selection balance: Ancestry, load, and maximum principle.
  Theoretical Population Biology {\bf 62}: 9--46.

\bibitem[{\sc Hill {\rm and} Robertson}(1966)]{hillrobertson66}
{\sc Hill, W. {\rm and} A.~Robertson}, 1966 The effect of linkage on limits to
  artificial selection. Genetical Research {\bf 8}: 269--294.

\bibitem[{\sc Hudson}(1990)]{hudson90}
{\sc Hudson, R.}, 1990 Gene genealogies and the coalescent process. Oxford
  Survey of Evolutionary Biology {\bf 7}: 1--44.

\bibitem[{\sc Hudson {\rm and} Kaplan}(1988)]{hudsonkaplan88}
{\sc Hudson, R. {\rm and} N.~Kaplan}, 1988 The coalescent process in models
  with selection and recombination. Genetics {\bf 120}: 831--840.

\bibitem[{\sc Hudson {\rm and} Kaplan}(1994)]{hudsonkaplan94}
{\sc Hudson, R. {\rm and} N.~Kaplan}, 1994 Gene trees with background
  selection. In {\em Non-neutral evolution: Theories and molecular data\/},
  edited by B.~Golding, pp. 140--153, Chapman and Hall, New York.

\bibitem[{\sc Hudson {\rm and} Kaplan}(1995)]{hudsonkaplan95}
{\sc Hudson, R. {\rm and} N.~Kaplan}, 1995 Deleterious background selection
  with recombination. Genetics {\bf 141}: 1605--1617.

\bibitem[{\sc Kaplan {\em et~al.\/}}(1988){\sc Kaplan, Darden, {\rm and}
  Hudson}]{kaplanhudson88}
{\sc Kaplan, N., T.~Darden, {\rm and} R.~Hudson}, 1988 The coalescent process
  in models with selection. Genetics {\bf 120}: 819--829.

\bibitem[{\sc Kim {\rm and} Stephan}(2002)]{kimstephan02}
{\sc Kim, Y. {\rm and} W.~Stephan}, 2002 Recent applications of diffusion
  theory to population genetics. In {\em Modern Developments in Theoretical
  Population Genetics: The Legacy of Gustave Malecot\/}, edited by M.~Slatkin
  {\rm and} M.~Veuille, Oxford University Press, Oxford, UK.

\bibitem[{\sc Kimura}(1955)]{kimura55c}
{\sc Kimura, M.}, 1955 Stochastic processes and distribution of gene
  frequencies under natural selection. Cold Spring Harbor Symposia on
  Quantitative Biology {\bf 20}: 33--53.

\bibitem[{\sc Kingman}(1982)]{kingman82a}
{\sc Kingman, J. F.~C.}, 1982 The coalescent. Stochastic Processes and their
  Applications {\bf 13}: 235--248.

\bibitem[{\sc Krone {\rm and} Neuhauser}(1997)]{kroneneuhauser97}
{\sc Krone, S.~M. {\rm and} C.~Neuhauser}, 1997 Ancestral processes with
  selection. Theoretical Population Biology {\bf 51}: 210--237.

\bibitem[{\sc McVean {\rm and} Charlesworth}(2000)]{mcveancharlesworth00}
{\sc McVean, G. A.~T. {\rm and} B.~Charlesworth}, 2000 The effects of
  hill-robertson interference between weakly selected mutations on patterns of
  molecular evolution and variation. Genetics {\bf 155}: 929--944.

\bibitem[{\sc Neuhauser {\rm and} Krone}(1997)]{neuhauserkrone97}
{\sc Neuhauser, C. {\rm and} S.~M. Krone}, 1997 The genealogy of samples in
  models with selection. Genetics {\bf 145}: 519--534.

\bibitem[{\sc Nordborg}(1997)]{nordborg97}
{\sc Nordborg, M.}, 1997 Structured coalescent processes on different
  timescales. Genetics {\bf 146}: 1501--1514.

\bibitem[{\sc O'Fallon {\em et~al.\/}}(2010){\sc O'Fallon, Seger, {\rm and}
  Adler}]{ofallonadler10}
{\sc O'Fallon, B.~D., J.~Seger, {\rm and} F.~R. Adler}, 2010 A continuous-state
  coalescent and the impact of weak selection on the structure of gene
  genealogies. Mol Biol Evol {\bf 27}: 1162--1172.

\bibitem[{\sc Przeworski {\em et~al.\/}}(1999){\sc Przeworski, Charlesworth,
  {\rm and} Wall}]{przeworski99}
{\sc Przeworski, M., B.~Charlesworth, {\rm and} J.~Wall}, 1999 Genealogies and
  weak purifying selection. Mol Biol Evol {\bf 16}: 246--252.

\bibitem[{\sc Sawyer {\rm and} Hartl}(1992)]{sawyerhartl92}
{\sc Sawyer, S.~A. {\rm and} D.~L. Hartl}, 1992 Population genetics of
  polymorphism and divergence. Genetics {\bf 132}: 1161--1176.

\bibitem[{\sc Seger {\em et~al.\/}}(2010){\sc Seger, Smith, Perry, Hunn,
  Kaliszewska, Sala, Pozzi, Rowntree, {\rm and} Adler}]{seger10}
{\sc Seger, J., W.~A. Smith, J.~J. Perry, J.~Hunn, Z.~A. Kaliszewska, L.~L.
  Sala, L.~Pozzi, V.~J. Rowntree, {\rm and} F.~R. Adler}, 2010 Gene genealogies
  strongly distorted by weakly interfering mutations in constant environments.
  Genetics {\bf 184}: 529--545.

\bibitem[{\sc Tavare}(2004)]{tavarebook}
{\sc Tavare, S.}, 2004 Ancestral inference in population genetics. In {\em
  Lectures on Probability Theory and Statistics\/}, edited by J.~Picard, volume
  1837, pp. 1--188, Springer, Berlin.

\bibitem[{\sc Wakeley}(2009)]{wakeleybook}
{\sc Wakeley, J.}, 2009 {\em Coalescent Theory, an Introduction\/}. Roberts and
  Company, Greenwood Village, CO.

\end{thebibliography}

\newpage

\clearpage

\newpage

\clearpage

\newpage

\clearpage

\newpage

\clearpage

\newpage

\begin{figure}
\includegraphics[width=3.0in]{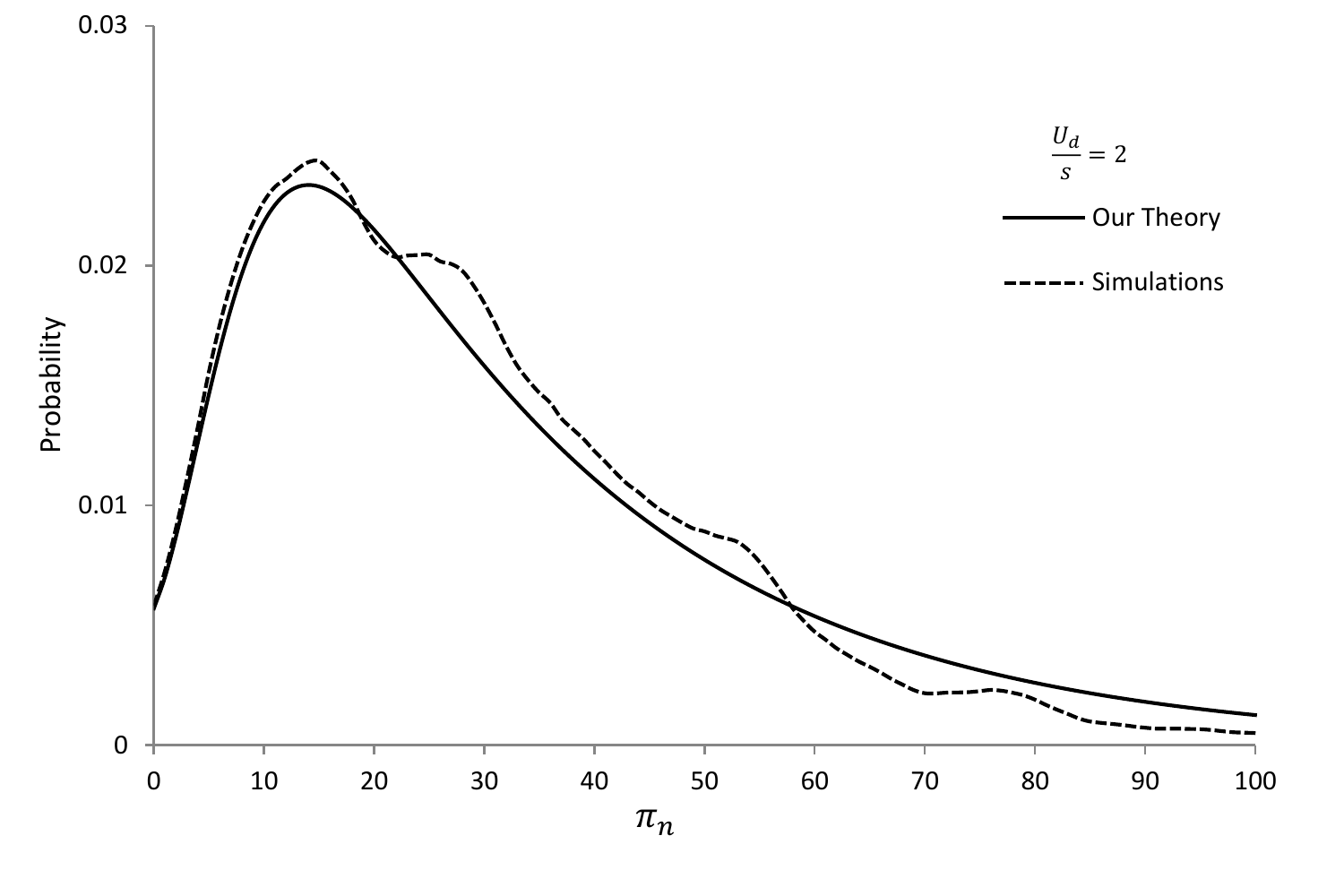}
\includegraphics[width=3.0in]{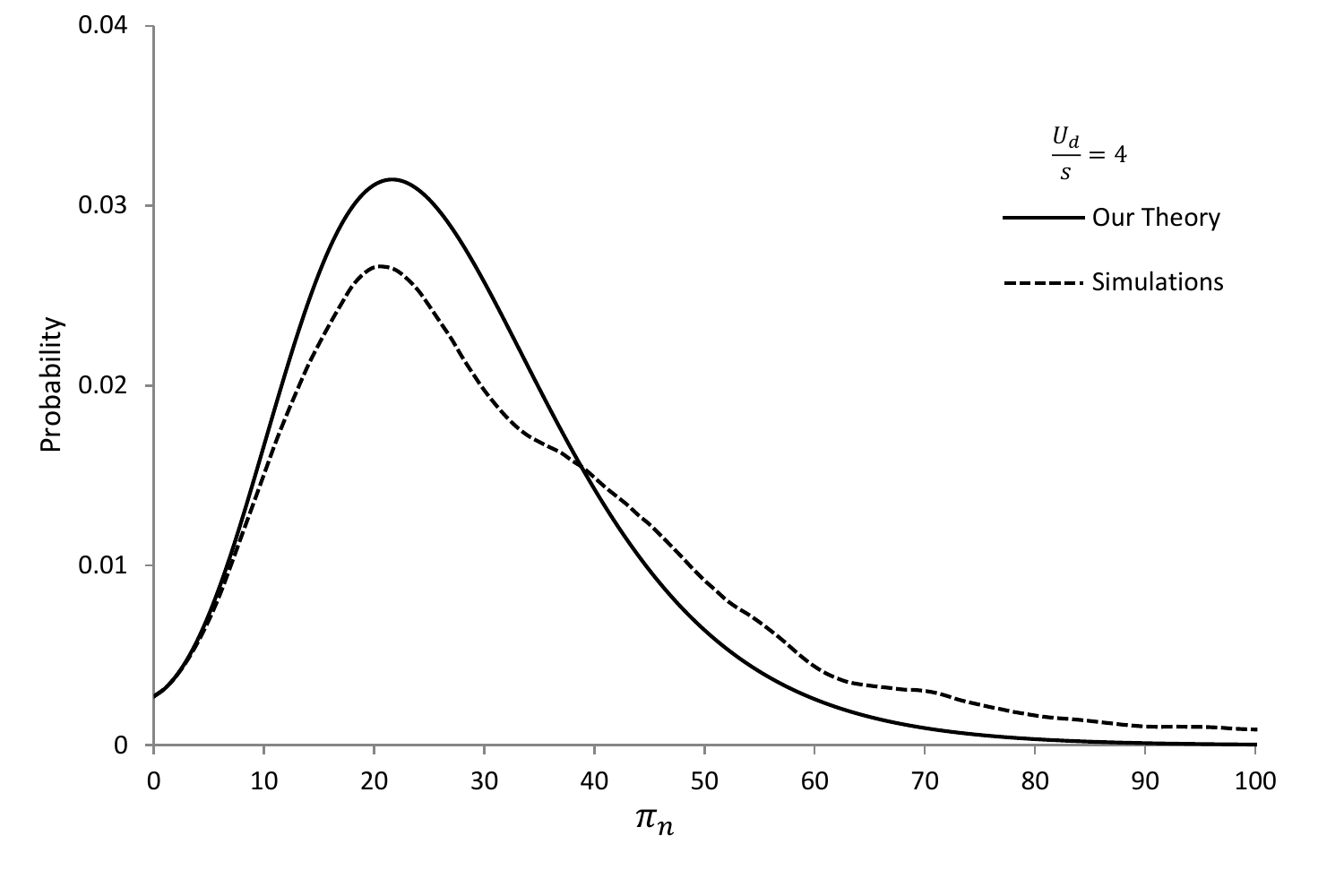}
\includegraphics[width=3.0in]{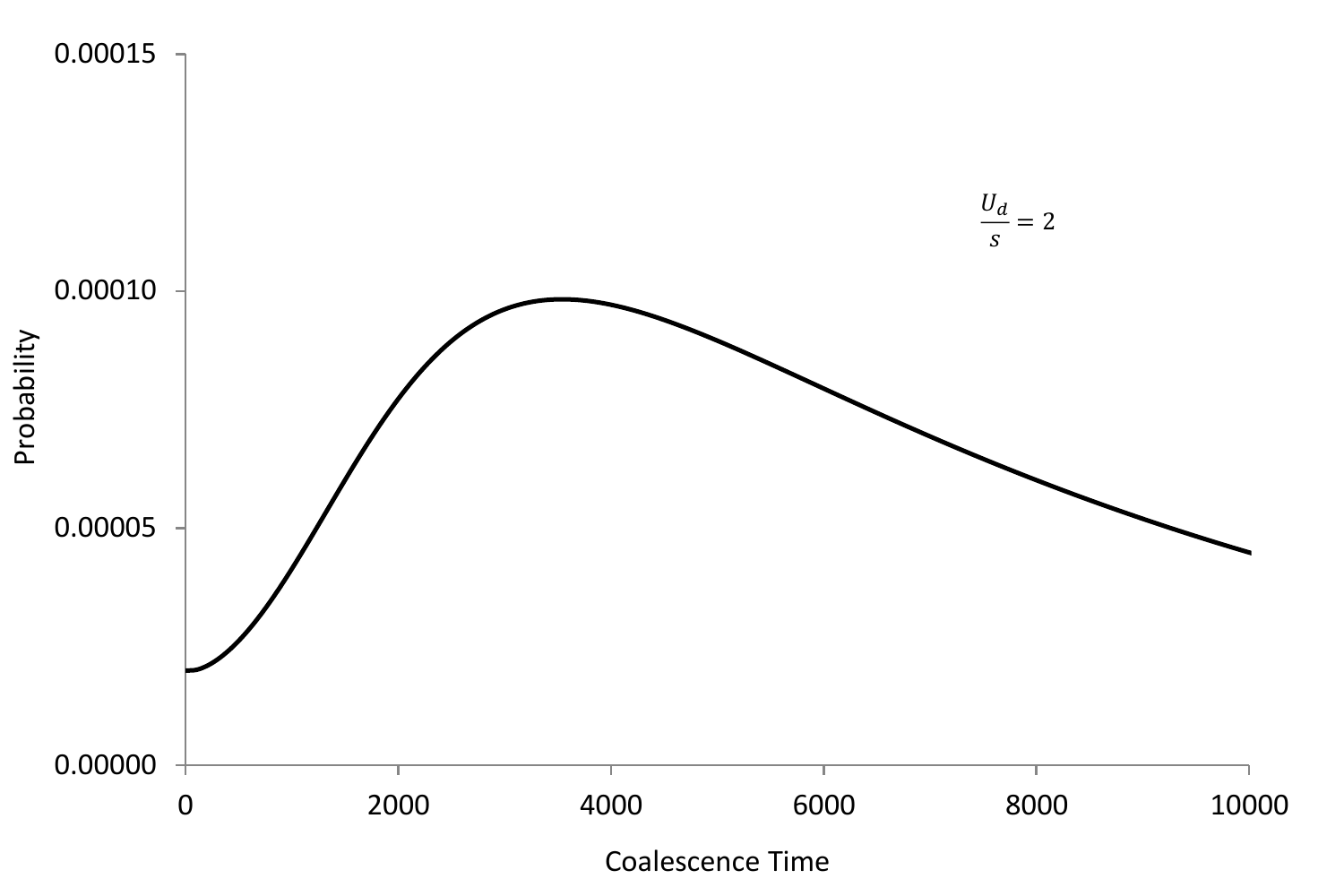}
\includegraphics[width=3.0in]{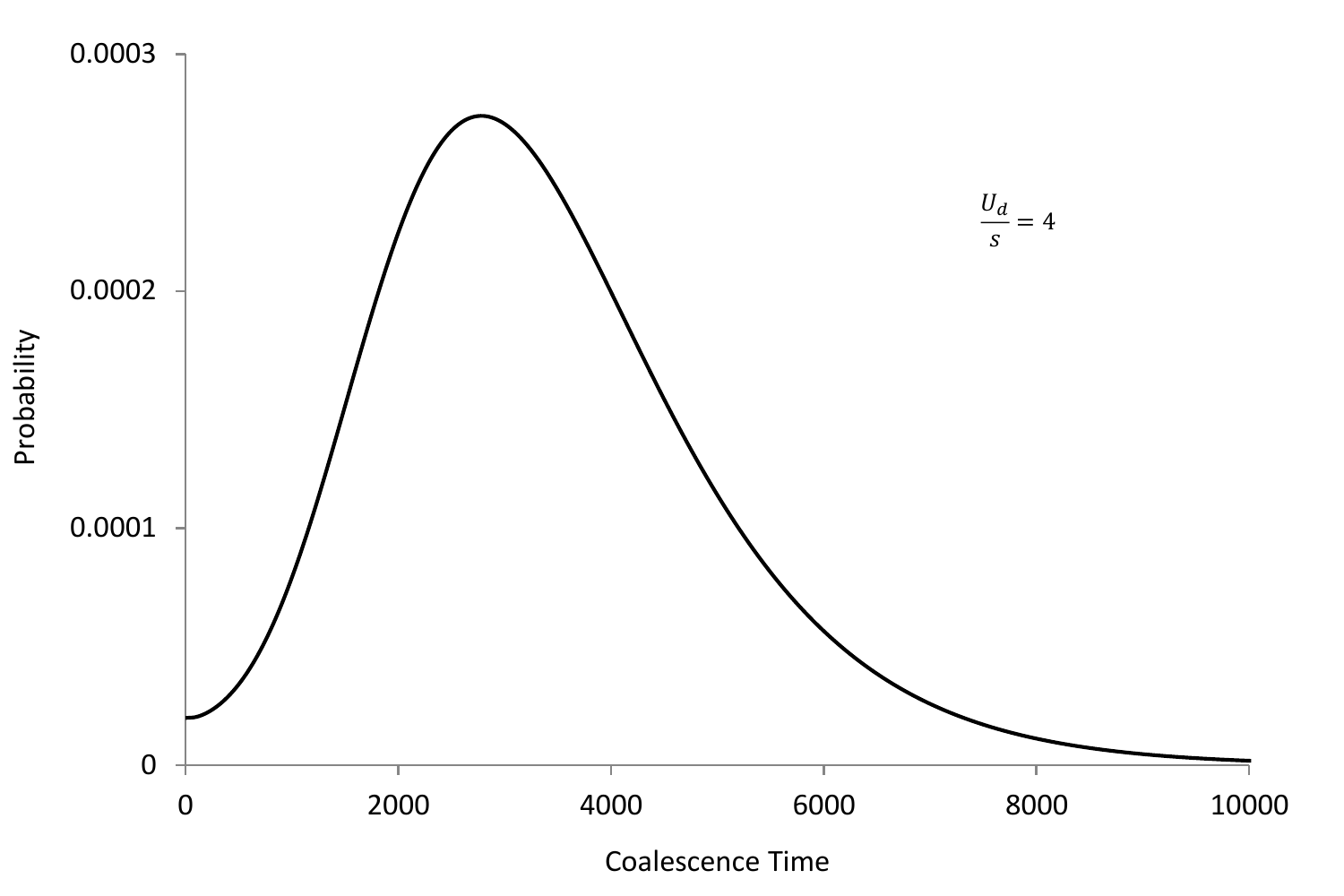}
\caption{Characteristic examples of the distributions of $\pi_n$ and the real coalescent times.  \textbf{(a)} Theoretical predictions for the distribution of $\pi_n$ for $\ud/s = 2$, compared to simulation results.  \textbf{(b)}  Theoretical predictions for the distribution of $\pi_n$ for $\ud/s=4$, compared to simulation results.  \textbf{(c)} Theoretical predictions for the distribution of real coalescence times for $\ud/s = 2$; note these simply mirror the distribution of $\pi_n$, as expected.  \textbf{(d)}  Theoretical predictions for the distribution of real coalescence times for $\ud/s=4$.  In all panels we have $N = 5 \times 10^4$ and $s = 10^{-3}$.  Our theory agrees well with the simulations, but note that, as with $\pi_d$, we tend to systematically underestimate $\pi_n$, and this tendency is worse for larger $\ud/s$.  This is due to Muller's ratchet, and as expected becomes more problematic for larger $\ud/s$.  This systematic underestimate becomes less severe (for all values of $\ud/s$) as we increase $N$, as expected, but comprehensive simulations for much larger $N$ are computationally prohibitive.  }
\label{fig7}
\end{figure}

\clearpage

\newpage

\begin{figure}
\includegraphics[width=5.0in]{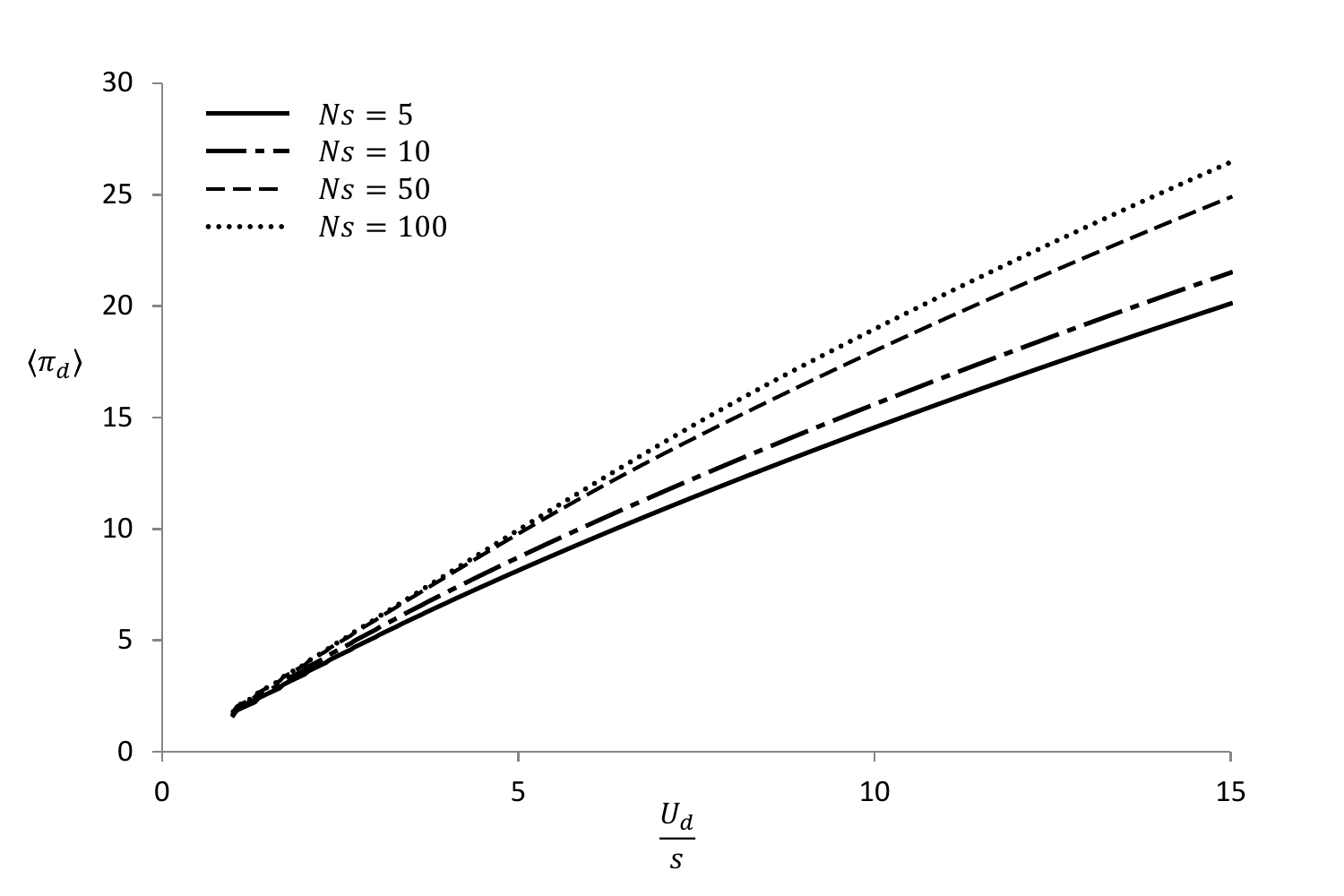}
\includegraphics[width=5.0in]{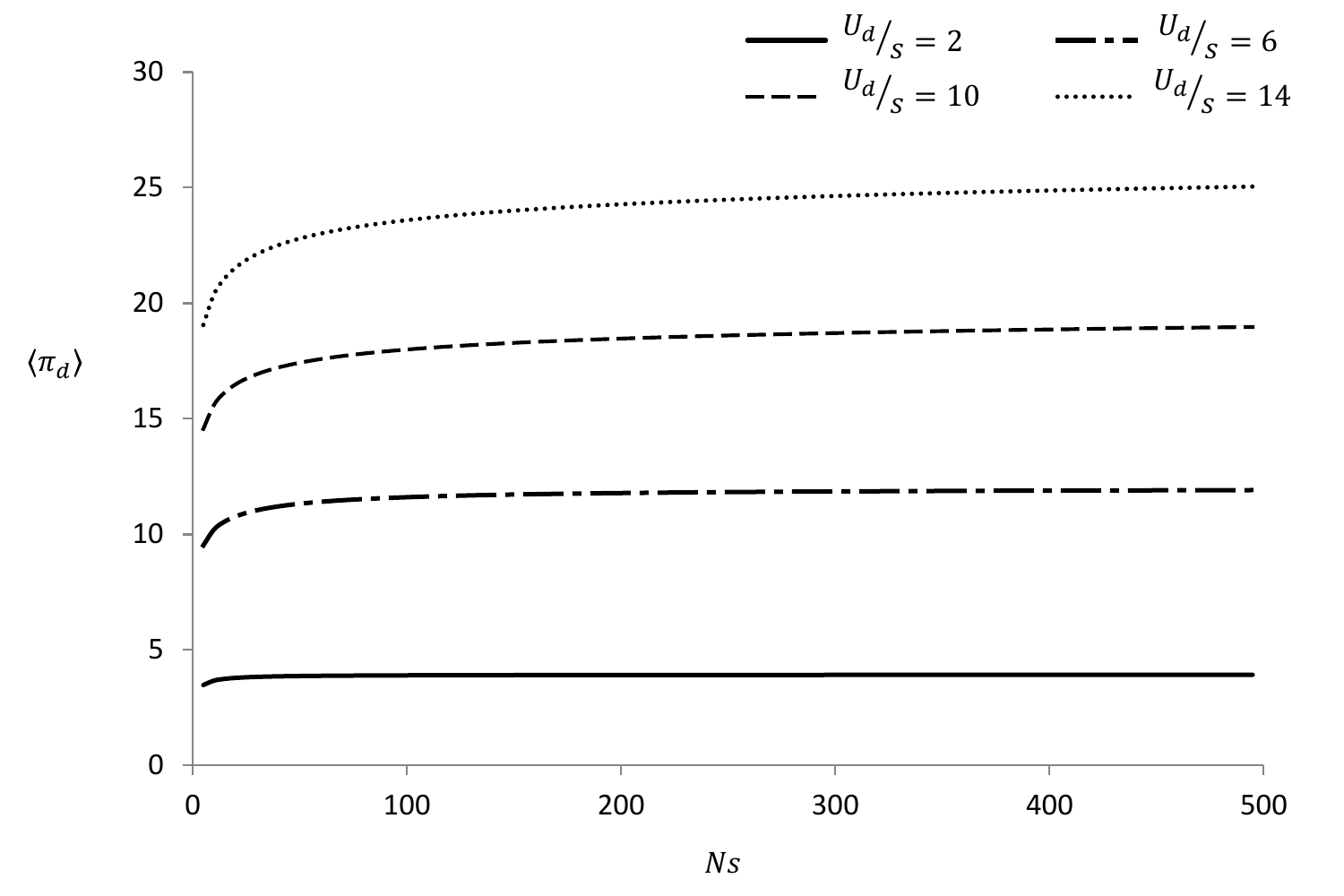}
\caption{Theoretical predictions for the mean pairwise heterozygosity at negatively selected sites, $\langle \pi_d \rangle$, as a function of the parameters.  \textbf{(a)} $\ev{\pi_d}$ as a function of $\ud/s$ for several values of $Ns$.  In the ``mutation-time'' approximation we expect this to be linear with a slope of $2$, since on average individuals are sampled from the mean class at $k = \ud/s$ and coalesce in the $0$-class, and hence have $\pi_d = 2 \ud/s$.  We see that as expected this approximation becomes more and more accurate as $Ns$ increases.  For smaller $N$, there is substantial probability of coalescence in the bulk of the fitness distribution, which is greater for larger $\ud/s$.  Thus the slope of $\ev{\pi_d}$ as a function of $\ud/s$ decreases as $Ns$ decreases, and has a downwards curvature.  \textbf{(b)} $\ev{\pi_d}$ as a function of $Ns$ for several values of $\ud/s$.  We see that as $Ns$ becomes large, $\ev{\pi_d}$ approaches $2 \ud/s$, again consistent with the mutation-time approximation.  As $Ns$ decreases, coalescence within the bulk of the fitness distribution becomes more likely, and hence $\ev{\pi_d}$ decreases.  }
\label{fig8}
\end{figure}

\clearpage

\newpage

\begin{figure}
\includegraphics[width=5.0in]{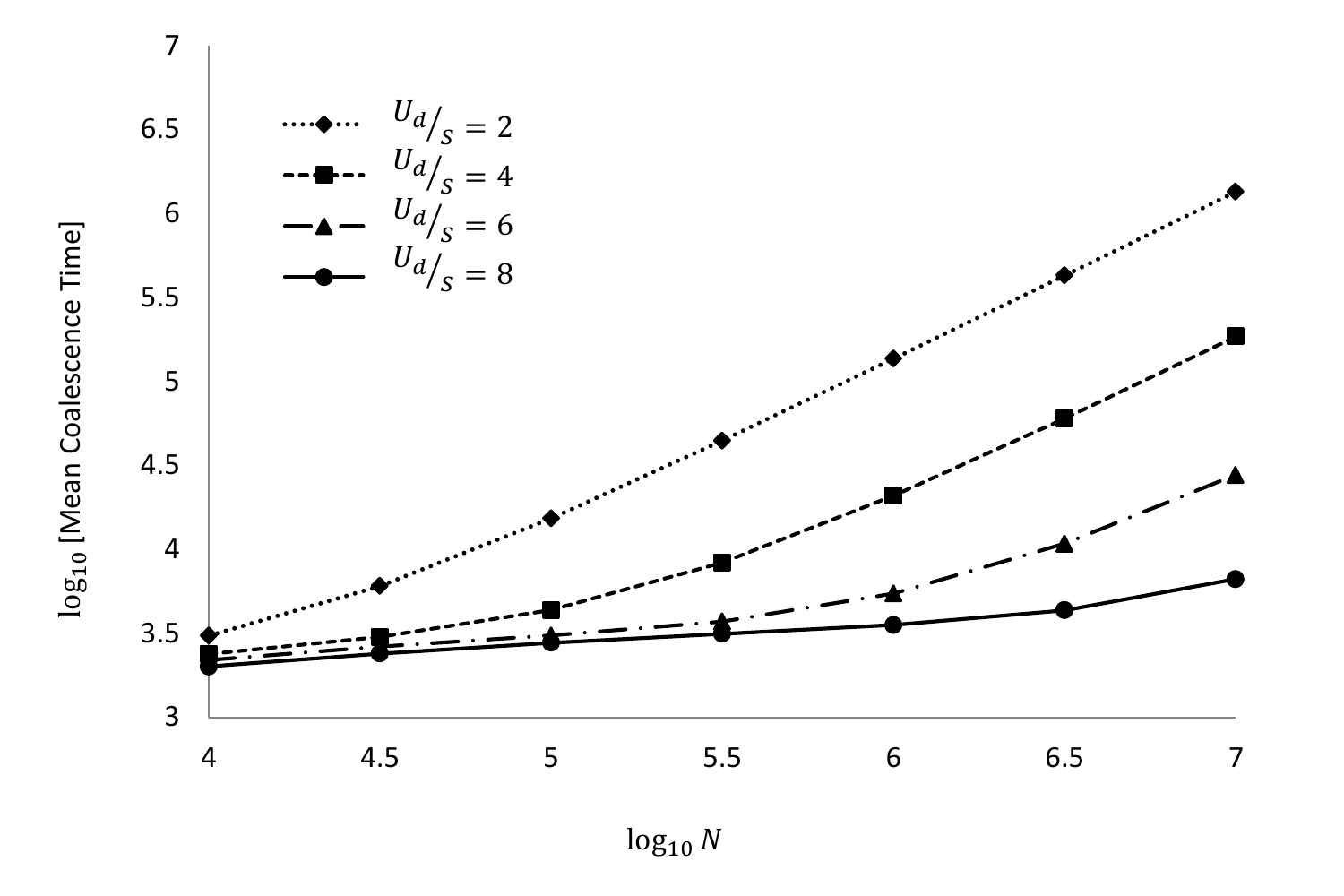}
\includegraphics[width=5.0in]{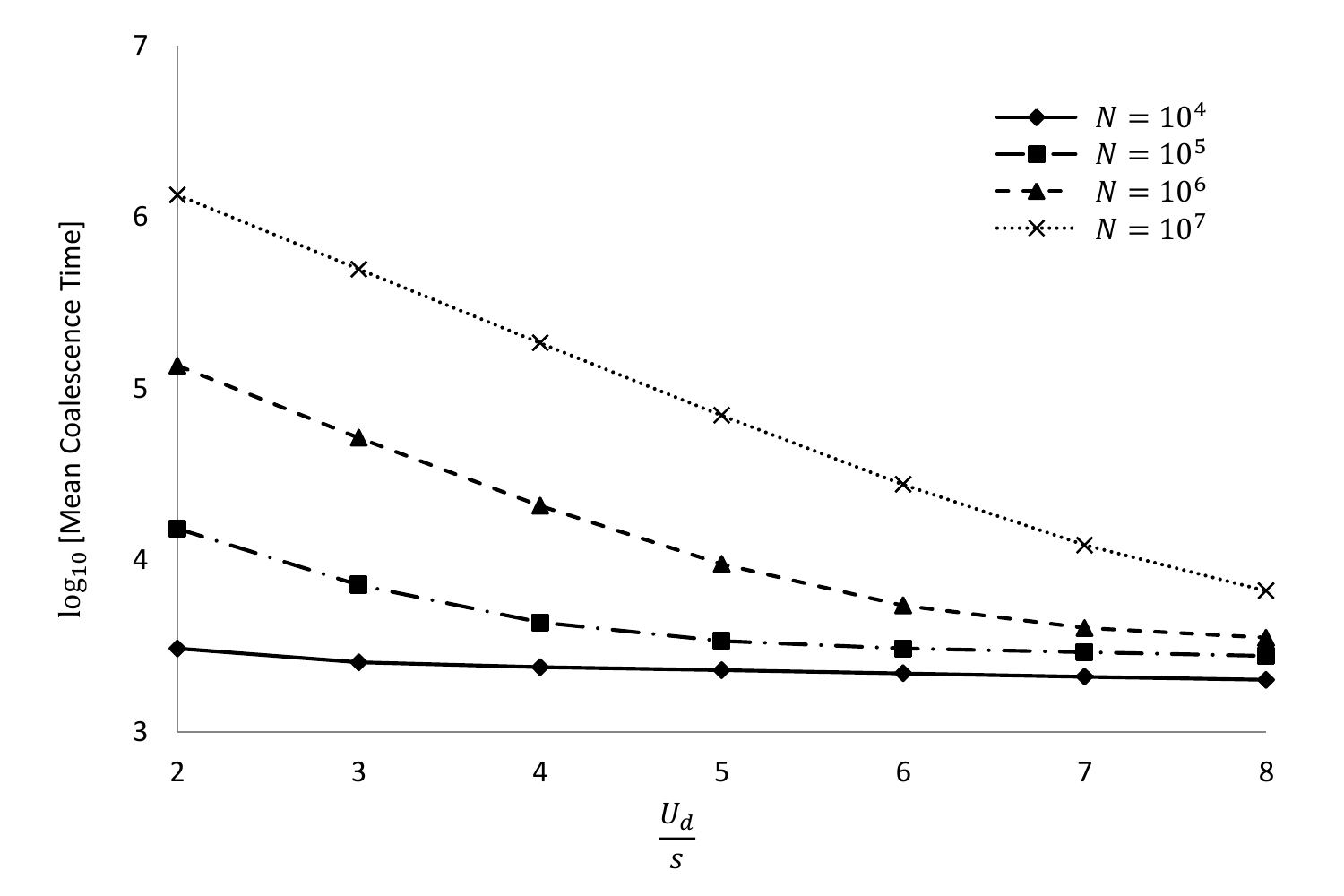}
\caption{Theoretical predictions for the mean real coalescence time $\ev{t}$.  All real coalescence times in our analysis scale linearly with $\frac{1}{s}$ (for fixed $N$ and $\ud/s$), so in this figure we fix $s = 10^{-3}$ and show the dependence of the mean pairwise heterozygosity on $N$ and on $\ud/s$.  The mean pairwise heterozygosity at neutral sites, $\ev{\pi_n}$ is simply $\ev{\pi_n} = 2 \un \ev{t}$.  \textbf{(a)} Mean coalescence time as a function of $N$ for various values of $\ud/s$.  We see that $\ev{t}$ increases slowly with $N$ until for large enough $N$ the EPS approximation applies and $\ev{t}$ becomes linear in $N$.  \textbf{(b)}  Mean coalescence time as a function of $\ud/s$ for several values of $N$.  For large $N$, the dependence is roughly linear, consistent with the EPS approximation.  For smaller $N$, coalescence can occur in the bulk of the fitness distribution, reducing the mean coalescence time.}
\label{fig9}
\end{figure}

\clearpage

\newpage

\begin{figure}
\includegraphics[scale=1]{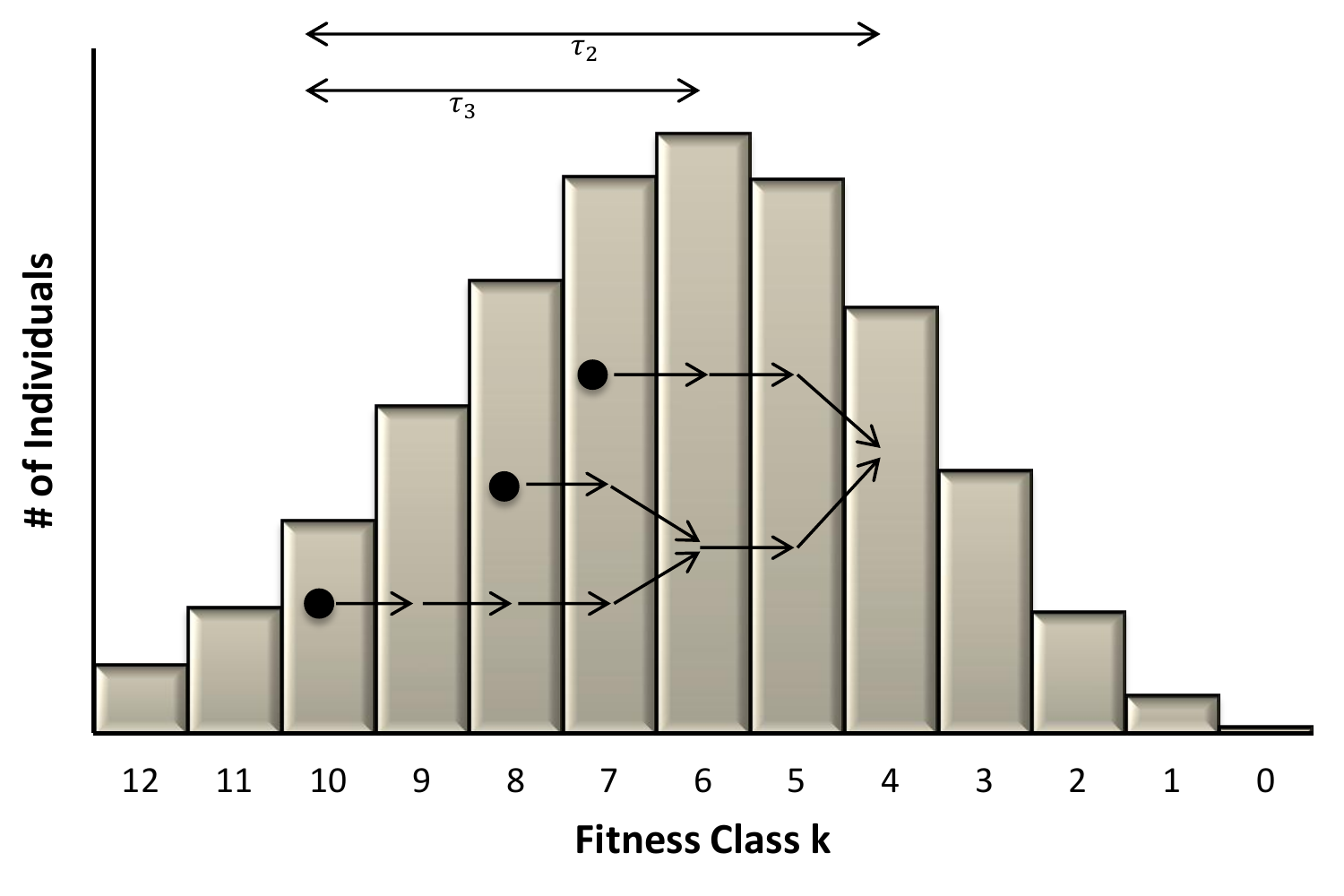}
\caption{The fitness-class coalescence process for three individuals, $A$, $B$ and $C$, where $A$ and $B$ coalesced $\tau_3$ steptimes ago and $C$ coalesced with the other two $\tau_2$ steptimes ago. }
\label{fig3}
\end{figure}

\clearpage

\newpage

\begin{figure}
\includegraphics[width=6.5in]{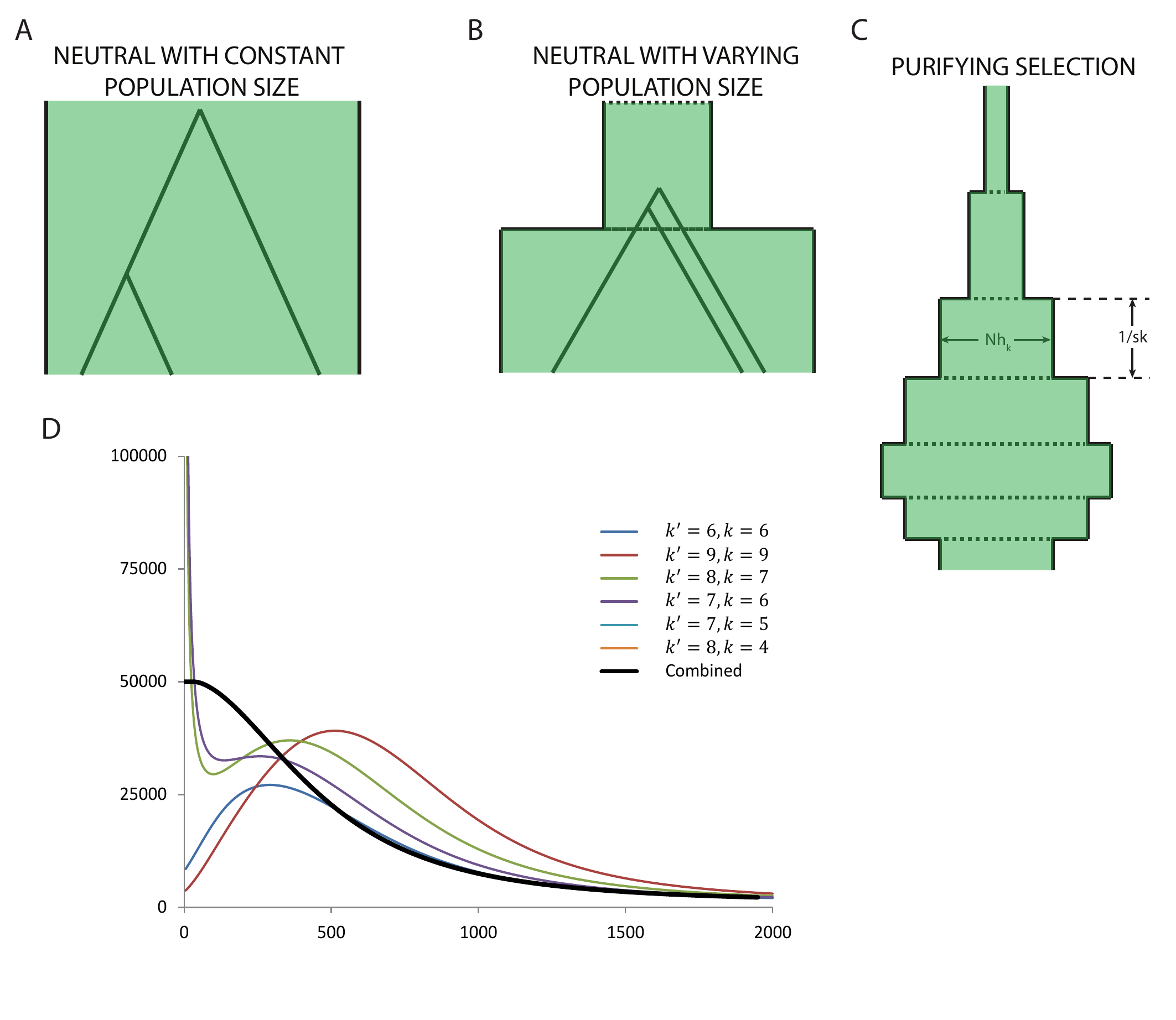}
\caption{Relationship between our results and an effective population size approximation.  \textbf{(a)} A typical coalescent tree in a neutral population of constant size.  The coalescent probability per generation between a random pair of individuals is the inverse population size.  Time runs from the past at the top to the present at the bottom. \textbf{(b)} An example of a neutral coalescent tree in a population which was smaller in the past than the present.  The population size is shown as the width in green. Coalescence events are more likely to occur when the population size is smaller.  \textbf{(c)} The effective population size history for an individual experiencing purifying selection according to our model.  The individual spends on average $\frac{1}{sk}$ generations in class $k$, which has a total size $N h_k$.  Note that pairs of individuals are sampled from different classes $k$ (i.e. they are not all sampled from the bottom of this picture).  Further, the coalescence probabilities also include a factor of $A/2$, which reflects the probability that two lineages are in the same class at the same time.  \textbf{(d)} The historically varying effective population size $N_e(t)$ for a pair of individuals sampled from classes $k$ and $k'$, as defined in the text, for several values of $k$ and $k'$.  The $N_e(t)$ for two individuals sampled at random from the whole population is also shown.  Here $N = 5 \times 10^4$, $\ud/s = 6$, and $s = 10^{-3}$.}
\label{newfig}
\end{figure}

\end{document}